\documentclass[prb,twocolumn,showpacs,amsmath,amssymb]{revtex4}

\usepackage{mathptmx}
\usepackage{graphicx}

\makeatletter
\renewcommand*{\p@subsection}{\thesection.}
\makeatother

\newcommand{\ix}[1]{\ensuremath{\text{#1}}}
\newcommand{\df}{\ensuremath{\text{d}}}
\newcommand{\im}{\ensuremath{\text{i}}}
\newcommand{\eu}{\ensuremath{\text{e}}}
\newcommand{\abs}[1]{\ensuremath{\left| #1 \right|}}
\DeclareMathOperator{\Tr}{Tr}
\DeclareMathOperator{\sign}{sign}
\DeclareMathOperator{\Real}{Re}
\DeclareMathOperator{\Imag}{Im}

\begin{document}

\title{Nonequilibrium functional RG with frequency dependent vertex
  function: A study of the single impurity Anderson model}

\author{Severin~G.~Jakobs}

\author{Mikhail~Pletyukhov}

\author{Herbert~Schoeller}

\affiliation{Institut f\"ur Theoretische Physik, RWTH Aachen
  University, and JARA--Fundamentals of Future Information
  Technology, D-52056 Aachen, Germany}

\date{12 March 2010}

\begin{abstract}
  We investigate nonequilibrium properties of the single impurity
  Anderson model by means of the functional renormalization group
  (fRG) within Keldysh formalism. We present how the level broadening
  $\Gamma/2$ can be used as flow parameter for the fRG. This choice
  preserves important aspects of the Fermi liquid behaviour that the
  model exhibits in case of particle-hole symmetry. An approximation
  scheme for the Keldysh fRG is developed which accounts for the
  frequency dependence of the two-particle vertex in a way similar but
  not equivalent to a recently published approximation to the
  equilibrium Matsubara fRG. Our method turns out to be a flexible
  tool for the study of weak to intermediate on-site interactions
  $U\lesssim 3\Gamma$. In equilibrium we find excellent agreement with
  NRG results for the linear conductance at finite gate voltage,
  magnetic field, and temperature. In nonequilibrium, our results for
  the current agree well with TD-DMRG. For the nonlinear conductance
  as function of the bias voltage, we propose reliable results at
  finite magnetic field and finite temperature. Furthermore, we
  demonstrate the exponentially small scale of the Kondo temperature
  to appear in the second order derivative of the self-energy. We show
  that the approximation is, however, not able to reproduce the
  scaling of the effective mass at large interactions.
\end{abstract}


\maketitle


\section{Introduction}
\label{sec:introduction}

Due to the enormous experimental progress in the investigation of
nanoelectronic and molecular systems, one of the central issues of
theoretical condensed matter physics is the understanding of
nonequilibrium phenomena in small strongly correlated quantum systems
coupled to an environment.  Traditionally such systems were
investigated in the context of dissipative quantum mechanics, where
energy exchange of a few-level system with a single bath of phonons
was considered.\cite{diss_QM} With the realization of various devices
involving metallic islands\cite{metallic}, quantum
dots\cite{coulomb_blockade}, single molecules\cite{molecules_exp}, or 
quantum wires\cite{semiconductor_quantum_wires,nanotubes_general}, a
tunneling coupling to leads was realized, which enabled the controlled
study of local quantum systems coupled via particle exchange to the
environment. Furthermore, in the presence of several leads, it became
possible to study such systems in the presence of a finite bias
voltage or temperature gradients (inhomogeneous boundary conditions),
i.e. a nonequilibrium and current-carrying state can be realized in
the stationary situation.

Whereas in equilibrium there are powerful numerical and analytical
techniques to describe linear transport or spectral properties of
mesoscopic systems, the available methods in nonequilibrium are quite
restrictive and still under development in the nonperturbative
regime. Exact solutions exist for some special
cases\cite{exact_general,boulat_saleur_schmitteckert_prl08,boulat_saleur_prb08}.
Scattering Bethe-Ansatz solutions have been applied to the interacting
resonant level model \cite{mehta_andrei_prl06} and the Anderson
impurity model \cite{andrei_private}, but the full understanding of
the stationary state in all regimes has not yet been reached. To
describe the steady state in the presence of a finite bias voltage,
numerical renormalization group (NRG) methods using scattering waves
\cite{anders}, time-dependent density matrix renormalization group
(TD-DMRG) \cite{heidrich-meisner}, quantum Monte Carlo (QMC) with
complex chemical potentials \cite{han}, and an approach based on
iterative summation of path-integrals (ISPI) (Ref.~\onlinecite{weiss})
have been developed, but the efficiency of these methods is still not
satisfactory in the regimes of strong Coulomb interaction and/or large
bias voltage.  To calculate the time evolution into the stationary
state, numerical techniques such as
TD-NRG,\cite{TD_NRG_general,Roosen_etal_08}
TD-DMRG,\cite{boulat_saleur_schmitteckert_prl08, heidrich-meisner, 
  TD-DMRG_general} ISPI,\cite{weiss} and QMC in
nonequilibrium\cite{Schmidt_etal_08, werner} have been used, as well
as a Fock space 
formulation of the multilayer multiconfiguration time-dependent
Hartree theory\cite{wang}. However, the description of finite bias or
the long-time limit still remains difficult.

Perturbation expansions in the coupling between the local system and
the reservoirs or in the Coulomb interaction on the local system can
not cover the most interesting regime of quantum fluctuations and
strong correlations at very low temperatures. To improve them
significantly, the most promising analytic approaches are perturbative
renormalization group (RG) methods for nonequilibrium systems. Several
methods have been proposed, which differ significantly concerning the
extent up to which nonequilibrium aspects are taken into account and
how the perturbative expansion in the renormalized vertices is set
up. One of the first formally exact nonequilibrium RG methods for
zero-dimensional quantum systems (quantum dots) was developed in
Refs.~\onlinecite{hs_koenig_PRL00, RTRG_review, keil_hs_prb01} and
is called the real-time renormalization group method (RTRG). This
technique is based on an expansion in the renormalized coupling
between quantum system and reservoir, whereas the correlations on the
local system are taken exactly into account.  Later on, this method
was formulated in pure frequency space and combined with a cutoff
procedure on the imaginary frequency axis (introduced in
Ref.~\onlinecite{jakobs2}), the so-called RTRG-FS method
\cite{hs_epj09}. Within this technique it was possible to solve
analytically the generic problem how RG flows are cut off by the
physics of transport rates. Furthermore, it was demonstrated how all
static and dynamic properties can be calculated analytically for the
nonequilibrium Kondo model in weak coupling
\cite{hs_reininghaus_prb09,schuricht_hs_prb09,pletyukhov_schuricht_hs_preprint09}
and for the interacting resonant level model (IRLM) in the scaling
limit \cite{andergassen_etal_preprint09}, which are two fundamental
models for the physics of spin and charge fluctuations, respectively.
Another perturbative RG method in nonequilibrium (called PRG-NE),
which also expands in the reservoir-system coupling, was developed in
Refs.~\onlinecite{rosch_kroha_woelfle_PRL01,
  rosch_paaske_kroha_woelfle_PRL03}, where the slave particle
approach was used in connection with Keldysh formalism and quantum
Boltzmann equations. In these works it was investigated for the first
time how the voltage and the magnetic field cuts off the RG flow for
the Kondo model and how the frequency dependence of the vertices
influences various logarithmic contributions for the susceptibility
and the nonlinear conductance.  A real-frequency cutoff was used and
the RG was formulated purely on one part of the Keldysh-contour
disregarding diagrams connecting the upper with the lower branch. This
procedure turns out to be sufficient for the Kondo model to calculate
logarithmic terms in leading order but the cutoff by relaxation and
decoherence rates was included intuitively (recently an improved
version of this method included parts of the spin relaxation and
decoherence rates via self-energy insertions
\cite{schmidt_woelfle_aop09}).  An alternative microscopic approach to
RTRG-FS for combining relaxation and decoherence rates within a
nonequilibrium RG method for the Kondo model was proposed in
Ref.~\onlinecite{kehrein_PRL05}, where flow-equation methods
\cite{flow_eq} were generalized to the nonequilibrium
situation. Within this method, it was shown for the Kondo model that
the cutoff of the RG flow by spin relaxation/decoherence rates occurs
due to a competition of 1-loop and 2-loop terms on the r.h.s.  of the
RG equation for the vertex.

Whereas all of these perturbative RG methods are very successful, they
have two important drawbacks. First, they expand all in the
reservoir-system coupling and thus can only be used reliably in the
regime of weak spin or orbital fluctuations (strong charge
fluctuations seem to be covered by RTRG-FS as was demonstrated
recently for the IRLM in
Ref.~\onlinecite{andergassen_etal_preprint09}). Secondly, the
RTRG-FS and the PRG-NE scheme work in a basis of the many-particle
eigenfunctions of the isolated local quantum system. As a consequence
larger systems such as multi-level quantum dots, large molecules, or the
crossover to the quantum wire case cannot be addressed due to an
exponentially large number of relevant many-particle
states. Therefore, another class of perturbative RG methods in
nonequilibrium have been developed which are based on an expansion in
the renormalized Coulomb interaction on the local quantum system by
combining the Keldysh formalism with a quantum-field theoretical
formulation of functional RG within the 1-particle irreducible
Wetterich scheme \cite{wetterich_morris}.  In the following we call
this approach the fRG-NE method. It has been proposed and applied to
transport through quantum dots and quantum wires in
Refs.~\onlinecite{jakobs_diplom, gezzi, jakobs2, gezzi_thesis,
  andergassen_etal_preprint09}. Similar implementations of the method
have been developed for the application to bulk systems, investigating
for instance quantum criticality in nonequilibrium\cite{mitra} or the
long-time evolution of a Bose gas starting from a nonequilibrium
state\cite{gasenzer}. Since the method is non-perturbative in the
reservoir-system coupling and is parametrized in terms of
single-particle states, it has the advantage that the case of low
Voltage and Temperature can be described and it can be applied to
multi-level quantum dots and quantum wires, at least if Coulomb
interactions are moderate.

So far, the fRG-NE scheme was used in approximation schemes with
frequency independent self-energy. This approach describes reliably
the leading effects of weak interactions in an effective
single-particle picture. At higher interactions the decay rates
generated by inelastic effects of the interaction are no longer
negligible and may drastically change the behaviour of the
system. These effects can only be described on the level of a
frequency dependent self-energy, which is the subject of the present
paper. We will address the problem of enhancing the level of the
truncation scheme used in fRG-NE, in order to enable it to describe
interaction induced decay rates. The flow equations for the vertex
functions are of such a structure that a frequency dependent
self-energy can only result from a frequency dependent two-particle
vertex. This is in turn a very complicated object, depending on four
frequency, state and Keldysh indices. Conservation laws for spin,
momentum, and frequency, and the symmetry properties of the vertex
function discussed in Ref.~\onlinecite{jakobs1} can be used to
reduce the complexity of the problem. Nevertheless, the two-particle
vertex function remains so complicated that it seems preferable to
develop the methodology first for a physical system with as few
interacting degrees of freedom as possible. This restricts at least
the dimensionality of the state dependence of the vertex
function. Insight gained at this level can later be used to tackle
larger systems.

We choose to investigate the single impurity Anderson
model\cite{anderson} (SIAM) which is considered to be a generic model
for strong local correlations\cite{hewson} and is a minimal model to
study the interplay of charge and spin fluctuations. It consists of a
single-level quantum dot with two spin states and on-site Coulomb
repulsion $U$, which is coupled to leads. Experimentally, it can be
realized for small quantum dots or molecules. In the Coulomb-blockade
regime, this model is equivalent to the Kondo model\cite{Kondo_theo}
and the unitarity limit of the Kondo effect has been
measured\cite{Kondo_exp}. The perturbation theory of this model in $U$
is completely regular\cite{yamada, yosida, yamada1, zlatic}.  However,
the low energy behaviour at large interaction $U$ is governed by the
so called Kondo scale\cite{hewson} which is exponentially small in $U$
and hence cannot be described in any finite order perturbation
theory. This makes the model an interesting object for RG
studies. Since the model can be described in equilibrium very
accurately by the numerical renormalization group (NRG)~\cite{bulla}
and even exact results for thermodynamic properties are available from
Bethe-Ansatz\cite{tsvelick} there is the possibility to
benchmark our results at least in equilibrium.

In case of thermal equilibrium the Matsubara fRG has so far been
applied to the SIAM in two distinct formulations. One of them is a
recent study based on Hubbard-Stratonovich fields representing spin
fluctuations\cite{bartosch}. The fRG-NE implementation presented in
this paper does not pursue this idea but is based on a weak coupling
expansion analogously to the second class of equilibrium fRG studies of
the SIAM\cite{karrasch1,andergassen,hedden,karrasch2}. These studies
in turn have been done in two different truncation
schemes.  One of those schemes reduces the flow of the two-particle
vertex to a renormalization of the static interaction strength and
produces a frequency independent self-energy. This approximation is
able to reproduce a Kondo scale exponentially small in $U$ which
appears e.g. in the pinning of the level to the chemical
potential\cite{karrasch1,andergassen}.  It turned out to be a quite
powerful tool for the description of static properties of diverse
quantum dot geometries even at fairly large
interactions~\cite{karrasch1}. The approximation is however unable to
describe finite frequency properties. Its restriction becomes manifest
very clearly in the shape of the spectral function: the approximation
of a static self-energy leads to a Lorentzian resonance peak. The
sharp Kondo resonance, the side bands, the suppression of the Kondo
resonance with temperature are features which cannot be described in
principle by the static approximation.

The second more elaborate approximation scheme which has been used
within the weak-coupling equilibrium fRG keeps the frequency
dependence of the two-particle vertex function and neglects the
three-particle vertex~\cite{hedden, karrasch2}. It yields
quantitatively good results for small to intermediate interaction
strengths.  While finite frequencies properties are accessible to this
frequency dependent fRG approximation for not too large interactions,
it has been observed that the large interaction asymptotics of this
refined version are worse than the ones found in the static fRG; no
Kondo scale exponentially small in $U$ has been
found~\cite{karrasch2}. Also a technical disadvantage related to the
use of the Matsubara formalism becomes apparent: while the self-energy
data are quite accurate on the imaginary frequency axis for small and
intermediate $U$, the analytic continuation to the real axis was
instable for nonzero temperatures, working only for certain parameter
sets. Therefore only observables which are accessible from the
imaginary frequency axis (without resorting to analytic continuation)
can be systematically studied at finite
temperature~\cite{karrasch2}. This finding is an additional motivation
to study a frequency dependent fRG within the Keldysh formalism, being
formulated on the real frequency axis from the outset.

Apart from the possibility to access real frequency properties at
finite temperature we envisage the opportunity to analyze the
non-equilibrium behaviour of the SIAM. This has been recently in the
focus of diverse publications. Among them is also a non-equilibrium
fRG study that is based on a real-frequency cut-off and a static
approximation scheme~\cite{gezzi, gezzi_thesis}; results known from
the Matsubara fRG could be partially reproduced by this method. It
suffers however from the violation of causality related to the choice
of the real-frequency cut-off. References~\onlinecite{hershfield, fujii}
present perturbative studies that apply to moderate interactions and
the special situation of particle-hole symmetry and vanishing magnetic
field.  The RTRG method was used in
Ref.~\onlinecite{hs_koenig_PRL00} to describe the mixed valence and
empty-orbital regime of the SIAM at finite bias. The results seem to
be reliable but it was not possible to study the Kondo regime since
essential processes describing spin fluctuations were neglected.  In
Ref.~\onlinecite{han} a time independent description of
non-equilibrium based on a density operator for the steady state is
used to make the problem accessible to QMC. The main challenge of this
method is the numerical analytic continuation of two imaginary
quantities to the real axis.  A recent improvement of nonequilibrium
QMC applied to the SIAM at zero magnetic field and zero gate
voltage\cite{werner} has given reliable results for $U < 3-5 \Gamma =
6-10 \Delta$ (in our notation $\Delta=\Gamma/2$ denotes the level
broadening, i.e. $\Gamma$ is the full width at half maximum of the
noninteracting spectral density) for the dot occupation and the
nonlinear current. The ISPI approach~\cite{weiss} provides an access
to the current through the system for moderate $U$ and not too low
temperatures. A very promising tool is the recently introduced
scattering states NRG~\cite{anders}. Unfortunately the results for the
conductance obtained by this method still bear a considerable
numerical uncertainty. Spataru et al.~\cite{spataru} generalized the
GW approximation to non-equilibrium and concentrate on the Coulomb
blockade regime of the model.  The TD-DMRG
technique~\cite{heidrich-meisner} tries to access the steady state
transport features of the model from the transient regime. Data of the
ISPI and the TD-DMRG methods for the current at moderate interactions
have been found to agree very well with our fRG results.\cite{eckel}

In contrast to many other methods, we propose in this paper a very
flexible approach which provides reliable results for moderate
interactions $U\lesssim 3\Gamma = 6\Delta$.  In equilibrium, we will
show that the linear conductance agrees very well with NRG data at
finite gate voltage, magnetic field, and temperature. In
nonequilibrium at finite bias $V$, the nonlinear current $I(V)$ is
found to agree very well with TD-DMRG data~\cite{heidrich-meisner}.
This provides the main evidence that our results can be trusted also
for the nonlinear conductance $G(V)$ at finite magnetic field $B$ and
finite temperature $T$. In particular, at $T=B=0$, we find that our
results for $G(V)$ do not show anomalous peaks as obtained within
fourth order perturbation theory~\cite{fujii}. Furthermore, we show
that even the exponentially small scale of the Kondo temperature can
be identified in the second order derivative of the self-energy and
the various Fermi-liquid relations are reproduced.  However, the
effective mass still does not contain an exponentially small scale
leading to a too large broadening of the spectral density for
interaction strengths $U>3\Gamma$, in line with the equilibrium
Matsubara fRG\cite{karrasch2}.

The paper is organized as follows. Section~\ref{sec:model} introduces
the model and its treatment within Keldysh formalism. In
section~\ref{sec:fRG} we recapitulate the core elements of the fRG for
irreducible vertex functions. A motivation and discussion of the
choice of hybridization as flow parameter follow in
section~\ref{sec:flow_parameter}. In section~\ref{sec:basic_approx} we
describe the more basic frequency independent approximation of the
flow equations and show that in the case of equilibrium and zero
temperature we reproduce exactly the flow equations known from the
Matsubara fRG. Section~\ref{sec:adv_approx} is then devoted to the fRG
in frequency dependent approximation. There are three channels
contributing to the flow of the two-particle vertex, and all three
need to be taken into account. We describe an approximation scheme
that simplifies the functional form in which the two-particle vertex
depends on frequencies. In section~\ref{sec:apprb} we demonstrate that
the fRG approach in the chosen truncation and approximation scheme
preserves the Fermi-liquid relations for the (imaginary part of the)
self-energy; furthermore we determine the fRG estimate for the
Fermi-liquid coefficients.  Section~\ref{sec:SIAM_results} finally
presents numerical results obtained from the fRG and compares them to
other methods. A conclusion is given in
section~\ref{sec:SIAM_conclusion}.  The Appendices present some mainly
technical considerations. In Appendix~\ref{sec:structure} we identify
a set of independent components which completely determine the
two-particle vertex function. The precise form of the flow equations
in terms of these components is given in
Appendices~\ref{sec:flow_eq_SE} and~\ref{sec:flow_eq_vert}.  In
Appendix~\ref{app:fermi_liquid} we recover within the fRG approach the
nonequilibrium Fermi-liquid relation known from
Ref.~\onlinecite{oguri}.


\section{SIAM and Keldysh formalism}
\label{sec:model}

The single impurity Anderson model\cite{anderson} under consideration
consists of a single electronic level with on-site repulsion, which is
coupled to two noninteracting reservoirs addressed as left (L) and
right (R). We denote the single particle states of the impurity by
$\sigma = \uparrow, \downarrow = +\frac{1}{2}, -\frac{1}{2}$ according
to the state of the electron spin. The matrix element of the on-site
two-particle interaction is
\begin{equation}
  \label{eq:2_part_int}
  \langle \sigma'_1 \sigma'_2 | v | \sigma_1 \sigma_2 \rangle
  =
  \begin{cases}
    U, & \text{if $\sigma'_1 = \sigma_1 = \overline \sigma'_2 = \overline
      \sigma_2$},
    \\
    0, & \text{else},
  \end{cases}
\end{equation}
with $U \ge 0$, where we used the notation $\overline \sigma = -
\sigma$. In standard notation of second quantization the Hamiltonian is
given by
\begin{subequations}
  \begin{align}
    \label{eq:Hamiltonian}
    H &= H_\ix{dot} + H_\ix{res} + H_\ix{coup},
    \\
    H_\ix{dot} &= \sum_{\sigma} \left(eV_\ix{g} - \sigma B -
      \frac{U}{2} \right) d^\dagger_\sigma d_\sigma + U
    d^\dagger_\uparrow d_\uparrow d^\dagger_\downarrow d_\downarrow,
    \\
    H_\ix{res} &= \sum_{r = \ix{L}, \ix{R}} H^{(r)}_\ix{res} = \sum_r
    \sum_\sigma \int \! \df k_r \, \epsilon_{k_r} c^\dagger_{k_r
      \sigma} c_{k_r \sigma},
    \\
    H_\ix{coup} &= \sum_r H^{(r)}_\ix{coup} = \sum_r \sum_\sigma \int
    \! \df k_r \, (V_{k_r} d^\dagger_\sigma c_{k_r \sigma} +
    \text{h.c.}),
  \end{align}
\end{subequations}
where the annihilators $d_\sigma, c_{k_r \sigma}$ and creators
$d^\dagger_\sigma, c^\dagger_{k_r \sigma}$ obey the usual fermionic
anti-commutation rules. The single-particle energies of the dot
\begin{equation}
  \label{eq:single_part_energy}
  \epsilon_\sigma = eV_\ix{g} - \sigma B - U/2
\end{equation}
depend on the gate voltage $V_\ix{g}$ and the magnetic field $B$ and
are shifted by $(-U/2)$ such that particle-hole symmetry is given when
$eV_\ix{g}$ equals the chemical potential.

We are interested in the stationary state which emerges a long time
after the system has been prepared in a product density matrix
\begin{equation}
  \rho(t_0) = \rho_0 = \rho_\ix{L} \otimes \rho_\ix{dot} \otimes
  \rho_\ix{R}, \quad t_0 \rightarrow - \infty.
\end{equation}
Here, $\rho_\ix{L}$ and $\rho_\ix{R}$ describe each an individual
grand-canonical equilibrium characterized by the Fermi function
\begin{equation}
  \label{eq:fermi}
  f_r(\omega) = \frac{1}{\eu^{(\omega - \mu_r)/T}+1}, \quad r =
  \ix{L}, \ix{R}.
\end{equation}
(We use units with $\hbar = 1$ and $k_\ix{B} = 1$ throughout this
paper.)  The temperature $T$ is assumed to be equal in both
reservoirs, whereas a possible difference between the two chemical
potentials accounts for a finite bias voltage,
\begin{equation}
  eV = \mu_\ix{L} - \mu_\ix{R}.
\end{equation}
We measure single-particle energies relative to the mean chemical
potential by setting
\begin{equation}
  \label{eq:mean_mu}
  \mu_\ix{L} + \mu_\ix{R} = 0.
\end{equation}

We describe the system in the framework of Keldysh
formalism\cite{langreth, lifshitz, rammer, haug}, where the
single-particle propagator between two states $q'$ and $q$ has four
components
\begin{subequations}
  \begin{align}
    G^{-|-}_{q|q'}(t|t') &= G^\ix{c}_{q|q'}(t|t') = - \im \Tr \rho_0 \mathcal{T}
    a_q(t)a_{q'}^\dagger(t')
    \\
    G^{-|+}_{q|q'}(t|t') &= G^<_{q|q'}(t|t') = \im \Tr \rho_0
    a_{q'}^\dagger(t')a_q(t)
    \\
    G^{+|-}_{q|q'}(t|t') &= G^>_{q|q'}(t|t') = - \im \Tr \rho_0
    a_q(t)a_{q'}^\dagger(t')
    \\
    G^{+|+}_{q|q'}(t|t') &= G^{\widetilde{\ix{c}}}_{q|q'}(t|t') = -
    \im \Tr \rho_0 \widetilde{\mathcal{T}} a_q(t)a_{q'}^\dagger(t')
  \end{align}
\end{subequations}
called chronologic, lesser, greater, and anti-chronologic,
respectively. Here $\mathcal{T}$ ($\widetilde{\mathcal{T}}$) denotes
the time (anti-time) ordering operator and $a_q(t)=d_\sigma(t)$ or
$c_{k_r \sigma}(t)$ is in the Heisenberg-picture at time $t$.  In the
time translational invariant stationary state $G(t|t') = G(t-t'|0)$
depends only on the difference of the two time arguments and we use
the Fourier transform
\begin{equation}
  G^{j|j'}_{q|q'}(\omega)
  =
  \int \! \df t \, \eu^{\im \omega t}  G^{j|j'}_{q|q'}(t|0),
  \quad
  j,j' = \mp.
\end{equation}
Instead of the contour basis with indices $j = \mp$ we use the Keldysh
basis\cite{keldysh} with indices $\alpha = 1,2$. The transformation to
this basis is given by
\begin{equation}
  G^{\alpha|\alpha'} = \sum_{j,j' = \mp} (D^{-1})^{\alpha|j} G^{j|j'}
  D^{j'|\alpha'},
\end{equation}
where
\begin{subequations}
  \begin{gather}
    D^{-| 1} = D^{\mp| 2} = (D^{-1})^{1|-} = (D^{-1})^{2|\mp} =
    \frac{1}{\sqrt{2}},  
    \\
    D^{+| 1} = (D^{-1})^{1|+} =  -\frac{1}{\sqrt{2}}.
  \end{gather}
\end{subequations}
The resulting components of the single particle Green function
\begin{equation}
  G^{1|1} = 0, \quad
  G^{1|2} = G^\ix{Av}, \quad
  G^{2|1} = G^\ix{Ret}, \quad
  G^{2|2} = G^\ix{K}
\end{equation}
are called advanced, retarded, and Keldysh, respectively. The
self-energy is transformed correspondingly,
\begin{equation}
  \Sigma^{\alpha'|\alpha} = \sum_{j',j = \mp} (D^{-1})^{\alpha'|j'} \Sigma^{j'|j}
  D^{j|\alpha},
\end{equation}
leading to
\begin{equation}
  \Sigma^{1|1} = \Sigma^\ix{K}, \quad
  \Sigma^{1|2} = \Sigma^\ix{Ret}, \quad
  \Sigma^{2|1} = \Sigma^\ix{Av}, \quad
  \Sigma^{2|2} = 0.
\end{equation}

The influence of the reservoirs on the dot Green function can be
described by reservoir self-energy contributions\cite{schrieffer}
\begin{equation}
  {\Sigma_\ix{res}^{(r)}}_\sigma^{\alpha'|\alpha}(\omega) = \int \! \df k_r \,
  V_{k_r} g^{\bar{\alpha'}|\bar{\alpha}}_{k_r}(\omega) V_{k_r}^\ast
\end{equation}
with $\bar 1 = 2$, $\bar 2 = 1$, and $g_{k_r}$ being the free
propagator in the reservoir state $k_r$ which is assumed to be
spin-independent. Explicitly, the individual Keldysh components are
\begin{subequations}
  \begin{align}
    {\Sigma_\ix{res}^{(r)}}_\sigma^\ix{Ret} (\omega)
    &=
    \frac{1}{2 \pi} \int \! \df \omega' \,
    \frac{\Gamma_r(\omega')}{\omega - \omega' + i \eta},
    \\
    {\Sigma_\ix{res}^{(r)}}_\sigma^\ix{Av} (\omega)
    &=
    {\Sigma_\ix{res}^{(r)}}_\sigma^\ix{Ret} (\omega)^\ast,
    \\
    {\Sigma_\ix{res}^{(r)}}_\sigma^\ix{K} (\omega)
    &=
    - \im \big[1 - 2 f_r(\omega) \big] \Gamma_r(\omega),
  \end{align}
\end{subequations}
where we made use of the hybridization function
\begin{equation}
  \Gamma_r (\omega) = 2 \pi \int \! \df {k_r} \,
  \abs{V_{k_r}}^2 \delta(\omega -
  \epsilon_{k_r}).
\end{equation}
As we are neither interested in the influence of the band structure of
the reservoirs nor in the momentum dependence of the hopping elements,
we linearize the reservoir dispersion, $\epsilon_{k_r} = w_r k_r$
(with $w_r = \df \epsilon_{k_r}/ \df k_r)$ being the inverse density
of states at the Fermi level, $k_r = k_r^\ix{F}$) and set the hopping
to momentum independent constants, $V_{k_r = k_r^\ix{F}} \equiv
V_r$. As a result, the hybridization functions are frequency
independent constants,
\begin{equation}
  \label{eq:Gamma_r_const}
  \Gamma_r(\omega)
  =
  \Gamma_r
  = 2 \pi \frac{\abs{V_r}^2}{w_r}.
\end{equation}
The results presented later apply to the special case of symmetric
coupling, $\Gamma_\ix{L} = \Gamma_\ix{R}$, which is technically
easier for the implementation of our approximations.

Defining the total hybridization $\Gamma$ via
\begin{equation}
  \label{eq:Gamma_const}
  \Gamma = \Gamma_\ix{L} + \Gamma_\ix{R},
\end{equation}
we find the reservoir-dressed noninteracting dot propagator
\begin{subequations}
  \label{eq:res_prop}
  \begin{align}
    g^\ix{Ret}_\sigma (\omega) &= \frac{1}{\omega - \epsilon_\sigma +
      \im \Gamma/2},
    \\
    g^\ix{Av}_\sigma (\omega) &= g^\ix{Ret}_\sigma (\omega)^\ast,
    \\
    g^\ix{K}_\sigma(\omega) &= \big[1 - 2 f_\ix{eff}(\omega) \big] \big[
    g^\ix{Ret}_\sigma(\omega) - g^\ix{Av}_\sigma(\omega) \big],
  \end{align}
\end{subequations}
with the effective distribution function
\begin{equation}
  \label{eq:eff_distr_func}
  f_\ix{eff}(\omega)
  =
  \sum_r \frac{\Gamma_r}{\Gamma} f_r(\omega).
\end{equation}
\begin{figure}
  \centering
  \includegraphics{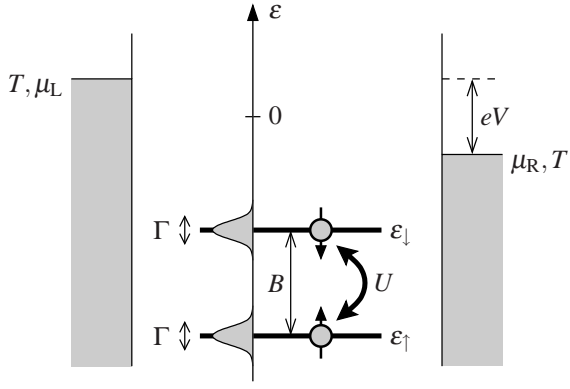}
  \caption{Sketch of the model with single particle energy in vertical
    direction.}
  \label{fig:model}
\end{figure}
A sketch of the model can be found in Fig.~\ref{fig:model}.

The current from the left reservoir through the dot to the right reservoir
is given by\cite{hershfield, meir}
\begin{equation}
  \label{eq:SIAM_current}
  I = e \frac{\Gamma_\ix{L}\Gamma_\ix{R}}{\Gamma}
  \int \! \df \omega \, \big[ f_\ix{L}(\omega) - f_\ix{R}(\omega)
  \big] \sum_\sigma \rho_\sigma(\omega)
\end{equation}
where the spectral density can be obtained from the interacting
single-particle dot Green function as
\begin{equation}
  \label{eq:spec_func}
  \rho_\sigma(\omega)
  =
  - \frac{1}{\pi} \Imag G^\ix{Ret}_\sigma(\omega).
\end{equation}
In the case $\Gamma_\ix{L} = \Gamma_\ix{R}$, which we focus on later,
the current is an odd function of the bias voltage: according to
equations~\eqref{eq:fermi}--\eqref{eq:mean_mu}, inverting the sign of
$V$ means interchanging $f_\ix{L}$ and $f_\ix{R}$. As a consequence,
the effective distribution function~\eqref{eq:eff_distr_func}, the
reservoir dressed propagator~\eqref{eq:res_prop}, the interacting
Green function, and the spectral function~\eqref{eq:spec_func} remain
unchanged, while the sign of the current~\eqref{eq:SIAM_current}
changes.

We note that the Hamiltonian is of such a type that the dot Green
function in the spin basis exhibits the special behavior under time
reversal described in section~5.5 of
Ref.~\onlinecite{jakobs1}. Therefore the Kubo-Martin-Schwinger
conditions which connect the components of the Green and vertex
functions in thermal equilibrium and which we apply later on take the
form of a generalized fluctuation dissipation theorem as stated in
section~5.6 of that reference.


\section{Vertex functions and functional renormalization group}
\label{sec:fRG}

A treatment of the problem within the fRG is set up by making the bare
propagator $g$ depend on a flow parameter $\lambda$.  Most commonly
$\lambda$ is chosen to suppress low energy degrees of freedom. Being
functionals of the bare propagator, the interacting Green and vertex
functions acquire a dependence on $\lambda$ as well, which is
described by an infinite hierarchy of coupled flow
equations.\cite{salmhofer1} The flow parameter is introduced in such a
way that two values of $\lambda$ are of particular importance: at
$\lambda = \lambda_\ix{start}$ the Green or vertex functions can be
determined exactly or in reasonable approximation; at
$\lambda=\lambda_\ix{stop}$ the free propagator takes its original
value, $g(\lambda_\ix{stop}) = g$, and so do the interacting Green or
vertex functions. Therefore the interacting Green or vertex functions
can be found by (approximately) computing their flow from
$\lambda_\ix{start}$ to $\lambda_\ix{stop}$. In this paper we focus on
the flow of the one-particle irreducible vertex
functions\cite{wetterich_morris, salmhofer2} which has proven to
provide a successful approach to the physics of diverse
low-dimensional correlated electron problems\cite{metzner, meden}.

The one-particle irreducible vertex functions can be derived from
generating functionals; details on this approach can be found for
Matsubara formalism, e.g., in Ref.~\onlinecite{negele}, and for
Keldysh formalism in Ref.~\onlinecite{gezzi}. Equivalently, the
one-particle irreducible $n$-particle vertex function
\begin{equation}
  \label{eq:vertex}
  \gamma_{q_1' \ldots q_n'|q_1 \ldots q_n}^{\alpha_1' \ldots
    \alpha_n'|\alpha_1 \ldots
    \alpha_n}(\omega_1', \ldots, \omega_n'|\omega_1, \ldots, \omega_n)
\end{equation}
can be defined diagrammatically as the sum of all one-particle
irreducible diagrams with $n$ amputated incoming lines (having states
$q_1, \ldots q_n$, Keldysh indices $\alpha_1, \ldots, \alpha_n$ and
frequencies $\omega_1, \ldots, \omega_n$) and $n$ amputated outgoing
lines (having states $q'_1, \ldots, q'_n$, Keldysh indices $\alpha'_1,
\ldots, \alpha'_n$ and frequencies $\omega'_1, \ldots, \omega'_n$).
Due to frequency conservation only $(2n-1)$ of the $2n$ frequencies
are independent. By the expression~\eqref{eq:vertex} we refer to a
function depending only on $(2n-1)$ frequency arguments, the last one
being redundant.

The vertex functions are anti-symmetric under exchange of
particles. In fact, the underlying diagrammatics is meant to be of the
Hugenholtz type~\cite{negele}, thus based on anti-symmetrized
interaction vertices
\begin{multline}
  \label{eq:bare_vertex}
  \overline v^{\alpha_1' \alpha_2'|\alpha_1 \alpha_2}_{q_1' q_2'|q_1 q_2}
  = \langle q_1' q_2'| v \big(|q_1 q_2 \rangle - |q_2 q_1 \rangle
  \big)
  \\
  \times
  \begin{cases}
    \frac{1}{2},
    & \text{if $\alpha_1 + \alpha_2 + \alpha'_1 + \alpha'_2$ is odd,}
    \\
    0,
    & \text{else},
  \end{cases}
\end{multline}
where $q, q_2$ ($q'_1, q'_2$) are incoming (outgoing) single particle
states and $\alpha_1, \alpha_2$ ($\alpha'_1, \alpha'_2$) incoming
(outgoing) Keldysh indices.

In order to evaluate a specific diagram contributing to a vertex
function, one determines the symmetry factor $S$, the number
$n_\ix{eq}$ of equivalent lines, the number $n_\ix{loop}$ of internal
loops and the permutation $P$ that describes which incoming index is
connected to which outgoing one. The value of the diagram is then
given by
\begin{equation}
  \label{eq:amp_diagram}
  \frac{(-1)^{n_\ix{loop}} (-1)^{P}}{2^{n_\ix{eq}} S}
  \left(\frac{2\pi}{\im}\right)^{n-1}
  \left[\prod \frac{\im}{2\pi} \overline v \right]
  \prod g,
\end{equation}
where one has to sum over all internal state and Keldysh indices and
to integrate over all independent internal frequencies. Details can be
found in Ref.~\onlinecite{jakobs1}.
Equation~\eqref{eq:amp_diagram} defines the prefactor of the vertex
functions in such a way, that they can be used themselves as vertices
in diagrams with the identical prefactor rules applicable to bare
$n$-particle interaction vertices.  The self-energy is equal to the
one-particle vertex, $\Sigma \equiv \gamma_{n=1}$.

A derivation of the flow equations for the vertex functions within the
generating functional approach in Keldysh formalism is given in
Ref.~\onlinecite{gezzi}. Reference~\onlinecite{jakobs2}
describes an equivalent derivation of the flow equations based on
diagrams and provides diagrammatic rules for their formulation. The
resulting flow equations form an infinite coupled hierarchy where the
flow of the $n$-particle vertex functions, $\df
\gamma_n^{\lambda}/ \df \lambda$, is a functional of
$\Sigma^{\lambda}, \gamma^{\lambda}_2, \dots,
\gamma^{\lambda}_{n+1}$. For example the flow equations for the one-
and two-particle vertex function read
\begin{equation}
  \label{eq:flow_self-energy}
  \frac{\df}{\df \lambda} \Sigma^{\lambda}_{1'|1} = -
  \frac{\im}{2 \pi} \gamma^{\lambda}_{1'2'|12}\,
  S^{\lambda}_{2|2'}
\end{equation}
and
\begin{subequations}
  \label{eq:flow_vertex}
  \begin{align}
    \label{eq:gamma3_gamma2}
    \frac{\df}{\df \lambda} \gamma^{\lambda}_{1'2'|12} = &-
    \frac{i}{2 \pi} \gamma^{\lambda}_{1'2'3'|123}\,
    S^{\lambda}_{3|3'}
    \\
    \label{eq:pp}
    & + \frac{\im}{2 \pi} 
    \gamma^{\lambda}_{1'2'|34}\,S^{\lambda}_{3|3'}\,G^{\lambda}_{4|4'}\,
    \gamma^{\lambda}_{3'4'|12}
    \\
    \label{eq:xph}
    &+ \frac{\im}{2 \pi} \gamma^{\lambda}_{1'4'|32}
    \left[S^{\lambda}_{3|3'}\,G^{\lambda}_{4|4'} +
      G^{\lambda}_{3|3'}\,S^{\lambda}_{4|4'} \right]
    \gamma^{\lambda}_{3'2'|14}
    \\
    \label{eq:dph}
    &- \frac{\im}{2 \pi} \gamma^{\lambda}_{1'3'|14}
    \left[S^{\lambda}_{3|3'}\,G^{\lambda}_{4|4'} +
      G^{\lambda}_{3|3'}\,S^{\lambda}_{4|4'} \right]
    \gamma^{\lambda}_{4'2'|32}.
  \end{align}
\end{subequations}
Here we use shorthand notation like
\begin{equation}
  \gamma^{\lambda}_{1'2'|12} \equiv
  (\gamma^\lambda_2)^{\alpha_1'\alpha_2'|\alpha_1 \alpha_2}_{q_1'
    q_2'|q_1 q_2}(\omega_1' \omega_2'|\omega_1 \omega_2),
\end{equation}
and indices occurring twice in a product implicate summation over state
and Keldysh indices and integration over independent frequencies.
Furthermore, 
\begin{equation}
  S_\lambda = G_\lambda g_\lambda^{-1} s_\lambda g_\lambda^{-1} G_\lambda 
\end{equation}
with
\begin{equation}
  s_\lambda = \frac{\df g_\lambda}{\df \lambda}
\end{equation}
denotes the so called single scale propagator. We call the
contributions~\eqref{eq:pp},~\eqref{eq:xph}, and~\eqref{eq:dph} to the
flow of $\gamma^{\lambda}_{1'2'|12}$ particle-particle, exchange
particle-hole, and direct particle-hole channel, respectively.

For practical computations the exact infinite set of flow equations is
reduced to a closed finite set of approximated flow equations. Typical
approximations are to neglect the flow of higher order vertex
functions by setting $\df \gamma_n^{\lambda} / \df \lambda \equiv 0$
for $n \ge n_0$ and to ascribe an effective parametrization to the
remaining ones. In the present investigation of the SIAM we neglect
the flow of $\gamma_3^{\lambda}$. Two different parametrizations of
$\gamma^{\lambda}_2$ will be discussed, a static (i.e. frequency
independent) and a dynamic (frequency dependent) one.


\section{Flow parameter}
\label{sec:flow_parameter}

In Ref.~\onlinecite{jakobs1} it has been shown how the choice
of the flow parameter determines whether certain exact properties of
the vertex functions are conserved by an approximated fRG flow. For
the study of the SIAM the conservation not only of causality but also
of the Kubo-Martin-Schwinger (KMS) conditions characterizing thermal
equilibrium is important. The correct description of thermal
equilibrium is necessary in order to capture fundamental aspects of
the low energy properties of the model. Consider for example the
expansion of the self-energy at $eV_\ix{g} = \mu = 0$, $B = 0$ in
leading order in frequency, temperature, and voltage,
\begin{multline}
  \Sigma^\ix{Ret}_\sigma(\omega, T, V) \simeq \frac{U}{2} + \left(1 -
    \frac{\tilde \chi_\ix{s} + \tilde \chi_\ix{c}}{2}\right) \omega
  \\
  - \im \frac{(\tilde \chi_\ix{s} - \tilde \chi_\ix{c})^2}{4\Gamma}
  \left[\omega^2 + (\pi T)^2 + \frac{3}{4}(eV)^2 \right],
  \label{eq:SIAM_SE_expansion}
\end{multline}
where $\tilde \chi_\ix{c}$ and $\tilde \chi_\ix{s}$ denote the reduced
charge and spin susceptibility~\cite{yamada, oguri}. The Fermi liquid
behavior~\ref{eq:SIAM_SE_expansion} determines the proper shape of the
spectral function, in particular its height and width. The derivation
of~\eqref{eq:SIAM_SE_expansion} is based on thermal equilibrium
particle statistics and hence requires the validity of the KMS
conditions for a treatment within Keldysh formalism.

Hybridization can be used as a flow parameter which conserves
causality and the KMS conditions~\cite{jakobs1}. For that purpose the
constants $\Gamma_{\ix{L},\ix{R}}$ and $\Gamma$ from
Eqs.~(\ref{eq:Gamma_r_const}) and (\ref{eq:Gamma_const}) are enhanced
artificially by setting
\begin{subequations}
  \begin{align}
    \label{eq:flow_parameter}
    \Gamma_\lambda &= \Gamma + \lambda,
    \\
    \Gamma_\lambda^{(r)} &=
    \Gamma_r + \frac{\Gamma_r}{\Gamma} \lambda,
    \quad
    r = \ix{L}, \ix{R},
  \end{align}
\end{subequations}
where $\lambda$ flows from $\infty$ to $0$. Via the hybridization, the
components of the noninteracting reservoir dressed propagator acquire
the $\lambda$-dependence
\begin{subequations}
  \label{eq:lambda_prop}
  \begin{align}
    (g_\lambda)^\ix{Ret}_\sigma(\omega) &= \frac{1}{\omega -
      \epsilon_\sigma + \im(\Gamma + \lambda)/2},
    \\
    (g_\lambda)^\ix{Av}_\sigma (\omega) &=
    (g_\lambda)^\ix{Ret}_\sigma(\omega)^\ast,
    \\
    (g_\lambda)^\ix{K}_\sigma(\omega) &= \big[1 - 2 f_\ix{eff}(\omega)
    \big] \left[(g_\lambda)^\ix{Ret}_\sigma(\omega) -
      (g_\lambda)^\ix{Av}_\sigma(\omega) \right],
  \end{align}
\end{subequations}
where $f_\ix{eff}^{\lambda}(\omega) \equiv f_\ix{eff}(\omega)$ does
not depend on $\lambda$ since $\Gamma_\lambda^{(r)}/\Gamma_\lambda =
\Gamma_r / \Gamma$, compare Eq.~\eqref{eq:eff_distr_func}. The
components of the free single scale propagator $s_\lambda=\df
g_\lambda / \df \lambda$ are given by
\begin{subequations}
  \begin{align}
    s_\lambda^\ix{Ret}(\omega) &= -\frac{\im}{2}
    \left[g_\lambda^\ix{Ret}(\omega)\right]^2, 
    \\
    s_\lambda^\ix{Av}(\omega) &= s_\lambda^\ix{Ret}(\omega)^\dagger
    \\
    s_\lambda^\ix{K}(\omega) &= \big[1 -  2 f_\ix{eff}(\omega) \big]
    \left[s_\lambda^\ix{Ret}(\omega) - s_\lambda^\ix{Av}(\omega)
    \right],
  \end{align}
\end{subequations}
and those of the full single scale propagator $S_\lambda = G_\lambda
g_\lambda^{-1} s_\lambda g_\lambda^{-1} G_\lambda$ by
\begin{subequations}
  \label{eq:single_scale_prop}
  \begin{align}
    \label{eq:single_scale_prop_Ret}
    S_\lambda^\ix{Ret}(\omega) =& -\frac{\im}{2}
    \left[G_\lambda^\ix{Ret}(\omega)\right]^2,
    \\
    \label{eq:single_scale_prop_Av}
    S_\lambda^\ix{Av}(\omega) =& S_\lambda^\ix{Ret}(\omega)^\dagger,
    \\
    S_\lambda^\ix{K}(\omega) =& -\frac{\im}{2}
    G_\lambda^\ix{Ret}(\omega) G_\lambda^\ix{K}(\omega) +
    \frac{\im}{2} G_\lambda^\ix{K}(\omega) G_\lambda^\ix{Av}(\omega)
    \nonumber
    \\
    & - \im \big[1 - 2 f_\ix{eff}(\omega) \big]
    G_\lambda^\ix{Ret}(\omega) G_\lambda^\ix{Av}(\omega).
    \label{eq:SK}
  \end{align}
\end{subequations}

A special situation occurs when
\begin{equation}
  \label{eq:simplification}
  G_\lambda^\ix{K}(\omega)
  =
  \big[1 - 2 f_\ix{eff}(\omega) \big] \left[G_\lambda^\ix{Ret}(\omega) -
    G_\lambda^\ix{Av}(\omega) \right].
\end{equation}
Then Eq.~\eqref{eq:SK} can be simplified to
\begin{equation}
  \label{eq:simplification_S}
  S_\lambda^\ix{K}(\omega)
  =
  \big[1 - 2 f_\ix{eff}(\omega) \big] \left[S_\lambda^\ix{Ret}(\omega) -
    S_\lambda^\ix{Av}(\omega) \right].
\end{equation}
The condition~\eqref{eq:simplification} is fulfilled for instance, due
to the fluctuation dissipation theorem, in thermal equilibrium, when
$\mu_\ix{L} = \mu_\ix{R}$ and $f_\ix{L}(\omega) = f_\ix{R}(\omega) =
f(\omega) = f_\ix{eff}(\omega)$. In non-equilibrium inelastic
interaction processes mediated by the two-particle interaction will in
general break the relation~\eqref{eq:simplification}: via
contributions to $\Sigma^\ix{K}$ and to the anti-Hermitian part of
$\Sigma^\ix{Ret}$ they tend to smoothen the effective
distribution. Later we will feed back only the static part of the
self-energy into the RG flow so that~\eqref{eq:simplification} is
fulfilled automatically, see section~\ref{subsec:feedback}.

The starting values of the vertex functions at $\lambda = \infty$ can
be determined as follows. For $\lambda \rightarrow \infty$ the
propagator $g^{\alpha|\alpha'}_\lambda$ vanishes as $1/\lambda$
whereas $\int \! \df \omega \, g^{\alpha|\alpha'}_\lambda(\omega)$
approaches a finite constant. Hence all diagrams having more internal
lines than integrations over independent frequencies vanish. For a
diagram with $m$ two-particle vertices which contributes to the
$n$-particle vertex function $\gamma_n$ the number of internal lines
is $2m-n$ while the number of integrations over independent
frequencies is $m-n+1$. Therefore only the first order contributions
to the vertex functions do not vanish for $\lambda \rightarrow
\infty$. This can also be understood in terms of time dependent
diagrammatics: Due to the diverging decay rate $\lambda \rightarrow
\infty$ only diagrams which are completely local in time do not
vanish. These are exactly the first order diagrams, namely the
Hartree-Fock diagram for the self-energy $\Sigma$ and the bare
interaction vertex for $\gamma_2$,
\begin{subequations}
  \label{eq:initial_cond_Gamma_flow}
  \begin{align}
    \gamma_n(\lambda=\infty) &= 0, \qquad n \ge 3,
    \\
    \gamma_2(\lambda=\infty) &= \overline v,
    \label{eq:gamma_2_initial_cond}
    \\
    \Sigma^{\ix{Ret},\ix{Av}}_\sigma(\lambda = \infty)
    &=
    \lim_{\lambda \rightarrow \infty} \left[- \frac{\im}{2 \pi}
      \sum_{\sigma'} \overline v_{\sigma \sigma'|\sigma \sigma'}
      \int \! \df \omega \, g^<_{\lambda \, \sigma'}(\omega) \right]
    \nonumber
    \\
    & = \frac{U}{2}
    \label{eq:initial_cond_Gamma_flow_Sigma_Ret}
    \\
    \Sigma^\ix{K}(\lambda=\infty) &= 0.
  \end{align}
\end{subequations}
In Eq.~\eqref{eq:gamma_2_initial_cond}, $\overline v$ is given
by
\begin{equation}
  \label{eq:bare_vertex_model} 
  \overline v^{\alpha'_1 \alpha'_2|\alpha_1 \alpha_2}_{\sigma'_1
    \sigma'_2|\sigma_1 \sigma_2} 
  =
  \begin{cases}
    \frac{1}{2} \overline v_{\sigma'_1 \sigma'_2|\sigma_1 \sigma_2},
    &\text{if $\alpha'_1 + \alpha'_2 + \alpha_1 + \alpha_2$ is odd},
    \\
    0,
    &\text{else},
  \end{cases}
\end{equation}
with
\begin{equation}
  \label{eq:antisym_vertex_SIAM}
  \overline v_{\sigma'_1 \sigma'_2|\sigma_1 \sigma_2} 
  =
  \begin{cases}
    U,
    &\text{if $\sigma'_1 = \sigma_1 = \overline \sigma'_2 = \overline \sigma_2$},
    \\
    -U,
    &\text{if $\sigma'_1 = \overline \sigma_1 = \overline \sigma'_2 = \sigma_2$},
    \\
    0
    &\text{else},
  \end{cases}
\end{equation}
compare~\eqref{eq:bare_vertex} and~\eqref{eq:2_part_int}. The integral
in~\eqref{eq:initial_cond_Gamma_flow_Sigma_Ret} has been evaluated by
\begin{multline}
  -\frac{\im}{2 \pi} \int \! \df \omega \,
  g^<_{\lambda \, \sigma'}(\omega) = \frac{1}{\pi} \int \! \df x \,
  \frac{f_\ix{eff}(\epsilon_{\sigma'} + x \Gamma_\lambda/2)}{x^2 + 1}
  \\
  \xrightarrow{\lambda \rightarrow \infty} \frac{1}{\pi} \int_{-
    \infty}^0 \! \df x \, \frac{1}{x^2 + 1} =\frac{1}{2},
  \label{eq:half_filled}
\end{multline}
where $x = 2 (\omega - \epsilon_{\sigma'}) / \Gamma_\lambda$. The
physical interpretation of~\eqref{eq:half_filled} is that the mean
occupation of an infinitely broadened level is $1/2$.

Hybridization as flow parameter has the advantage to conserve
causality and the KMS conditions (the latter being vio\-la\-ted for
example by the imaginary frequency cut-off introduced
in~\onlinecite{jakobs2}) and to be applicable to zero-dimensional
systems (as opposed to the momentum cut-off). Since it changes the
free propagator in a smooth way, $S_\lambda$ is not restricted sharply
to a single scale by a delta function, as it happens for step-function
cut-offs, e.g. the momentum cut-off. Such a delta function simplifies
the flow equations since it cancels one of the necessary
integrations. The hybridization-flow thus has the disadvantage of an
extra integration in the flow equations compared to step function
cut-offs.


\section{FRG in static approximation}
\label{sec:basic_approx}

Before discussing the more elaborate frequency dependent truncation
scheme, we describe a basic static approximation to the flow: the flow
of the $n$-particle vertex functions for $n \ge 3$ is neglected and
the flow of the two-particle vertex function is reduced to the flow of
a frequency-independent effective interaction strength.  As a
consequence the self-energy remains frequency independent which means
that the influence of the interaction on single-particle properties of
the system is described by a mere shift of the single-particle
levels. This approximation has been studied at $T=0$ within a
fRG treatment based on Matsubara formalism using an
imaginary frequency cut-off~\cite{karrasch1, andergassen}.
There it was shown to produce a Kondo scale exponentially small in
$U/\Gamma$ which occurs in the pinning of the renormalized level
position to the chemical potential.  Results for the dependence of the
linear conductance on the gate voltage for different magnetic fields
have been found to be in very good agreement with Bethe-Ansatz and
numerical renormalization group computations. It turns out that in
thermal equilibrium at $T=0$ the hybridization flow produces the
identical flow equations.  Hence this Keldysh approach incorporates
all features found in Ref.~\onlinecite{andergassen} for the
Matsubara fRG.

Apart from its success in describing the linear conductance the
applicability of the approximation in question is restricted.  Given a
frequency independent self-energy, the spectral density
$\rho_\sigma(\omega)$ is a Lorentzian that is centered at the
renormalized level position $\tilde \epsilon_\sigma$ and has fixed
width and height,
\begin{equation}
  \rho_\sigma(\omega) = \frac{1}{\pi} \frac{\Gamma / 2}{(\omega
      - \tilde \epsilon_\sigma)^2 + \Gamma^2/4}.
\end{equation}
Therefore the approximation cannot describe any features connected to
details of the spectral function such as the formation of a Kondo
resonance with side bands and its dependence on temperature or
voltage. That's why we do not elaborate in detail on this
approximation but merely show that at zero temperature and in
equilibrium the resulting flow equation are identical to those of
Ref.~\onlinecite{andergassen}.

Reducing the two-particle vertex function to a frequency independent
effective interaction means setting
\begin{equation}
  \label{eq:static_vertex}  
  (\gamma^\lambda_2)^{\alpha'_1 \alpha'_2|\alpha_1 \alpha_2}_{\sigma'_1
    \sigma'_2|\sigma_1 \sigma_2} (\omega'_1 \omega'_2|\omega_1
  \omega_2)
  =
  (\overline v_\lambda)^{\alpha'_1 \alpha'_2|\alpha_1 \alpha_2}_{\sigma'_1
    \sigma'_2|\sigma_1 \sigma_2},
\end{equation}
where $\overline v_\lambda$ is obtained from $\overline v$ by
replacing the bare interaction $U$ by a renormalized one $U_\lambda$
in~\eqref{eq:antisym_vertex_SIAM}.  Here the effective interaction
$U_\lambda$ is a real number whose flow starts at
\begin{equation}
  \label{eq:interaction_init_cond}
  U_{\lambda=\infty} = U.
\end{equation}
As a consequence of that approximation to the vertex function, the
flowing self-energy remains frequency independent and real, so that we
can speak of a flowing effective level position
$\epsilon_\sigma^{\lambda} = \epsilon_\sigma +
\Sigma^\ix{Ret}_{\lambda \, \sigma}$. The initial condition for the
  level position is according to
  Eqs.~\eqref{eq:single_part_energy}
  and~\eqref{eq:initial_cond_Gamma_flow_Sigma_Ret}
\begin{equation}
  \label{eq:level_init_cond}
  \epsilon_\sigma^{\lambda=\infty} 
  =
  \epsilon_\sigma + \Sigma^\ix{Ret}_\sigma(\lambda=\infty)
  =
  eV_\ix{g} - \sigma B.
\end{equation}
The starting values~\eqref{eq:interaction_init_cond}
and~\eqref{eq:level_init_cond} are identical to those of
Ref.~\onlinecite{andergassen}. Note that in the framework used there
the initial condition for the level position is the result of a first
stage of flow from $\lambda = \infty$ to $\lambda = \lambda_0
\rightarrow \infty$, see e.g. Ref.~\onlinecite{karrasch1}.

From Eq.~\eqref{eq:flow_self-energy} we derive the flow equation
for the level position
\begin{align}
  \frac{\df \epsilon_\sigma^{\lambda}}{\df \lambda} =& - \frac{\im}{2
    \pi} \frac{U_\lambda}{2} \int \! \df \omega \, S^\ix{K}_{\lambda
    \, \overline \sigma}(\omega) \nonumber
  \\
  =& - \frac{U_\lambda}{8 \pi} \int \! \df \omega \, \sign(\omega)
  \Big\{\big[G^\ix{Ret}_{\lambda \, \overline \sigma}(\omega) \big]^2
  + \big[G^\ix{Av}_{\lambda \, \overline
    \sigma}(\omega) \big]^2 \Big\},
  \label{eq:static_level_flow}
\end{align}
where we used Eq.~\eqref{eq:simplification_S} with 
\begin{equation}
  \label{eq:fermi_sign}
  1 - 2 f_\ix{eff}(\omega)
  =
  1 - 2 f(\omega)
  =
  \sign(\omega)
  \quad
  \text{for}
  \quad T = 0, \; \mu=0.
\end{equation}
Inserting
\begin{equation}
  \label{eq:GRet}
  G_\lambda^\ix{Ret}(\omega)
  =
  \frac{1}{\omega - \epsilon^{\lambda} + \im
    (\Gamma + \lambda)/2}
  =
  G_\lambda^\ix{Av}(\omega)^\dagger
\end{equation}
we can evaluate the integral in Eq.~\eqref{eq:static_level_flow}
and obtain
\begin{equation}
  \label{eq:level_flow}
  \frac{\df \epsilon_\sigma^{\lambda}}{\df \lambda}
  =
  \frac{U_\lambda}{2 \pi}
  \frac{\epsilon_{\overline \sigma}^{\lambda}}{{\epsilon_{\overline
        \sigma}^{\lambda}}^2  +  (\Gamma + \lambda)^2/4}.
\end{equation}

The flow equation~\eqref{eq:flow_vertex} for $\gamma_2$ generates a
frequency dependent vertex function which has a richer structure in
terms of Keldysh and spin indices than indicated in
Eq.~\eqref{eq:static_vertex}. It can be shown that after
neglecting the contribution from the three-particle vertex
function~\eqref{eq:gamma3_gamma2} this structure can be reduced
to the simple form given in Eq.~\eqref{eq:static_vertex} by
setting all outer frequency arguments equal to the chemical potential
$\mu$. We note that this projection onto the Fermi surface is the
  core idea of the static approximation.  The resulting flow equation
acquires the form
\begin{multline}
  \frac{\df U_\lambda}{\df \lambda} = \frac{U_\lambda^2}{2}
  \Big[ (I^\ix{\,pp}_\lambda)_{\uparrow \downarrow}^{22|12}(0)
  + (I^\ix{\,pp}_\lambda)_{\uparrow \downarrow}^{22|21}(0)
  \\
  + (I^\ix{\,ph}_\lambda)_{\uparrow \downarrow}^{21|22}(0) +
  (I^\ix{\,ph}_\lambda)_{\uparrow \downarrow}^{22|12}(0) \Big],
\end{multline}
where we abbreviated frequency integrals appearing in the
particle-particle and particle-hole channel by
\begin{widetext}
  \begin{subequations}
    \label{eq:channel_integrals}
    \begin{align}
      (I^\ix{\,pp}_\lambda)_{\sigma_1 \sigma_2}^{\alpha_1
        \alpha_2|\alpha'_1 \alpha'_2}(\omega) &= \frac{\im}{2 \pi}
      \int \! \df \omega' \Big[
      G^{\alpha_1|\alpha'_1}_{\lambda \, \sigma_1}(\tfrac{\omega}{2} +
      \omega')
      S^{\alpha_2|\alpha'_2}_{\lambda \, \sigma_2}(\tfrac{\omega}{2} -
      \omega') +
      S^{\alpha_1|\alpha'_1}_{\lambda \, \sigma_1}(\tfrac{\omega}{2} +
      \omega')
      G^{\alpha_2|\alpha'_2}_{\lambda \, \sigma_2}(\tfrac{\omega}{2} -
      \omega') \Big],
      \\
      (I^\ix{\,ph}_\lambda)_{\sigma_1 \sigma_2}^{\alpha_1
        \alpha_2|\alpha'_1 \alpha'_2}(\omega) &= \frac{\im}{2 \pi}
      \int \! \df \omega' \Big[
      G^{\alpha_1|\alpha'_1}_{\lambda \, \sigma_1}(\omega' -
      \tfrac{\omega}{2})
      S^{\alpha_2|\alpha'_2}_{\lambda \, \sigma_2}(\omega' +
      \tfrac{\omega}{2}) +
      S^{\alpha_1|\alpha'_1}_{\lambda \, \sigma_1}(\omega' -
      \tfrac{\omega}{2})
      G^{\alpha_2|\alpha'_2}_{\lambda \, \sigma_2}(\omega' +
      \tfrac{\omega}{2}) \Big].
    \end{align}
  \end{subequations}
\end{widetext}
Making use of Eqs.~\eqref{eq:single_scale_prop_Ret},
\eqref{eq:single_scale_prop_Av}, \eqref{eq:simplification},
\eqref{eq:simplification_S}, \eqref{eq:fermi_sign} and~\eqref{eq:GRet}
we evaluate the integrals and obtain
\begin{equation}
  \label{eq:interaction_flow}
  \frac{\df U_\lambda}{\df \lambda}
  =
  \frac{U_\lambda^2}{\pi}
  \,
  \frac{\epsilon_\uparrow^{\lambda}}{{\epsilon_\uparrow^{\lambda}}^2 +
    (\Gamma + \lambda)^2/4}
  \,
  \frac{\epsilon_\downarrow^{\lambda}}{{\epsilon_\downarrow^{\lambda}}^2 +
    (\Gamma + \lambda)^2/4}.
\end{equation}
The substitution $\lambda \rightarrow 2\lambda$ maps the flow
equations~\eqref{eq:level_flow} and~\eqref{eq:interaction_flow} onto
those of Ref.~\onlinecite{andergassen}. We note in passing, that the
static approximation scheme can be carried out as well with the
imaginary frequency cut-off from Ref.~\onlinecite{jakobs2} instead
of the hybridization flow parameter; the special advantage of the
hybridization flow to conserve the KMS relations becomes relevant only
in the dynamic approximation scheme introduced in
section~\ref{sec:adv_approx}. The flow equations of the static scheme
based on the imaginary frequency cut-off are again identical to
Eqs.~\eqref{eq:level_flow} and~\eqref{eq:interaction_flow} and to
the flow equations of Ref.~\onlinecite{andergassen}. The real
frequency cut-off used in Ref.~\onlinecite{gezzi} on the other hand
produces different flow equations which nevertheless lead again to the
same final solution\cite{gezzi}.

It is easy to generalize the static approximation scheme to
non-equilibrium. Since the Keldysh component of the self-energy is not
renormalized we merely need to replace~\eqref{eq:fermi_sign} by
\begin{equation}
  1 - 2 f_\ix{eff}(\omega) =  \sum_r
  \frac{\Gamma^{(r)}}{\Gamma} \sign(\omega - \mu_r),
\end{equation}
which leads to a superposition of the flow equations found for thermal
equilibrium in the form
\begin{subequations}
  \begin{align}
    \frac{\df \epsilon^{\lambda}_\sigma}{\df \lambda} &=
    \frac{U_\lambda}{2 \pi} \sum_r \frac{\Gamma_r}{\Gamma}
    \frac{\epsilon^{\lambda}_{\overline \sigma} -
      \mu_r}{(\epsilon^{\lambda}_{\overline \sigma} - \mu_r)^2 +
      (\Gamma + \lambda)^2/4},
    \\
    \frac{\df U_\lambda}{\df \lambda} &=
    \frac{U_\lambda^2}{\pi} \sum_r \frac{\Gamma_r}{\Gamma}
    \frac{(\epsilon_\uparrow^{\lambda} -
      \mu_r)}{\big(\epsilon_\uparrow^{\lambda} - \mu_r \big)^2 +
      (\Gamma + \lambda)^2/4}
    \nonumber
    \\
    & \hspace{6em}
    \times
    \frac{(\epsilon_{\downarrow}^{\lambda} -
      \mu_r)}{\big(\epsilon_{\downarrow}^{\lambda} - \mu _r\big)^2 +
      (\Gamma + \lambda)^2/4}. 
  \end{align}
\end{subequations}
This approach, however, does not provide a reliable description of the
influence of bias voltage, since it restricts the spectral function by
construction to a single Lorentzian peak; a splitting or
damping of the peak is principally impossible.


\section{FRG in dynamic approximation}
\label{sec:adv_approx}

In order to overcome the restrictions of the approximation described
in section~\ref{sec:basic_approx}, a frequency dependent self-energy
is required. The flow equation~\eqref{eq:flow_self-energy} produces a
frequency dependent self-energy only if the two-particle vertex
function is frequency dependent. Therefore our aim is to extend the
scheme described in section~\ref{sec:basic_approx} in a way that a
frequency dependent two-particle vertex function is generated. We seek
for an extension which is as basic as possible; in particular we
neglect the influence of the three-particle vertex function further
on, such that the flow of the two-particle vertex function is induced
by the three contributions~(\ref{eq:pp}--\ref{eq:dph}).


\subsection{Ladder approximations}
\label{subsec:ladders}

Recent investigations of the SIAM by diagrammatic techniques have
revealed the importance of the exchange particle-hole ladder (RPA
series) for the emergence of Kondo physics\cite{logan, janis}. Let us
therefore in a first step neglect all contributions to the flow of the
two-particle vertex function except for the exchange particle-hole
channel~\eqref{eq:xph}. If we suppress additionally the feedback of
the self energy into the propagators except for the initial Hartree
term~\eqref{eq:initial_cond_Gamma_flow_Sigma_Ret} then the flow
equation for the two-particle vertex functions reads
\begin{align}
  \frac{\df}{\df \lambda}
  (\gamma^\ix{\,x}_\lambda)_{1'2'|12} =& \frac{\im}{2 \pi}
  (\gamma^\ix{\,x}_\lambda)_{1'4'|32}
  \left[s^{\lambda}_{3|3'}\,g^{\lambda}_{4|4'} +
    g^{\lambda}_{3|3'}\,s^{\lambda}_{4|4'} \right]
  \nonumber
  \\ & \hspace{12em}
  \times
  (\gamma^\ix{\,x}_\lambda)_{3'2'|14}
  \nonumber
  \\
  =& \frac{\im}{2 \pi} (\gamma^\ix{\,x}_\lambda)_{1'4'|32}
  \frac{\df g^{\lambda}_{3|3'}\,g^{\lambda}_{4|4'}}{\df \lambda}
  (\gamma^\ix{x}_\lambda)_{3'2'|14},
  \label{eq:xph_flow}
\end{align}
where $g^{\lambda}$ is obtained from Eq.~\eqref{eq:lambda_prop}
by replacing $\epsilon_\sigma$ with $\epsilon_\sigma + U/2$. We
denoted the vertex function in this approximation with a superscript
``$\ix{x}$'' which refers to ``exchange''. The solution of the flow
equation~\eqref{eq:xph_flow} with the initial condition
$\gamma^\ix{\,x}_{\lambda=\infty}=\overline v$ is given by the
exchange particle-hole ladder (RPA series)
\begin{align}
  (\gamma^\ix{\,x}_\lambda)_{1'2'|12}
  &=
  \overline v_{1'2'|12}
  +
  \frac{\im}{2\pi}
  \overline v_{1'4'|32}\,g^{\lambda}_{3|3'}\,g^{\lambda}_{4|4'}\,
  \overline v_{3'2'|14}
  + \ldots
  \nonumber
  \\
  &=
  \overline v_{1'2'|12}
  +
  \frac{\im}{2\pi}
  \overline v_{1'4'|32}\,g^{\lambda}_{3|3'}\,g^{\lambda}_{4|4'}\,
  (\gamma^\ix{\,x}_\lambda)_{3'2'|14}.
  \label{eq:xph_def}
\end{align}
Due to frequency conservation at each interaction vertex, the
frequency dependence of the ladder is reduced to a single bosonic
combination of the external frequencies,
\begin{equation}
  \gamma^\ix{\,x}_\lambda(\omega_1', \omega_2'|\omega_1, \omega_2)
  =
  (\gamma^\ix{\,x}_\lambda)(X),
\end{equation}
with
\begin{equation}
  X = \omega_2' - \omega_1 = \omega_2 - \omega_1'.
\end{equation}

We evaluate the RPA series and obtain
\begin{equation}
  \label{eq:xph_eval}
  (\gamma^\ix{x}_\lambda)^{12|22}_{\sigma \overline \sigma|\sigma \overline
    \sigma}(X)
  =
  \frac{U}{2} +
  \frac{U^2}{4} \frac{(B^\ix{x}_\lambda)_{\sigma \overline \sigma}(X)}{1 -
    \frac{U}{2} (B^\ix{x}_\lambda)_{\sigma \overline \sigma}(X)},
\end{equation}
where 
\begin{multline}
  (B^\ix{x}_\lambda)_{\sigma \overline \sigma}(X) = \frac{\im}{2 \pi}
  \int \!  \df \omega \, \big[ g^\ix{Ret}_{\lambda \, \sigma}(\omega -
  \tfrac{X}{2}) g^\ix{K}_{\lambda \, \overline \sigma}(\omega +
  \tfrac{X}{2})
  \\
  + g^\ix{K}_{\lambda \, \sigma}(\omega - \tfrac{X}{2})
  g^\ix{Av}_{\lambda \, \overline \sigma}(\omega + \tfrac{X}{2}) \big]
\end{multline}
denotes the particle-hole polarization operator. In the
special case $T=0$, $B=0$, $V = 0$, $eV_\ix{g} = \mu = 0$ we find
\begin{equation}
  \label{eq:Bxph}
  (B^\ix{x}_\lambda)_{\sigma \overline \sigma}(X)
  =
  -\frac{2}{\pi} \frac{\Gamma_\lambda}{X(X - \im \Gamma_\lambda)} \ln
  \left(1 + \im \tfrac{X}{\Gamma_\lambda/2} \right). 
\end{equation}
Expanding $1/(B^\ix{x}_\lambda)_{\sigma \overline \sigma}(X)$ in powers of
$X/\Gamma_\lambda$ results in
\begin{equation}
  \label{eq:xph_sing}
  (\gamma^\ix{x}_\lambda)^{12|22}_{\sigma \overline \sigma|\sigma \overline
    \sigma}(X)
  \simeq
  \frac{U}{2}
  - \frac{\im}{2 \pi}
  \frac{U^2}{X - \im (\Gamma_\lambda/2 - U/\pi)},
  \quad
  X \ll \Gamma_\lambda,
\end{equation}
which features a singularity when $\lambda$ reaches the value
\begin{equation}
  \lambda = \lambda_\ix{c}
  = 2 U / \pi - \Gamma.
\end{equation}
Additionally, for $\lambda < \lambda_\ix{c}$ the function
$(\gamma^\ix{x}_\lambda)^{12|22}_{\sigma \overline \sigma|\sigma
  \overline \sigma}(X)$ exhibits the wrong analytic behaviour, being
analytic in the upper half plane of $X$ instead of the lower one as
required by causality\cite{jakobs1}. Therefore the flow $\lambda
\rightarrow 0$ can only be finished if
\begin{equation}
  U < U_\ix{c} 
  =
  \pi \Gamma / 2.
\end{equation}
This limitation is not even overcome when the full self energy is fed
back into the flow, replacing $s$ and $g$ in
Eq.~\eqref{eq:xph_flow} by $S$ and $G$.

Hence we also have to take into account the other two
channels~(\ref{eq:pp},\ref{eq:dph}). A parquet summation based
procedure mixing the exchange particle-hole and the particle-particle
channel has been set up in Ref.~\onlinecite{janis}, in which
the authors study the SIAM within Matsubara formalism. Essentially the
particle-particle channel serves them to renormalize the interaction
$U$ which enters the exchange channel to an effective value
$U_\ix{eff}$ which is always lesser than $U_\ix{c}$. The difference
$U_\ix{c} - U_\ix{eff}$ is then used as a measure for the Kondo scale.

We aim to implement a similar proceeding within the framework of the
functional RG. We consider it appropriate to include also
contributions from the direct particle-hole channel, because this
channel bears the same singularity as the exchange channel. In order
to see this let us define the particle-particle and the direct
particle-hole ladder on the analogy of Eq.~\eqref{eq:xph_flow} by
the flow equations
\begin{subequations}
  \begin{align}
    \label{eq:pp_flow}
    \frac{\df}{\df \lambda}
    (\gamma^\ix{\,p}_\lambda)_{1'2'|12} 
    &= \frac{i}{4 \pi}
    (\gamma^\ix{\,p}_\lambda)_{1'2'|34}
    \frac{\df g^{\lambda}_{3|3'}\,g^{\lambda}_{4|4'}}{\df \lambda}
    (\gamma^\ix{\,p}_\lambda)_{3'4'|12},
    \\
    \label{eq:dph_flow}
    \frac{\df}{\df \lambda}
    (\gamma^\ix{\,d}_\lambda)_{1'2'|12} 
    &= -\frac{i}{2 \pi}
    (\gamma^\ix{\,d}_\lambda)_{1'3'|14}
    \frac{\df g^{\lambda}_{3|3'}\,g^{\lambda}_{4|4'}}{\df \lambda}
    (\gamma^\ix{\,d}_\lambda)_{4'2'|32}.
  \end{align}
\end{subequations}
In correspondence with Eq.~\eqref{eq:xph_def} the solutions of these
flow equations satisfy
\begin{subequations}
  \label{eq:ladders}
  \begin{align}
    (\gamma^\ix{\,p}_\lambda)_{1'2'|12}
    &=
    \overline v_{1'2'|12}
    +
    \frac{i}{4\pi}
    \overline v_{1'2'|34}\,g_{3|3'}\,g_{4|4'}\,
    (\gamma^\ix{\,p}_\lambda)_{3'4'|12},
    \label{eq:pp_def}
    \\
    (\gamma^\ix{\,d}_\lambda)_{1'2'|12}
    &=
    \overline v_{1'2'|12}
    -
    \frac{i}{2\pi} 
    \overline v_{1'3'|14}\,g_{3|3'}\,g_{4|4'}\,
    (\gamma^\ix{\,d}_\lambda)_{4'2'|32}.
    \label{eq:dph_def}
  \end{align}
\end{subequations}
Their frequency dependence takes the form
\begin{subequations}
  \begin{align}
    (\gamma^\ix{\,p}_\lambda)(\omega_1', \omega_2'|\omega_1, \omega_2)
    &=
    (\gamma^\ix{\,p}_\lambda)(\Pi),
    \\
    (\gamma^\ix{\,d}_\lambda)(\omega_1', \omega_2'|\omega_1, \omega_2)
    &=
    (\gamma^\ix{\,d}_\lambda)(\Delta),
  \end{align}
\end{subequations}
with
\begin{subequations}
  \label{eq:bos_freq}
  \begin{align}
    \Pi &= \omega_1 + \omega_2 = \omega_1' + \omega_2',
    \\
    \Delta &= \omega_1' - \omega_1 = \omega_2 - \omega_2'.
  \end{align}
\end{subequations}
Evaluating~\eqref{eq:pp_def} one finds
\begin{equation}
  (\gamma^\ix{\,p}_\lambda)^{12|22}_{\sigma \overline \sigma | \sigma
    \overline\sigma}(\Pi)
  =
  \frac{U}{2} +
  \frac{U^2}{4} \frac{(B^\ix{p}_\lambda)_{\sigma \overline \sigma}(\Pi)}{1 -
    \frac{U}{2} (B^\ix{p}_\lambda)_{\sigma \overline \sigma}(\Pi)},
\end{equation}
where at $T=0$, $B=0$, $V = 0$, $eV_\ix{g} = \mu=0$ the
particle-particle polarization operator $B^\ix{p}_\lambda$ takes the form
\begin{equation}
  (B^\ix{p}_\lambda)_{\sigma \overline \sigma}(\Pi)
  =
  \frac{2}{\pi} \frac{\Gamma_\lambda}{\Pi (\Pi + \im \Gamma_\lambda)}
  \ln \left(1 - \im \tfrac{\Pi}{\Gamma_\lambda/2} \right),
\end{equation}
such that $\gamma^\ix{\,p}_\lambda$ is a regular function on the real
$\Pi$-axis for all values of $U$.  In contrast,
$(\gamma^\ix{d}_\lambda)^{12|22}$ satisfies
\begin{align}
  (\gamma^\ix{d})^{12|22}_{\sigma \overline \sigma|\sigma \overline
    \sigma}(\Delta) &= \frac{1}{2}
  \left[(\gamma^\ix{p})^{12|22}_{\sigma \overline \sigma|\sigma
      \overline \sigma}(\Delta) + (\gamma^\ix{x})^{12|22}_{\sigma
      \overline \sigma|\sigma \overline \sigma}(\Delta)^\ast \right],
  \\
  (\gamma^\ix{d})^{12|22}_{\sigma \sigma|\sigma \sigma}(\Delta) &=
  \frac{1}{2} \left[(\gamma^\ix{p})^{12|22}_{\sigma \overline
      \sigma|\sigma \overline \sigma}(\Delta) -
    (\gamma^\ix{x})^{12|22}_{\sigma \overline \sigma|\sigma \overline
      \sigma}(\Delta)^\ast \right],
\end{align}
and thus possesses the same singularity as the exchange particle-hole
channel.


\subsection{Approximated mixing of the channels}
\label{subsec:approx_mix}

In an approximation taking into account all three channels
(\ref{eq:pp}--\ref{eq:dph}), the two-particle vertex function will
depend on all three frequencies, $\gamma(\Pi, X, \Delta)$.  Due to
frequency conservation these three ones are indeed sufficient to
express the general frequency dependence. In the following the three
bosonic frequency arguments of the vertex function will always be
indicated in the order $(\Pi, X, \Delta)$. When all three
contributions~(\ref{eq:pp}--\ref{eq:dph}) to the flow are taken into
account simultaneously, then the dependence on $\Pi, X, \Delta$ is
mixed through the feedback of the vertex function on the right hand
side of the flow equation that reads
\begin{widetext}
  \begin{subequations}
    \label{eq:flow_vertex_3freq}
    \begin{align}
      &\frac{\df}{\df \lambda} \gamma^{\lambda}_{1'2'|12}(\Pi, X,
      \Delta)
      = \frac{\im}{2 \pi} \int \! \df \omega
      \bigg\{
      \gamma^{\lambda}_{1'2'|34}\left(\Pi, \omega + \tfrac{X-\Delta}{2},
        \omega - \tfrac{X - \Delta}{2}\right)
      S^{\lambda}_{3|3'}\left(\tfrac{\Pi}{2}-\omega\right)
      G^{\lambda}_{4|4'}\left(\tfrac{\Pi}{2}+\omega\right) 
      \gamma^{\lambda}_{3'4'|12}\left(\Pi, \tfrac{X+\Delta}{2} + \omega,
        \tfrac{X + \Delta}{2} - \omega \right)
      \label{eq:pp_freq}
      \\
      &
      + 
      \gamma^{\lambda}_{1'4'|32}\left(\tfrac{\Pi + \Delta}{2} + \omega,
        X, \tfrac{\Pi + \Delta}{2} - \omega \right)
      \Big[S^{\lambda}_{3|3'}\left(\omega - \tfrac{X}{2}\right)
      G^{\lambda}_{4|4'}\left(\omega + \tfrac{X}{2}\right)
      + G^{\lambda}_{3|3'}\left(\omega -
        \tfrac{X}{2}\right) S^{\lambda}_{4|4'}\left(\omega +
        \tfrac{X}{2}\right) \Big] \gamma^{\lambda}_{3'2'|14}\left(\omega
        + \tfrac{\Pi - \Delta}{2}, X, \omega - \tfrac{\Pi - \Delta}{2}
      \right)
      \label{eq:xph_freq}
      \\
      &
      - 
      \gamma^{\lambda}_{1'3'|14}\left(\omega + \tfrac{\Pi-X}{2}, \omega
        - \tfrac{\Pi - X}{2}, \Delta \right)
      \Big[S^{\lambda}_{3|3'}\left(\omega - \tfrac{\Delta}{2}\right)
      G^{\lambda}_{4|4'}\left(\omega + \tfrac{\Delta}{2} \right) 
      +G^{\lambda}_{3|3'}\left(\omega - \tfrac{\Delta}{2}\right)
      S^{\lambda}_{4|4'}\left(\omega + \tfrac{\Delta}{2}\right) \Big]
      \gamma^{\lambda}_{4'2'|32}\left(\tfrac{\Pi+X}{2} + \omega,
        \tfrac{\Pi + X}{2} - \omega, \Delta \right)
      \bigg\}.
      \label{eq:dph_freq}
    \end{align}
  \end{subequations}
\end{widetext}
Here we use shorthand notation like
\begin{equation}
  \gamma^{\lambda}_{1'2'|12} \equiv
  (\gamma_2^\lambda)^{\alpha_1'\alpha_2'|\alpha_1 \alpha_2}_{\sigma_1'
    \sigma_2'|\sigma_1 \sigma_2},
\end{equation}
and indices occurring twice in a product implicate summation over state
and Keldysh indices. The flow equation~\eqref{eq:flow_vertex_3freq}
leads to a complicated dependence of the vertex function on the three
frequencies $(\Pi,X,\Delta)$. As a consequence a numerical solution of
the flow equation requires a sampling of three dimensional frequency
space which constitutes a high computational effort. The structure of
the flow equation suggests to approximate the frequency dependence of
the two-particle vertex function by
\begin{equation}
  \label{eq:sum_structure}
  \gamma_\lambda(\Pi,X,\Delta)
  \simeq
  \overline v + \varphi^\ix{p}_\lambda(\Pi) + \varphi^\ix{x}_\lambda(X) +
  \varphi^\ix{d}_\lambda(\Delta),
\end{equation}
where the bare interaction vertex $\overline v$ is the initial value
at the beginning of the flow, $\overline v = \gamma_{\lambda=\infty}$,
and where $\varphi^\ix{p}_\lambda(\Pi)$, $\varphi^\ix{x}_\lambda(X)$,
$\varphi^\ix{d}_\lambda(\Delta)$ are approximations to the parts of
$\gamma$ produced by the three channels~\eqref{eq:pp_freq},
\eqref{eq:xph_freq}, \eqref{eq:dph_freq}, respectively. An analogous
approximation has been investigated in studies of the SIAM based on
the equilibrium Matsubara fRG\cite{karrasch2}. The results achieved
with and without this approximation were of the same quality. Reliable
results were obtained for small and intermediate $U/\Gamma$. The
limiting effect for large $U/\Gamma$ seems to stem from omitting the
three-particle vertex function. For small $U/\Gamma$ the system can be
described by second order perturbation theory which also produces a
two-particle vertex function of the form~\eqref{eq:sum_structure}. Our
fRG treatment will fully comprise the diagrams of second order
perturbation theory for the self-energy and for the vertex function.
Therefore, for $U \rightarrow 0$ the fRG results asymptotically
approach those of second order perturbation theory.

In order to achieve the form~\eqref{eq:sum_structure} where the
frequency dependence is split up into three functions we have to
eliminate $X$ and $\Delta$ from~\eqref{eq:pp_freq}, $\Pi$ and $\Delta$
from~\eqref{eq:xph_freq}, and $\Pi$ and $X$
from~\eqref{eq:dph_freq}. It is obvious that any manipulation of this
type can only yield convincing results if the frequency dependence of
the vertex functions is not very pronounced. This limits the range of
applicability of the approximation to small and intermediate
interaction strengths. Kondo physics emerging for large $U/\Gamma$
cannot be described in general: we expect $\gamma(\Pi, X, \Delta)$ to
exhibit a sharp resonance as function of $X$ in this regime
(cf. Ref.~\onlinecite{janis} and section~\ref{subsec:ladders});
furthermore omitting the three particle vertex is not justified for
large $U/\Gamma$.

Different approaches in order to get rid of the frequency mixing on
the right hand side of~\eqref{eq:flow_vertex_3freq} are
conceivable. The variety of reasonable replacements is however
restricted by the necessity that the components of the vertex function
maintain their fundamental features concerning exchange of particles,
causality and the KMS conditions as described in
Ref.~\onlinecite{jakobs1}. We have tested different
possibilities. Here we present only that one which provided the most
convincing results. It is based on a procedure to assign a static
vertex to the functions $\varphi^{\ix{p},\ix{x},\ix{d}}_\lambda$,
\begin{subequations}
  \label{eq:replacement}
  \begin{align}
    \varphi^\ix{p}_\lambda(\Pi) &\rightarrow 
    \Phi^\ix{p}_\lambda,
    \\
    \varphi^\ix{x}_\lambda(X) &\rightarrow 
    \Phi^\ix{x}_\lambda,
    \\
    \varphi^\ix{d}_\lambda(\Delta) &\rightarrow 
    \Phi^\ix{d}_\lambda.
  \end{align}
\end{subequations}
The $\Phi^\ix{p,x,d}_\lambda$ are frequency independent and represent
effective static interactions between particles of either opposite or
identical spin state. Since permutations of the incoming indices map
the exchange particle-hole channel onto the direct particle-hole
channel and vice versa, the $\Phi^\ix{x,d}$ have to satisfy
\begin{subequations}
  \begin{align}
    (\Phi^\ix{x}_\lambda)^{\alpha'_1 \alpha'_2|\alpha_2
      \alpha_1}_{\sigma \overline \sigma|\sigma \overline \sigma} &= -
    (\Phi^\ix{d}_\lambda)^{\alpha'_1 \alpha'_2|\alpha_1
      \alpha_2}_{\sigma \overline \sigma|\overline \sigma \sigma},
    \\
    (\Phi^\ix{x}_\lambda)^{\alpha'_1 \alpha'_2|\alpha_1
      \alpha_2}_{\sigma \overline \sigma|\overline \sigma \sigma} &= -
    (\Phi^\ix{d}_\lambda)^{\alpha'_1 \alpha'_2|\alpha_2
      \alpha_1}_{\sigma \overline \sigma|\sigma \overline \sigma},
    \\
    (\Phi^\ix{x}_\lambda)^{\alpha'_1 \alpha'_2|\alpha_1
      \alpha_2}_{\sigma \sigma|\sigma \sigma} &=
    -(\Phi^\ix{d}_\lambda)^{\alpha'_1 \alpha'_2|\alpha_2
      \alpha_1}_{\sigma \sigma|\sigma \sigma}.
  \end{align}
\end{subequations}
This is achieved in the parameterization
\begin{subequations}
  \label{eq:eff_vertex}
  \begin{align}
    \label{eq:eff_vertex_p}
    (\Phi^\ix{p}_\lambda)_{\sigma'_1 \sigma'_2|\sigma_1 \sigma_2}   
    &=
    \begin{cases}
      U^\ix{p}_\lambda,
      &\text{if $\sigma'_1 = \sigma_1 = \overline \sigma'_2 = \overline
        \sigma_2$},
      \\
      -U^\ix{p}_\lambda,
      &\text{if $\sigma'_1 = \overline \sigma_1 = \overline \sigma'_2 =
        \sigma_2$},
      \\
      0, &\text{else},
    \end{cases}
    \displaybreak[0]
    \\
    \label{eq:eff_vertex_x}
    (\Phi^\ix{x}_\lambda)_{\sigma'_1 \sigma'_2|\sigma_1 \sigma_2}   
    &=
    \begin{cases}
      U^\ix{x}_\lambda,
      &\text{if $\sigma'_1 = \sigma_1 = \overline \sigma'_2 = \overline
        \sigma_2$},
      \\
      -U^\ix{d}_\lambda,
      &\text{if $\sigma'_1 = \overline \sigma_1 = \overline \sigma'_2 =
        \sigma_2$},
      \\
      -W^\ix{d}_{\lambda \, \sigma_1},
      &\text{if $\sigma'_1 =  \sigma_1 =  \sigma'_2 = \sigma_2$},
      \\
      0, &\text{else},
    \end{cases}
    \displaybreak[0]
    \\
    \label{eq:eff_vertex_d}
    (\Phi^\ix{d}_\lambda)_{\sigma'_1 \sigma'_2|\sigma_1 \sigma_2}   
    &=
    \begin{cases}
      U^\ix{d}_\lambda,
      &\text{if $\sigma'_1 = \sigma_1 = \overline \sigma'_2 = \overline
        \sigma_2$},
      \\
      -U^\ix{x}_\lambda,
      &\text{if $\sigma'_1 = \overline \sigma_1 = \overline \sigma'_2 =
        \sigma_2$},
      \\
      W^\ix{d}_{\lambda \, \sigma_1},
      &\text{if $\sigma'_1 =  \sigma_1 =  \sigma'_2 = \sigma_2$},
      \\
      0, &\text{else},
    \end{cases}  
  \end{align}
\end{subequations}
and
\begin{multline}
  \label{eq:eff_vertex1}
  (\Phi^{\ix{p}, \ix{x}, \ix{d}}_\lambda)_{\sigma'_1 \sigma'_2|\sigma_1
    \sigma_2}^{\alpha'_1 \alpha'_2|\alpha_1  \alpha_2}
  =
  \\
  \begin{cases}
    \frac{1}{2}(\Phi^{\ix{p}, \ix{x}, \ix{d}}_\lambda)_{\sigma'_1 \sigma'_2|\sigma_1
    \sigma_2},
  & \text{if $\alpha'_1 + \alpha'_2 + \alpha_1 + \alpha_2$ is odd}
  \\
  0, & \text{else},
  \end{cases}
\end{multline}
with $U^{\ix{p},\ix{x},\ix{d}}_\lambda$ and $W^\ix{d}_{\lambda \,
  \sigma}$ being real numbers, compare~~\eqref{eq:bare_vertex_model}
and~\eqref{eq:antisym_vertex_SIAM}. The detailed procedure to
determine appropriate constants $U^\ix{p,x,d}_\lambda$,
$W^\ix{d}_{\lambda \, \sigma}$ has to be consistent with the flow
equations and is discussed below in section~\ref{subsec:effect_int}.
In~\eqref{eq:eff_vertex_p} we exclude the possibility that an
effective interaction between two particles of identical spin state
could result from the particle-particle channel. The reason is that a
static interaction vertex between particles of identical spin state
vanishes when being anti-symmetrized since the direct and exchange
terms cancel each other. This is consistent with the fact that such a
contribution is not generated by the flow
equation~\eqref{eq:approxB_pp} below, since $\overline v +
\Phi_\lambda^\ix{x} + \Phi_\lambda^\ix{d}$ allows only interactions of
particles in opposite spin states. Also the fact that $U^{\ix{p},
  \ix{x}, \ix{d}}_\lambda$ are independent of of the spin (while
$W^\ix{d}_{\lambda \, \sigma}$ is not) is consistent with the
structure of an effective interaction vertex,
compare~\eqref{eq:antisym_vertex_SIAM}, and will be reproduced by the
flow equations below.

When inserting now the form~\eqref{eq:sum_structure} on the right hand
side of the flow equation~\eqref{eq:flow_vertex_3freq}, the sum
structure is reproduced by the flow if we replace
$\varphi^\ix{x}_\lambda(X)$ and $\varphi^\ix{d}_\lambda(\Delta)$ in
the particle-particle channel~\eqref{eq:pp_freq} by the constant
vertices $\Phi^\ix{x}_\lambda$, $\Phi^\ix{d}_\lambda$ and proceed
analogously for the other channels: in the flow of any channel the
other two channels are reduced to constant vertices. The flow
equations for $\varphi^{\ix{p},\ix{x},\ix{d}}_\lambda$ are then given
by
\begin{subequations}
  \label{eq:approxB}
  \begin{align}
    \label{eq:approxB_pp}
    &\frac{\df}{\df \lambda}
    (\varphi^\ix{p}_\lambda)_{1'2'|12}(\Pi) = \frac{1}{2}
    \big(\overline v + \varphi^\ix{p}_\lambda(\Pi) +
    \Phi^\ix{x}_\lambda + \Phi^\ix{d}_\lambda\big)_{1'2'|34}
    \nonumber
    \\
    & \hspace{4em} \times
    (I^\ix{pp}_\lambda)_{34|3'4'}(\Pi)
    \big(\overline v + \varphi^\ix{p}_\lambda(\Pi) +
    \Phi^\ix{x}_\lambda + \Phi^\ix{d}_\lambda\big)_{3'4'|12},
    \displaybreak[0]
    \\
    \label{eq:approxB_xph}
    &\frac{\df}{\df \lambda}
    (\varphi^\ix{x}_\lambda)_{1'2'|12}(X)
    = 
    \big(\overline v + \Phi^\ix{p}_\lambda +
    \varphi^\ix{x}_\lambda(X) + \Phi^\ix{d}_\lambda\big)_{1'4'|32} 
    \nonumber
    \\
    & \hspace{4em} \times
    (I^\ix{ph}_\lambda)_{34|3'4'}(X)
    \big(\overline v + \Phi^\ix{p}_\lambda + 
    \varphi^\ix{x}_\lambda(X) + \Phi^\ix{d}_\lambda\big)_{3'2'|14},
    \displaybreak[0]
    \\
    \label{eq:approxB_dph}
    &\frac{\df}{\df \lambda}
    (\varphi^\ix{d}_\lambda)_{1'2'|12}(\Delta) = - \big(\overline v +
    \Phi^\ix{p}_\lambda + \Phi^\ix{x}_\lambda +
    \varphi^\ix{d}_\lambda(\Delta) \big)_{1'3'|14}
    \nonumber
    \\
    & \hspace{4em} \times
    (I^\ix{ph}_\lambda)_{34|3'4'}(\Delta) 
    \big(\overline v +
    \Phi^\ix{p}_\lambda + \Phi^\ix{x}_\lambda +
    \varphi^\ix{d}_\lambda(\Delta) \big)_{4'2'|32},
  \end{align}
\end{subequations}
where we used a summation convention for index numbers representing
the state $\sigma$ and the Keldysh index $\alpha$, and
\begin{equation}
  \label{eq:Integral_Indices}
  (I^{\ix{pp}, \ix{ph}}_\lambda)_{34|3'4'}
  =
  \delta_{\sigma_3 \sigma'_3} \delta_{\sigma_4 \sigma'_4}
  (I^{\ix{pp}, \ix{ph}}_\lambda)_{\sigma_3 \sigma_4}^{\alpha_3 
    \alpha_4|\alpha'_3 \alpha'_4}
\end{equation}
refers to~\eqref{eq:channel_integrals}.

The different spin and Keldysh components of the functions
$\varphi^\ix{p,x,d}$ are not completely independent of each other:
there exist identities connecting them which stem from the invariance
under exchange of particles and complex
conjugation\cite{jakobs1}. Further identities follow from spin
conservation and from the special structure of the flow
equations~\eqref{eq:approxB}. In Appendix~\ref{sec:structure} we
determine a set of independent spin and Keldysh components of the
functions $\varphi^\ix{p,x,d}$ from which all other components can be
derived. It is then sufficient to compute the RG flow of these
independent components; the corresponding flow equations are presented
in the Appendices~\ref{sec:flow_eq_SE} and~\ref{sec:flow_eq_vert}.


\subsection{How to determine $U^\ix{p,x,d}_\lambda$ and
  $W^\ix{d}_{\lambda \, \sigma}$}
\label{subsec:effect_int}

The approximation scheme described above requires a procedure which
assigns effective constant interaction vertices
$\Phi^\ix{p,x,d}_\lambda$ to the functions
$\varphi^\ix{p}_\lambda(\Pi)$, $\varphi^\ix{x}_\lambda(X)$,
$\varphi^\ix{d}_\lambda(\Delta)$. Hence it is a central question how
to determine appropriate real numbers $U^\ix{p,x,d}_\lambda$,
$W^\ix{d}_{\lambda \, \sigma}$ which characterize the vertices
$\Phi^\ix{p,x,d}_\lambda$ in Eq.~\eqref{eq:eff_vertex}. We
propose a very simple scheme for the case of thermal equilibrium. In
order to generalize this approach to non-equilibrium however we will
need to formulate additional constraints.

Let us first assume thermal equilibrium. Then we can show that
\begin{subequations}
  \begin{align}
    \Phi^\ix{P}_\lambda &= \varphi^\ix{p}_\lambda(\Pi = 0),
    \\
    \Phi^\ix{x}_\lambda &= \varphi^\ix{x}_\lambda(X = 0),
    \\
    \Phi^\ix{d}_\lambda &= \varphi^\ix{d}_\lambda(\Delta = 0)
  \end{align}
\end{subequations}
have exactly the structure given in Eqs.~\eqref{eq:eff_vertex}
and~\eqref{eq:eff_vertex1}. For the proof we make use of the special
spin and Keldysh structure of $\varphi^\ix{p,x,d}$ that is described
in Appendix~\ref{sec:structure}.

We first discuss the particle-hole channel. From
Eqs.~\eqref{eq:ad_symm} and~\eqref{eq:add_symm} we infer that
$(a^\ix{d}_\lambda)_{\sigma \overline \sigma}(0)$ and
$(a^\ix{d}_\lambda)_{\sigma \sigma}(0)$ are real numbers. Using the
KMS conditions~\eqref{eq:d_KMS} and~\eqref{eq:dd_KMS} combined with
Eq.~\eqref{eq:ad_symm1} we conclude further that
$(b^\ix{d}_\lambda)_{\sigma \overline \sigma}(0) = 0$ and
$(b^\ix{d}_\lambda)_{\sigma \sigma}(0) = 0$. In thermal equilibrium we
can use the fluctuation dissipation theorem
\begin{subequations}
  \label{eq:fluct_diss}
  \begin{align}
    G^\ix{K}_\lambda(\omega) &= [1 - 2 f(\omega)]
    \left[G^\ix{Ret}_\lambda(\omega) - G^\ix{Av}_\lambda(\omega)
    \right],
    \\
    S^\ix{K}_\lambda(\omega) &= [1 - 2 f(\omega)]
    \left[S^\ix{Ret}_\lambda(\omega) - S^\ix{Av}_\lambda(\omega)
    \right],
  \end{align}
\end{subequations} to show that
\begin{multline}
  \label{eq:Iph_real}
  (I^\ix{\,ph}_\lambda)_{\sigma \overline \sigma}^{21|22}(0) +
  (I^\ix{\,ph}_\lambda)_{\sigma \overline \sigma}^{22|12}(0) =
  \\
  - 2 \Real \bigg\{ \frac{\im}{2 \pi} \int \! \df \omega \, \left[
      G^\ix{Av}_\sigma(\omega) S^\ix{Av}_{\overline \sigma}(\omega) +
      S^\ix{Av}_\sigma(\omega) G^\ix{Av}_{\overline \sigma}(\omega)
    \right] 
    \\
    \times \left[ 1 - 2 f(\omega) \right]\bigg\}
\end{multline}
is a real number. From the flow equation~\eqref{eq:ax_floweq_B} it
follows that $(a^\ix{x}_\lambda)_{\sigma \overline \sigma}(0)$ is
real. The KMS condition~\eqref{eq:x_KMS} then entails that
$(b^\ix{x}_\lambda)_{\sigma \overline \sigma}(0) = 0$.  In total this
means that we can set
\begin{equation}
  \Phi^\ix{d}_\lambda = \varphi^\ix{d}_\lambda(0)
  \quad \text{and}\quad
  \Phi^\ix{x}_\lambda = \varphi^\ix{x}_\lambda(0),
\end{equation}
with
\begin{subequations}
  \label{eq:Uxd_Wd}
  \begin{align}
    U^\ix{x}_\lambda &= 2(a^\ix{x}_\lambda)_{\sigma \overline
      \sigma}(0),
    \\
    U^\ix{d}_\lambda &= 2(a^\ix{d}_\lambda)_{\sigma \overline
      \sigma}(0),
    \label{eq:Ud_def}
    \\
    W^\ix{d}_{\lambda \, \sigma} &= 2(a^\ix{d}_\lambda)_{\sigma
      \sigma}(0).
    \label{eq:W_def}
  \end{align}
\end{subequations}
When we insert~\eqref{eq:W_def} into~\eqref{eq:a_d_floweq_B} this
yields in particular $\df (a^\ix{d}_\lambda)_{\sigma \overline
  \sigma} / \df \lambda = 0$. Combining this with the initial
condition $(a^\ix{d}_{\lambda = \infty})_{\sigma \overline \sigma} =
0$ we conclude from~\eqref{eq:Ud_def} that
\begin{equation}
  U^\ix{d}_\lambda \equiv 0.
  \label{eq:Ud_zero}
\end{equation}

For the particle-particle channel we combine the fluctuation
dissipation theorem~\eqref{eq:fluct_diss} with
\begin{equation}
  1 - 2 f(\omega) = - \left[1 - 2 f(-\omega) \right]
\end{equation}
(mind the convention $\mu=0$) to find that
\begin{multline}
  \label{eq:Ipp_real}
  (I^\ix{\,pp}_\lambda)_{\sigma \overline \sigma}^{22|12}(0) +
  (I^\ix{\,pp}_\lambda)_{\sigma \overline \sigma}^{22|21}(0) =
  \\
  2 \Real
  \bigg\{\frac{\im}{2 \pi} \int \! \df \omega \, \Big[
  G^\ix{Ret}_\sigma(\omega) S^\ix{Av}_{\overline \sigma}(- \omega)
  + S^\ix{Ret}_\sigma(\omega) G^\ix{Av}_{\overline \sigma}(-\omega)
  \Big]
  \\
  \times\big[ 1 - 2 f(\omega) \big] \bigg\}
\end{multline}
is a real number. Hence it follows from the flow
equation~\eqref{eq:ap_floweq_B} that
$(a^\ix{p}_\lambda)_{\sigma \overline \sigma}(0)$ is real. The KMS
condition~\eqref{eq:p_KMS} then demands that
$(b^\ix{p}_\lambda)_{\sigma \overline \sigma}(0) = 0$. Therefore
we can set
\begin{equation}
  \Phi^\ix{p}_\lambda = \varphi^\ix{p}_\lambda(0)
\end{equation}
with
\begin{equation}
  U^\ix{p}_\lambda = 2(a^\ix{p}_\lambda)_{\sigma \overline
    \sigma}(0).
  \label{eq:Up}
\end{equation}
(Note that $(\overline v + \Phi^\ix{p}_\lambda + \Phi^\ix{x}_\lambda +
\Phi^\ix{d}_\lambda) = \overline v_\lambda$ is exactly the flowing
effective interaction vertex used in the static approximation scheme
in section~\ref{sec:basic_approx}.)

In the case of non-equilibrium with $\mu_\ix{L} + \mu_\ix{R} = 0$ we
will also use~\eqref{eq:Uxd_Wd} and~\eqref{eq:Up}. While
$U^\ix{d}_\lambda$ and $W^\ix{d}_\lambda$ are then again guaranteed to
be real valued by~\eqref{eq:ad_symm} and~\eqref{eq:add_symm},
additional assumptions are necessary to ensure this for
$U^\ix{x}_\lambda$ and $U^\ix{p}_\lambda$. If the propagator which we
feed back into the flow of the two-particle vertex satisfies
Eq.~\eqref{eq:simplification}, then Eq.~\eqref{eq:Iph_real}
is still valid in non-equilibrium, with $f_\ix{eff}(\omega)$ replacing
$f(\omega)$; this warrants that $U^\ix{x}_\lambda$ is real. Under the
additional assumptions
\begin{equation}
  \Gamma_\ix{L} = \Gamma_\ix{R}
  \quad \text{and} \quad
  T_\ix{L} = T_\ix{R}
\end{equation}
also Eq.~\eqref{eq:Ipp_real} can be maintained; in that case also
$U^\ix{p}_\lambda$ is real.


\subsection{Self-energy feedback}
\label{subsec:feedback}

The propagators $S, G$ appearing in the flow equation for the
self-energy and for the vertex function are full propagators, which
means that they are dressed with the current value of the self-energy
$\Sigma^\lambda(\omega)$. In this way the self-energy feeds back into
its own flow and into the flow of the vertex function. In the
preceeding section we described that our flow scheme requires the
validity of the fluctuation dissipation theorem
Eq.~\eqref{eq:simplification}, or, equivalently,
\begin{equation}
  \label{eq:fluc_diss_Sigma}
  \Sigma^\ix{K}_\lambda(\omega) = \left[1 - 2 f_\ix{eff}(\omega) \right]
  \left[\Sigma^\ix{Ret}_\lambda(\omega) -
    \Sigma^\ix{Av}_\lambda(\omega) \right].
\end{equation}
While this identity is guaranteed in equilibrium by the KMS conditions,
it constitutes an approximation in non-equilibrium; interaction effects
that lead to a redistribution of particle statistics are neglected. 

The results discussed in section~\ref{sec:SIAM_results} below show
that the feedback of the full frequency dependent self-energy into the
flow leads to an artificial smoothening of spectral features. The
Kondo resonance at $\omega=0$ for instance acquires a much broader
shape than expected. Better results are achieved with a different
feedback scheme for the self-energy which reduces its back-coupling
into the flow to a simple shift of the single-particle levels. In
analogy to the way of determining an effective static interaction
vertex from the vertex function in section~\ref{subsec:effect_int}, we
assign a real and static level shift $E$ to the self-energy by setting
\begin{subequations}
  \begin{align}
    \label{eq:level_shift}
    E^{\ix{Ret} \, \lambda}_\sigma &= E^{\ix{Av} \, \lambda}_\sigma =
    \Real \Sigma^{\ix{Ret} \, \lambda}_\sigma (\omega =0),
    \\
    E^{\ix{K} \, \lambda}_\sigma &= 0.
  \end{align}
\end{subequations}
Only this static renormalization enters the propagators on the right
hand side of the flow equations for $\Sigma^\lambda$ and
$\gamma^\lambda$. Due to the missing Keldysh component of
$E^\lambda_\sigma$ the condition~\eqref{eq:simplification} is then
readily fulfilled.

It turns out that this static self-energy feedback produces results
distinctly better than in case of the full frequency dependent
self-energy feedback. We also studied mixed schemes, for instance, to
feed the static self-energy into the flow of $\gamma^\lambda$ while
the full self-energy is used for the flow of $\Sigma^\lambda$, or vice
versa. However, static feedback into both flow equations performed
better. Hence all results shown in section~\ref{sec:SIAM_results} are
computed with this scheme. The fact that static feedback performs best
highlights the difference between the Keldysh fRG and its Matsubara
counterpart, where the full self-energy feedback is
used\cite{karrasch2}. Note that in frequency-dependent approximations
the two methods are not equivalent (in equilibrium), even if the same
feedback scheme is used. We suspect that the feedback of the full
self-energy into the flow becomes favorable in a truncation scheme
taking into account contributions from the flow of the three-particle
vertex. Such a scheme, allowing for a renormalization of the static
part of the two-particle vertex, might lead to an enhanced effective
interaction strength, higher values of the effective mass, and a
sharpening of spectral features -- following the tendency expected
from the Ward identities\cite{yamada, yosida, yamada1}.


\section{Approximate solution in the particle-hole symmetric point}
\label{sec:apprb}

In this section we discuss the special case of vanishing magnetic
field, $B=0$, and particle-hole symmetry; the latter is given for
$eV_\ix{g} = (\mu_\ix{L} + \mu_\ix{R})/2 = 0$. Note that we already
introduced before the restriction $\Gamma_\ix{L} = \Gamma_\ix{R}$. In
that symmetric situation the RG flow does not generate any
contribution to $\Real \Sigma^\ix{Ret}_\lambda(\omega = 0)$ which
hence remains at its initial value $U/2$ given in
Eq.~\eqref{eq:initial_cond_Gamma_flow_Sigma_Ret}. The
renormalized effective level position is then given by
\begin{equation}
  \epsilon^\lambda_\sigma = \epsilon_\sigma + \Real
  \Sigma^\ix{Ret}_{\lambda \, \sigma} (\omega = 0) = 0,
\end{equation}
compare Eqs.~\eqref{eq:single_part_energy}
and~\eqref{eq:level_shift}. Hence the effective single-particle levels
for the two spin-states are degenerate and fixed in the middle of the
two chemical potentials throughout the flow.

This section is organized as follows. At first, we discuss the
structure of the vertex functions in the particle-particle (PP) and
exchange particle-hole (PH(e)) channels. At second, we focus on the
vertex functions in the direct particle-hole (PH(d)) channel. We
distinguish this case from the other two channels because of a bit
more involved spin structure. At third, we formulate a simplified
version of the flow equation~\eqref{eq:flow_self-energy} for the
self-energy.  On its basis we recover the Fermi-liquid relations
\cite{yamada,oguri} for the (imaginary part of) self-energy providing
an analytical fRG estimate for a characteristic energy scale of the
quasiparticle resonance at zero frequency.


\subsection{Particle-particle and exchange particle-hole channels}

Let us first discuss the structure of the vertex functions in the
particle-particle and exchange particle-hole channels characterized by
energy exchange frequencies $\Pi$ and $X$, respectively.

From Eqs.~\eqref{eq:Ud_def} and \eqref{eq:Ud_zero} we establish the
property
\begin{equation}
  (a^{\ix{d}}_{\lambda})_{\sigma \overline{\sigma}} (0) \equiv
  U^{\ix{d}}_{\lambda}/2 = 0 ,
\label{eq:simpdir}
\end{equation}
which leads, as we will show below, to decoupling of PP and PH(e)
channels from PH(d) channel.

Taking into account \eqref{eq:simpdir} we introduce for convenience
the following shorthand notations (cf. Eqs.~\eqref{eq:eff_vertex},
\eqref{eq:abp} and \eqref{eq:abx})
\begin{eqnarray}
  a_1 (\Pi)  &=& (a^\ix{p}_\lambda)_{\sigma
    \overline{\sigma}}(\Pi) + \frac{U + U^\ix{x}_\lambda }{2}, 
   \label{eq:a1_redef}  \\ 
  F_1 (\Pi) &=& 
  (I^\ix{\,pp}_\lambda)_{\sigma \overline{\sigma}}^{22|12}(\Pi) 
  + (I^\ix{\,pp}_\lambda)_{\sigma \overline{\sigma}}^{22|21}(\Pi), 
  \label{eq:F1} \\
 b_1 (\Pi) &=& (b_{\lambda}^{\ix{p}})_{\sigma \overline{\sigma}} (\Pi),
\label{eq:b1_redef} \\
  H_1 (\Pi) &=& 2 \im \Imag \left[ (I^\ix{\,pp}_\lambda)_{\sigma
      \overline{\sigma}}^{22|11}(\Pi )  
    + \frac{1}{2} (I^\ix{\,pp}_\lambda)_{\sigma
      \overline{\sigma}}^{22|22}(\Pi ) \right], 
\end{eqnarray}
and
\begin{eqnarray}
  a_3 (X) &=& \frac{U + U^\ix{p}_\lambda}{2} +
  (a^\ix{x}_\lambda)_{\sigma \overline{\sigma}}(X), \label{eq:a3_redef} \\
  F_3 (X) &=& (I^\ix{\,ph}_\lambda)_{\sigma \overline{\sigma}}^{21|22}(X)
  + (I^\ix{\,ph}_\lambda)_{\sigma \overline{\sigma}}^{22|12}(X),
  \label{eq:F3} \\
    b_3 (X) &=&   (b_{\lambda}^{\ix{x}})_{\sigma \overline{\sigma}} (X),
\label{eq:b3_redef} \\ 
  H_3 (X) &=& 2 \im \Imag \left[ (I^\ix{\,ph}_\lambda)_{\sigma
      \overline{\sigma}}^{12|21}(X) + \frac{1}{2}
    (I^\ix{\,ph}_\lambda)_{\sigma \overline{\sigma}}^{22|22}(X) 
  \right] ,
\end{eqnarray}
where the functions \eqref{eq:a1_redef} and \eqref{eq:a3_redef} are
specifically defined to fulfill $a_1 (0) = a_3 (0)$.  The flow
equations~\eqref{eq:ap_floweq_B} and \eqref{eq:ax_floweq_B} can be
then transformed into
\begin{eqnarray}
  \frac{\df a_1 (\Pi)}{\df \lambda} &=& [
  a_1 (\Pi)]^2 F_1 (\Pi)  +
  \left( \frac{\tilde{U}_{\lambda}}{2} \right)^2 F_3 (0)
  , \label{eq:ppfull} \\ 
  \frac{\df a_3 (X)}{\df \lambda} &=& \left(
    \frac{\tilde{U}_{\lambda}}{2} \right)^2 F_1 (0 )+ [ a_3 (X)]^2 F_3 (X), 
   \label{eq:phfull} 
\end{eqnarray}
where the static part
\begin{equation}
  \tilde{U}_{\lambda} = U + U_{\lambda}^{\ix{p}} +  U_{\lambda}^{\ix{x}}
  =2 a_1 (0) = 2 a_3 (0)
\end{equation}
obeys the equation
\begin{equation}
  \frac{\df \tilde{U}_{\lambda}}{\df \lambda} =
  \frac{\tilde{U}^2_{\lambda}}{2} \left[ F_1 (0)  +  F_3 (0)
  \right]. 
\end{equation}
In case of particle-hole symmetry and zero magnetic field the
following property holds
\begin{equation}
  F_3 (X) = - F_1 (-X),
\end{equation}
which, in particular, implies $F_1 (0) + F_3 (0) = 0$. Therefore the
static component $\tilde{U}_{\lambda} \equiv U$ does not flow within
the approximation introduced in subsection~\ref{subsec:approx_mix}.

The flow equations~\eqref{eq:bp_floweq_B} and \eqref{eq:bx_floweq_B}
for the functions \eqref{eq:b1_redef} and \eqref{eq:b3_redef} read in
the new notations
\begin{eqnarray}
  \frac{\df b_1 (\Pi)}{\df \lambda} &=& 
    | a_1 (\Pi)|^2 H_1 (\Pi) \nonumber \\
    &+& 2 \im
  \Imag \left\{ a_1 (\Pi) b_1 (\Pi) F_1 (\Pi)\right\}, 
   \label{eq:eqb1} \\ 
  \frac{\partial b_3 (X)}{\partial \lambda} &=& 
    |a_3 (X)|^2 H_3 (X) \nonumber \\    &+& 
    2  \im \Imag \left\{ a_3 (X) b_3 (X) F_3 (X) \right\}. \label{eq:eqb3}
\end{eqnarray}

The functions $F_{1,3}$ and $H_{1,3}$ are given by
\begin{eqnarray}
  F_1 (\Pi) &=& \frac{\im}{2 \pi} \frac{\partial}{\partial \lambda}  
  \int d \omega' \left[G_{\sigma}^{\ix{Ret}} (\frac{\Pi}{2} +\omega')
    G_{\overline{\sigma}}^{\ix{K}} (\frac{\Pi}{2} -\omega')
  \right. \nonumber \\ 
  & & \left. \qquad + G_{\sigma}^{\ix{K}} (\frac{\Pi}{2} +\omega')
    G_{\overline{\sigma}}^{\ix{Ret}} (\frac{\Pi}{2} -\omega') \right], 
 \label{eq:f1gg} \\ 
  F_3 (X) &=& \frac{\im}{2 \pi} \frac{\partial}{\partial \lambda}
  \int d \omega' \left[G_{\sigma}^{\ix{Ret}} (\omega' - \frac{X}{2})
    G_{\overline{\sigma}}^{\ix{K}} (\omega' + \frac{X}{2}) \right. \nonumber
  \\ 
  & & \left. \qquad + G_{\sigma}^{\ix{K}} (\omega' - \frac{X}{2})
    G_{\overline{\sigma}}^{\ix{Av}} (\omega' + \frac{X}{2}) \right] 
\label{eq:f3gg}
\end{eqnarray}
and
\begin{eqnarray}
  H_1 (\Pi ) \! &=& \! \frac{\im}{2 \pi} \frac{\partial}{\partial \lambda}  
  \int d \omega' \nonumber \\
&\times &\left[G_{\sigma}^{\ix{Ret}} G_{\overline{\sigma}}^{\ix{Ret}} +
    G_{\sigma}^{\ix{Av}} G_{\overline{\sigma}}^{\ix{Av}} + G_{\sigma}^{\ix{K}}
    G_{\overline{\sigma}}^{\ix{K}} \right], 
\label{eq:h1gg} \\ 
  H_3 (X ) \! &=& \! \frac{\im}{2 \pi} \frac{\partial}{\partial \lambda}  
  \int d \omega' \nonumber \\
 &\times & \left[G_{\sigma}^{\ix{Ret}} G_{\overline{\sigma}}^{\ix{Av}} +
    G_{\sigma}^{\ix{Av}} G_{\overline{\sigma}}^{\ix{Ret}} + G_{\sigma}^{\ix{K}}
    G_{\overline{\sigma}}^{\ix{K}} \right], 
\label{eq:h3gg}
\end{eqnarray}
where the partial derivative $\partial/ \partial \lambda$ is supposed
to act only on explicitly $\lambda$-dependent term within $G$'s, and
the arguments of Green functions in \eqref{eq:h1gg} and
\eqref{eq:h3gg} correspond to those in \eqref{eq:f1gg} and
\eqref{eq:f3gg}, respectively. (For brevity we also denote
$G_{\lambda} \equiv G$.) By a straightforward evaluation one can check
that in equilibrium the following identities hold
\begin{eqnarray}
  H_1 (\Pi) &=& \coth \left( \frac{
      \Pi}{2 T}\right) \left[ F_1 (\Pi)
    - F_1^* (\Pi) \right], 
\label{eq:h1_f1_eq} \\ 
  H_3 (X) &=& - \coth \left(\frac{X}{2 T} \right) \left[ F_3 (X)
    - F_3^* (X) \right]. 
\label{eq:h3_f3_eq}
\end{eqnarray}
With their help we can establish that the combinations
\begin{eqnarray}
  c_1 (\Pi) &=& b_1 (\Pi) -
  \coth \left( \frac{\Pi}{2 T}\right) \left[
    a_1 (\Pi) -  a_1^* (\Pi)\right], \\ 
  c_3 (X) &=& b_3 (X) +\coth \left( \frac{X}{2T} \right)
  \left[ a_3 (X) - a_3^* (X) \right]
\end{eqnarray}
obey the equations
\begin{eqnarray}
  \frac{\df c_1 }{\df \lambda} &=& 2 \Real \{ a_1 F_1 \} c_1 , 
\label{eq:c1_floweq} \\ 
  \frac{\df c_3 }{\df \lambda} &=& 2 \Real \{ a_3 F_3 \} c_3 ,  
\label{eq:c3_floweq}
\end{eqnarray}
respectively. Since initially $c_1 (\Pi) = c_3 (X)
\equiv 0$, they also remain zero during the flow, accordingly to 
\eqref{eq:c1_floweq} and \eqref{eq:c3_floweq}. This proves that the KMS 
relations \eqref{eq:p_KMS} and \eqref{eq:x_KMS} inherent to the equilibrium 
state are exactly preserved under the fRG flow in the chosen approximation.

The derivative of the function $F_3$ at zero temperature reads
\begin{eqnarray}
  & & \frac{\partial F_3 (X)}{\partial X}\bigg|_{X=0} \equiv F'_3 (0)
  = -\frac{1}{\pi} \frac{\partial}{\partial \lambda} \int d \omega'  
\label{eq:f3_deriv} \\
  & \times & \bigg\{ 2 \delta ( \omega' )\left[ G_{\sigma}^{\ix{Ret}} (\omega')  
    \Imag G_{\overline{\sigma}}^R (\omega') - \Imag
    G_{\sigma}^{\ix{Ret}} (\omega')  G_{\overline{\sigma}}^{\ix{Av}} (\omega')
   \right] \nonumber \\ 
  &+&  h_F (\omega') \left[ G_{\sigma}^{\ix{Ret}} (\omega') \Imag
    \frac{\partial G_{\overline{\sigma}}^{\ix{Ret}} (\omega')}{\partial
      \omega'} - \Imag \frac{\partial G_{\sigma}^{\ix{Ret}}
      (\omega')}{\partial \omega'}  G_{\overline{\sigma}}^{\ix{Av}}
    (\omega')\right] \bigg\}, 
  \nonumber
\end{eqnarray}
where $h_F (\omega) \equiv 1 - 2 f (\omega) \stackrel{T \to 0}{=} {\rm
  sign} (\omega)$. Under constraints adopted in this section the
formula \eqref{eq:f3_deriv} simplifies to
\begin{eqnarray}
  & & F'_3 (0) = -\frac{\im}{\pi} 
  \frac{\partial}{\partial \lambda} \int d \omega'  \nonumber \\
  & \times & \left\{ 4 \delta (\omega' ) 
     \left[ \Imag G_{\sigma}^{\ix{Ret}} (\omega') \right]^2
    + h_F (\omega') \frac{\partial }{\partial \omega'} 
      \left[ \Imag G_{\sigma}^{\ix{Ret}} (\omega') \right]^2 \right\} 
\nonumber \\ 
  &=&  -\frac{2 \im}{\pi} 
  \frac{\partial}{\partial \lambda} \left[ \Imag G_{\sigma}^{\ix{Ret}} (0)
\right]^2
   = \frac{16 \im}{\pi  \Gamma_{\lambda}^3}. 
  \label{eq:fprime0}
\end{eqnarray}

In the approximation achieved by neglecting the self-energy feedback
$G_{\lambda} \to g_{\lambda}$ we can replace in \eqref{eq:f3gg} the partial 
derivative $\partial/\partial \lambda$ by the full one $\df/\df \lambda$. 
Then we find that
\begin{equation}
  F_3 (X) = \frac{\df}{\df \lambda} \left[ - \frac{2}{\pi}
    \frac{\Gamma_{\lambda}}{X (X - \im \Gamma_{\lambda})}  \ln \left( 1
      + \im \frac{X}{\Gamma_{\lambda}/2} \right) \right]
\end{equation}
is given by the full $\lambda$-derivative of the particle-hole polarization
operator \eqref{eq:Bxph}.  In particular, we obtain $F_3 (0) = -4/(\pi
\Gamma_{\lambda}^2)$, and at small $X \leq \Gamma_{\lambda}$ we
approximately have
\begin{equation}
  F_3 (X) \approx \frac{4}{\pi (2 X - \im \Gamma_{\lambda})^2}.
\end{equation}
This leads to an approximate solution of
Eq.~\eqref{eq:phfull} 
\begin{equation}
  a_3 (X) \approx -\frac{U}{2} \frac{X+ \im B (U,\lambda)}{X - \im B
    (U,\lambda)}, 
\end{equation}
where the width of the Lorentzian is given by
\begin{equation}
  B (U,\lambda) = \frac{4 U}{\pi} e^{-\frac{4 U}{\pi \Gamma_{\lambda}}}.
  \label{eq:widthB}
\end{equation}
It explicitly manifests that the exponential scale develops in the
broadening of the renormalized frequency-dependent vertex function in
the exchange particle-hole channel.


\subsection{Direct particle-hole channel}

The flow equations for the vertex functions in this channel are split
off those considered in the previous subsection. We analogously
introduce the new notations (cf. \eqref{eq:adud} and \eqref{eq:aduu})
\begin{eqnarray}
  a_{20} (\Delta) &=& \frac{U+U_{\lambda}^{\ix{p}} +
    U_{\lambda}^{\ix{x}}}{2} + (a_{\lambda}^{\ix{d}})_{\sigma
    \overline{\sigma}} (\Delta), \\ 
  a_{2 \sigma} (\Delta) &=& (a_{\lambda}^{\ix{d}})_{\sigma \sigma}
  (\Delta) -\frac{W_{\lambda \sigma}^{\ix{d}}}{2}, \\ 
  F_{2 \sigma} (\Delta) &=& (I^\ix{\,ph}_\lambda)^{22|21}_{\sigma
    \sigma}(\Delta) 
  +(I^\ix{\,ph}_\lambda)^{12|22}_{\sigma \sigma}(\Delta), \\
   b_{20} (\Delta) &=& (b_{\lambda}^{\ix{d}})_{\sigma
    \overline{\sigma}} (\Delta), 
  \\
  b_{2 \sigma} (\Delta) &=& (b_{\lambda}^{\ix{d}})_{\sigma \sigma}
  (\Delta), 
  \\
  H_{2 \sigma} (\Delta) &=&  2 \im \Imag
  \left\{(I^\ix{\,ph}_\lambda)_{\sigma \sigma}^{12|21}(\Delta) +
    \frac{1}{2}(I^\ix{\,ph}_\lambda)_{\sigma \sigma}^{22|22}(\Delta)
  \right\} , 
\end{eqnarray}
and rewrite the corresponding Eqs.~\eqref{eq:d_floweq_B} and
\eqref{eq:dd_floweq_B}
\begin{eqnarray}
  \frac{\df a_{20} (\Delta)}{\df \lambda} &=& -a_{20}
  (\Delta ) \sum_{s= \uparrow, \downarrow} a_{2s} (\Delta) F_{2s}
  (\Delta ) \nonumber \\ 
  & & + \left( \frac{\tilde{U}_{\lambda}}{2}\right)^2 \left[
    F_1 (0)  +  F_3 (0) \right], \label{eq:ph1full} \\ 
  \frac{\df a_{2 \sigma} (\Delta)}{\df \lambda}&=& -[ a_{20}
  (\Delta)]^2 F_{2 \overline{\sigma}} (\Delta)-[ a_{2 \sigma}
  (\Delta)]^2 F_{2 \sigma} (\Delta) \nonumber \\ 
  & & +\left(\frac{\tilde{U}_{\lambda}}{2} \right)^2 F_{2
    \overline{\sigma}} (0),
  \label{eq:ph2full}
\end{eqnarray}
\begin{eqnarray}
  \frac{\df b_{20} (\Delta)}{\df \lambda} &=& - a_{20}
  (\Delta) a_{2 \overline{\sigma}}^* (\Delta) H_{2 \overline{\sigma}}
  (\Delta) - a_{20}^* (\Delta)  a_{2 \sigma} (\Delta) H_{2 \sigma}
  (\Delta) \nonumber \\ 
  & & - a_{20} (\Delta) b_{2 \overline{\sigma}} (\Delta) F_{2
    \overline{\sigma}} (\Delta) - a_{20}^* (\Delta) b_{2 \sigma}
  (\Delta) F_{2 \sigma}^* (\Delta) \nonumber \\ 
  & & -a_{2 \overline{\sigma}}^* (\Delta) b_{20} (\Delta) F_{2
    \overline{\sigma}}^* (\Delta) - a_{2 \sigma} (\Delta) b_{20}
  (\Delta) F_{2 \sigma} (\Delta), \nonumber \\ 
  \label{eq:eqb20} \\
  \frac{\df b_{2 \sigma} (\Delta)}{\df \lambda} &=& -
  |a_{20} (\Delta)|^2  H_{2 \overline{\sigma}} (\Delta) - |a_{2
    \sigma} (\Delta)|^2 H_{2 \sigma} (\Delta) \nonumber \\ 
  & & - 2 \Real \{a_{2 \sigma} (\Delta) F_{2 \sigma} (\Delta)\}
  b_{2\sigma} (\Delta)  \nonumber \\ 
  & & - 2 \im \Imag \{ a_{20}^* (\Delta) b_{20} (\Delta) F_{2
    \overline{\sigma}}^* (\Delta) \} . \label{eq:eqb2s} 
\end{eqnarray}
We note that $a_{20} (0) = \tilde{U}_{\lambda}/2$ and $a_{2 \sigma} (0)
=0$ as well as $F_{2 \sigma} (\Delta) = F_3
(-\Delta)$ and  $H_{2 \sigma} (\Delta) = H_3 (-\Delta)$. One can also observe
 that $b_{20} (\Delta)$ contains both real $b_{20}^{\ix{r}}
(\Delta) = \Real b_{20} (\Delta)$ and imaginary $b_{20}^{\ix{i}}
(\Delta) = \Imag b_{20} (\Delta)$ parts, while $b_{2 \sigma} (\Delta)$
is purely imaginary.

Analogously to \eqref{eq:h3_f3_eq} the following identity holds
in equilibrium
\begin{equation}
  H_{2 \sigma} (\Delta) = \coth \left( \frac{\Delta}{2 T} \right)
  \left[ F_{2 \sigma} (\Delta) - F_{2 \sigma}^* (\Delta) \right]. 
\end{equation}
It allows us to express the equations for $b_{20}^{\ix{r}} (\Delta)$
and for the  combinations
\begin{eqnarray}
  c_{20} (\Delta ) &=& \im b_{20}^{\ix{i}} (\Delta ) -\coth \left(
    \frac{\Delta}{2 T} \right) \left[ a_{20} (\Delta) - a_{20}^*
    (\Delta) \right], \\ 
  c_{2 \sigma} (\Delta) &=& b_{2 \sigma} (\Delta) -\coth \left(
    \frac{\Delta}{2 T} \right) \left[ a_{2 \sigma} (\Delta ) -
    a_{2 \sigma}^* (\Delta ) \right], 
\end{eqnarray}
in the following form
\begin{eqnarray}
  \frac{\df b_{20}^{\ix{r}}}{\df \lambda} &=& - \Real \{
  F_{2 \sigma} a_{2 \sigma} + F_{2 \overline{\sigma}} a_{2
    \overline{\sigma}} \} b_{20}^{\ix{r}} 
  \nonumber \\
  & & - \im \Imag \{ F_{2 \sigma} a_{2 \sigma} - F_{2
    \overline{\sigma}} a_{2 \overline{\sigma}} \} c_{20} 
  \nonumber \\
  & & + \im \Imag \{ F_{2 \sigma} a_{20}\} c_{2 \sigma} - \im \Imag \{
  F_{2 \overline{\sigma}} a_{20}\} c_{2 \overline{\sigma}}, \\ 
  \frac{\df c_{20}}{\df \lambda} &=& - \im \Imag \{ F_{2
    \sigma} a_{2 \sigma} - F_{2 \overline{\sigma}} a_{2
    \overline{\sigma}} \} b_{20}^{\ix{r}} 
  \nonumber \\ 
  & & - \Real \{ F_{2 \sigma} a_{2 \sigma} + F_{2 \overline{\sigma}}
  a_{2 \overline{\sigma}} \} c_{20} 
  \nonumber \\
  & & - \Real \{ F_{2 \sigma} a_{20}\} c_{2 \sigma} - \Real \{ F_{2
    \overline{\sigma}} a_{20}\} c_{2 \overline{\sigma}}, \\ 
  \frac{\df c_{2 \sigma}}{\df \lambda} &=& 2 \im \Imag \{
  F_{2 \overline{\sigma}} a_{20} + F_{2 \sigma} a_{2 \sigma} \}
  b_{20}^{\ix{r}} \nonumber \\ 
  & & -2 \Real \{ F_{2 \overline{\sigma}}  a_{20} \} c_{20}  -2 \Real
  \{ F_{2 \sigma} a_{2 \sigma}  \} c_{2 \sigma}. 
\end{eqnarray}
Since initially $b_{20}^{\ix{r}} (\Delta) = c_{20} (\Delta) = c_{2
  \sigma} (\Delta) =0$, they remain zero during the flow
[cf. Eq.~\eqref{eq:d_KMS}]. This observation  completes the proof of the 
statement that the KMS relations are preserved in the chosen approximation 
scheme.


\subsection{Fermi-liquid relations for the self-energy}
\label{sec:fermiliq}

In the case of particle-hole symmetry it is convenient to introduce 
the notations
\begin{eqnarray}
  \bar{a}_1 (\Pi) &=& - a_1 (-\Pi), \\
  \bar{a}_{2 \sigma} (\Delta) &=& - a_{2 \sigma} (-\Delta),
  \label{eq:bar_2sigma}
\end{eqnarray}
and rewrite an equilibrium version of the Eq.~\eqref{eq:flow_SigmaRet}
for the retarded self-energy in a simplified form
\begin{eqnarray}
  \frac{\df}{\df \lambda} \Sigma_{\sigma}^{\ix{Ret}} (\omega) &=&  
  \frac{1}{\pi} \int d \tilde{\omega} \left[ - \frac{U}{2} h_F
    (\tilde{\omega}) 
    \Imag s_{\overline{\sigma}}^{\ix{Ret}} (\tilde{\omega})\right. \nonumber \\ 
  & & + h_F (\tilde{\omega}) \bar{a}_1 (\tilde{\omega} - \omega )  
  \Imag s_{\overline{\sigma}}^{\ix{Ret}} (-\tilde{\omega})
  \nonumber \\
  & & + h_B (\tilde{\omega} - \omega) \Imag \bar{a}_1 (\tilde{\omega}
  -\omega)  
  s_{\overline{\sigma}}^{\ix{Av}} (-\tilde{\omega}) \nonumber \\
  & & + h_F (\tilde{\omega}) a_3 (\tilde{\omega} - \omega) 
  \Imag s_{\overline{\sigma}}^{\ix{Ret}} (\tilde{\omega}) \nonumber \\
  & & - h_B (\tilde{\omega} - \omega) \Imag a_3 (\tilde{\omega} -
  \omega)  
  s_{\overline{\sigma}}^{\ix{Ret}}  (\tilde{\omega}) \nonumber \\
  & &  + h_F (\tilde{\omega}) \bar{a}_{2 \sigma} (\tilde{\omega} -
  \omega) 
  \Imag s_{\sigma}^{\ix{Ret}} (\tilde{\omega}) \nonumber \\
  & & \left. - h_B (\tilde{\omega} - \omega) \Imag \bar{a}_{2 \sigma} 
    (\tilde{\omega} - \omega) s_{\sigma}^{\ix{Ret}}  (\tilde{\omega})\right],
  \label{eq:se1} 
\end{eqnarray}
where $h_B (\omega) = \coth \frac{\omega}{2 T}$. In this equation the
self-energy feedback is neglected as well by replacing $S_{\lambda}
\to s_{\lambda} = \frac{\partial g_{\lambda}}{\partial \lambda}$.

Comparing Eqs.~\eqref{eq:ppfull} and \eqref{eq:phfull} with 
\eqref{eq:ph1full} and \eqref{eq:ph2full}, one
can observe that the relation
\begin{equation}
  \bar{a}_{2 \sigma} (\omega) = \frac{\bar{a}_1 (\omega) + a_3 (\omega) }{2}
\end{equation}
holds at every $\lambda$.  The identities $\Imag s^{\ix{Ret}} (\omega)
= \Imag s^{\ix{Ret}} (-\omega)$, $s^{\ix{Av}} (-\omega) = -
s^{\ix{Ret}} (\omega)$ are fulfilled as well.  Therefore
\eqref{eq:se1} reduces to the form
\begin{eqnarray}
  \frac{\df}{\df \lambda} \Sigma_{\sigma}^{\ix{Ret}} (\omega) &=& 
  \frac{3}{2 \pi} \int d \tilde{\omega} \nonumber \\
  &\times & \left[ h_F (\tilde{\omega}) \bar{a}_1 (\tilde{\omega} - \omega )  
    \Imag s_{\overline{\sigma}}^{\ix{Ret}} (\tilde{\omega}) \right. \nonumber \\
  &-&  h_B (\tilde{\omega} - \omega) \Imag \bar{a}_1 (\tilde{\omega}
  -\omega) s_{\overline{\sigma}}^{\ix{Ret}} (\tilde{\omega}) \nonumber \\
  &+&   h_F (\tilde{\omega}) a_3 (\tilde{\omega} - \omega) 
  \Imag s_{\overline{\sigma}}^{\ix{Ret}} (\tilde{\omega}) \nonumber \\
  &-& \left. h_B (\tilde{\omega} - \omega) \Imag a_3 (\tilde{\omega} - \omega) 
    s_{\overline{\sigma}}^{\ix{Ret}}  (\tilde{\omega}) \right].
  \label{eq:se2}
\end{eqnarray}
In particular, its imaginary part reads
\begin{eqnarray}
  \frac{\df}{\df \lambda} \Imag \Sigma_{\sigma}^{\ix{Ret}} (\omega)
  &=& 
  \frac{3}{2 \pi} \int d \tilde{\omega} \left[  h_F (\tilde{\omega}) -
    h_B (\tilde{\omega} - \omega) \right] \label{eq:se3} \\
  &\times & \Imag \left[ \bar{a}_1 (\tilde{\omega} -\omega ) + a_3
    (\tilde{\omega} - \omega)  \right] 
  \Imag s_{\overline{\sigma}}^{\ix{Ret}} (\tilde{\omega}),  \nonumber
\end{eqnarray}
which at $T=0$ amounts to
\begin{eqnarray}
  & & \frac{\df}{\df \lambda} \Imag \Sigma_{\sigma}^{\ix{Ret}} (\omega) 
  \label{eq:se4} \\ 
  & & \quad = \frac{3}{\pi} \int_0^{\omega} d \tilde{\omega} 
  \Imag \left[ \bar{a}_1 (\tilde{\omega} -\omega ) + a_3
    (\tilde{\omega} - \omega)  \right] 
  \Imag s_{\overline{\sigma}}^{\ix{Ret}} (\tilde{\omega}).  \nonumber
\end{eqnarray}
Differentiating the latter expression with respect to $\omega$ and
taking into account symmetries of integrands we obtain
\begin{eqnarray}
  \frac{\df}{\df \lambda} \Imag \Sigma'' =  
  -\frac{3}{\pi} \Imag \left[ \frac{\partial \bar{a}_1}{\partial
      \omega}  
    + \frac{\partial a_3}{\partial \omega}  \right]_{\omega =0}
  \Imag s_{\overline{\sigma}}^{\ix{Ret}} (0),  \nonumber
\end{eqnarray}
where $\Imag \Sigma'' = \Imag \frac{\partial^2
  \Sigma_{\sigma}^{\ix{Ret}}}{\partial \omega^2} |_{\omega =0}$. Note
that the first derivative $\Imag \Sigma' = \Imag \frac{\partial
  \Sigma_{\sigma}^{\ix{Ret}}}{\partial \omega} |_{\omega =0}$
vanishes. It means that $\omega$-expansion of $\Imag \Sigma^{\ix{Ret}}
(\omega) \approx \frac12 \Imag \Sigma'' \omega^2$ starts from the
quadratic term, in compliance with the Fermi-liquid expression
\eqref{eq:SIAM_SE_expansion}.

In order to find an fRG estimate for $\Imag \Sigma''$, we need to 
establish equations for $\bar{a}'_1 = \Imag \frac{\partial
  \bar{a}_1}{\partial \omega}|_{\omega=0}$ and $a'_3 = \Imag
\frac{\partial a_3}{\partial \omega}|_{\omega=0}$. This is achieved by 
differentiating \eqref{eq:ppfull} and \eqref{eq:phfull}
\begin{eqnarray}
  \frac{\df \bar{a}'_1}{\df \lambda} &=& - U
  \bar{a}'_1 F_3 (0) 
  + \frac{U^2}{4}  \Imag F'_3 (0) , \label{eq:ap1} \\
  \frac{\df a'_3}{\df \lambda} &=&  U a'_3 F_3 (0)
  + \frac{U^2}{4}  \Imag F'_3 (0) . \label{eq:ap3}
\end{eqnarray}
Using  \eqref{eq:f3_deriv} and $\Imag s^{\ix{Ret}} (0) = 2/\Gamma_{\lambda}^2$, 
we obtain the following set of equations
\begin{eqnarray}
  \frac{\df \Imag \Sigma''}{\df \lambda}  &=& 
  - (\bar{a}'_1 + a'_3) \frac{6}{\pi \Gamma_{\lambda}^2}, \label{eq:sedd}\\
  \frac{\df \bar{a}'_1}{\df \lambda} &=&  \bar{a}'_1  
  \frac{4 U}{\pi \Gamma_{\lambda}^2} + \frac{4 U^2}{\pi \Gamma_{\lambda}^3}  , \\
  \frac{\df a'_3}{\df \lambda} &=& - a'_3  
  \frac{4 U}{\pi \Gamma_{\lambda}^2} +  \frac{4 U^2}{\pi \Gamma_{\lambda}^3}. 
\end{eqnarray}
The solution of the last equations is given by
\begin{eqnarray}
  \bar{a}'_1 &=& \frac{\pi}{4} \left[ 1 - \frac{4 U}{\pi \Gamma_{\lambda}} -
    e^{-\frac{4 U}{\pi \Gamma_{\lambda}}}\right], \\ 
  a'_3 &=& \frac{\pi}{4} \left[ 1 + \frac{4 U}{\pi \Gamma_{\lambda}} -
    e^{+\frac{4 U}{\pi \Gamma_{\lambda}}}\right], 
\end{eqnarray}
and therefore
\begin{equation}
  \bar{a}'_1 + a'_3 = \frac{\pi}{2} \left[ 1 
    - \cosh \frac{4 U}{\pi \Gamma_{\lambda}} \right].
\end{equation}
Integrating the equation for $\Imag \Sigma''$, we obtain at $\lambda =0$
\begin{equation}
  \label{eq:Sigma_second_scaling}
  \Imag \Sigma'' = \frac{3}{\Gamma} 
      \left[ 1 - \frac{\pi \Gamma}{4U} 
       \sinh \frac{4 U}{\pi \Gamma} \right] . 
\end{equation}
At large $U \gg \Gamma$ an exponentially large scale emerges in
$a'_3 \approx -\frac{\pi}{4} e^{\frac{4 U}{\pi \Gamma}}$ and in
$\Imag \Sigma'' \approx -\frac{3 \pi}{8 U} e^{\frac{4 U}{\pi \Gamma}}$. 
We also note that $a'_3$ determines 
the width $B=-U/a'_3$ given in Eq.~\eqref{eq:widthB}.

In a similar way we can establish the Fermi-liquid coefficient
corresponding to small-temperature expansion. Let us set $\omega=0$ in
Eq.~\eqref{eq:se3} and rescale the integration variable
$\tilde{\omega} = x T$. Then
\begin{eqnarray}
  & & \frac{\df}{\df \lambda} \Imag \Sigma_{\sigma}^{\ix{Ret}} (0) 
  \label{eq:se5} \\ 
  &=& - \frac{3}{\pi} \int \frac{d \tilde{\omega}}{\sinh
    \tilde{\omega}/T} 
  \Imag \left[ \bar{a}_1 (\tilde{\omega}) + a_3 (\tilde{\omega})  
  \right] \Imag s_{\overline{\sigma}}^{\ix{Ret}} (\tilde{\omega})  \nonumber \\
  &=& 
  - \frac{3 T}{\pi} \int \frac{d x}{\sinh x} 
  \Imag \left[ \bar{a}_1 (x T) + a_3 (x T)  
  \right] \Imag s_{\overline{\sigma}}^{\ix{Ret}} (x T).  \nonumber
\end{eqnarray}
Differentiating it twice with respect to $T$ we obtain
\begin{eqnarray}
  & & \frac{\df}{\df \lambda} \Imag 
  \frac{\partial^2 \Sigma_{\sigma}^{\ix{Ret}} (0)}{\partial T^2} \bigg|_{T=0}
  \\  
  &=& 
  - \frac{3}{\pi}  \left( \int \frac{2 x d x}{\sinh x} \right)
  \Imag \left[ \frac{\partial \bar{a}_1 }{\partial \omega} 
    + \frac{\partial a_3 }{\partial \omega} \right]_{\omega =0} 
  \Imag s_{\overline{\sigma}}^{\ix{Ret}} (0),  \nonumber
\end{eqnarray}
where $\int_{-\infty}^{\infty} \frac{2 x d x}{\sinh x} = \pi^2$. Comparing this
expression with \eqref{eq:sedd} we establish [cf. Eq.~\eqref{eq:SIAM_SE_expansion}]
\begin{equation}
  \Imag 
  \frac{\partial^2 \Sigma_{\sigma}^{\ix{Ret}}}{\partial T^2}
  \bigg|_{\omega,T=0} = \pi^2 \Imag \Sigma''. 
\end{equation}

Calculation of the Fermi-liquid coefficient corresponding to
small-bias expansion is presented in the
Appendix~\ref{app:fermi_liquid}.


\section{Results of the frequency dependent fRG}
\label{sec:SIAM_results}

In this section we discuss numerical results obtained from the
frequency dependent Keldysh fRG and compare them to those found by
other methods. The underlying approximation scheme is the one
described in section~\ref{sec:adv_approx}; in particular, only the
static part of the self-energy is fed back into the flow of
$\Sigma^\lambda$ and $\gamma_2^\lambda$, as explained in
section~\ref{subsec:feedback}. In the figures, results of this method
are labelled by ``fRG''.

We start with the particle-hole symmetric model in thermal equilibrium
at $B=0$.  The low energy expansion of the retarded self-energy is
then given by Eq.~\eqref{eq:SIAM_SE_expansion}. The reduced
susceptibilities serving as coefficients of this expansion are known
from Bethe-Ansatz and are~\cite{zlatic}
\begin{subequations}
  \label{eq:chi_exact}
  \begin{align}
    \tilde \chi_\ix{s} &= \exp\left(\tfrac{\pi}{4} \tfrac{U}{\Gamma}\right)
    \sqrt{\frac{\Gamma}{U}} \int_0^\infty \! \df x \, \exp\left(-
      \tfrac{\pi}{4} \tfrac{\Gamma}{U} x^2\right) \frac{\cos(\pi x/2)}{1 -
      x^2},
    \\
    \tilde \chi_\ix{c} &= \exp\left(-\tfrac{\pi}{4} \tfrac{U}{\Gamma}\right)
    \sqrt{\frac{\Gamma}{U}} \int_0^\infty \! \df x \, \exp\left(-
      \tfrac{\pi}{4} \tfrac{\Gamma}{U} x^2\right) \frac{\cosh(\pi x/2)}{1 +
      x^2}.
  \end{align}
\end{subequations}
For $U \gtrsim \pi \Gamma$ the reduced susceptibilities acquire their
asymptotic form where~\cite{zlatic}
\begin{equation}
  \label{eq:sus_asmyptot}
  \tilde \chi_\ix{s} \pm \tilde \chi_\ix{c}
  \;\simeq\;
  \tilde \chi_\ix{s}
  \;\simeq\;
  \frac{\pi}{2} \sqrt{\frac{\Gamma}{U}} \,
  \exp \left(\tfrac{\pi}{4}\left[\tfrac{U}{\Gamma} - \tfrac{\Gamma}{U}
    \right]\right), 
  \qquad
  U \gtrsim \pi \Gamma.
\end{equation}
This exponential behaviour governs the width of the Kondo resonance,
which can be seen as follows. At $eV_\ix{g} = 0 \;(=\mu)$, $T=0$, $B=0$,
$V=0$ the retarded Green function can be approximated for $\omega$
close to $0$ by
\begin{equation}
  G^\ix{Ret}_\sigma(\omega) \;\simeq\; \frac{1}{m^\ast\omega +
    \im \Gamma/2},
\end{equation}
compare~\eqref{eq:single_part_energy}
and~\eqref{eq:SIAM_SE_expansion}, the effective mass $m^\ast$ being
defined as
\begin{equation}
  m^\ast = 1 - \left.\frac{\partial \Sigma^\ix{Ret}}{\partial \omega}
  \right|_{\omega=0}.
\end{equation} 
Hence
\begin{equation}
  \rho_\sigma(\omega)
  = - \frac{1}{\pi} \Imag G^\ix{Ret}_\sigma(\omega)
  \;\simeq\; \frac{1}{\pi} \frac{\Gamma/2}{(m^\ast \omega)^2 + \Gamma^2/4},
\end{equation}
so that the full width at half maximum of the peak within this
approximation is given by
\begin{equation}
  T_\ix{K} = \frac{\Gamma}{m^\ast}.
  \label{eq:T_K}
\end{equation}
From~\eqref{eq:SIAM_SE_expansion} and~\eqref{eq:sus_asmyptot} we
deduce
\begin{subequations}
  \label{eq:eff_mass_chi1}
  \begin{align}
    \label{eq:eff_mass_chi}
    m^\ast &= \frac{\tilde \chi_\ix{s} + \tilde \chi_\ix{c}}{2} 
    \\
    &\simeq \;
    \frac{\pi}{4} \sqrt{\frac{\Gamma}{U}} \,
    \exp \left(\tfrac{\pi}{4}\left[\tfrac{U}{\Gamma} - \tfrac{\Gamma}{U}
      \right] \right), 
    \qquad
    U \gtrsim \pi \Gamma,
  \end{align}
\end{subequations}
and thus
\begin{equation}
  T_\ix{K} = \frac{4}{\pi} \sqrt{U \Gamma} \, \exp \left(-
    \tfrac{\pi}{4}\left[\tfrac{U}{\Gamma} - \tfrac{\Gamma}{U}\right] \right).
  \label{eq:T_K1}
\end{equation}

\begin{figure}
  \begin{center}
    \includegraphics{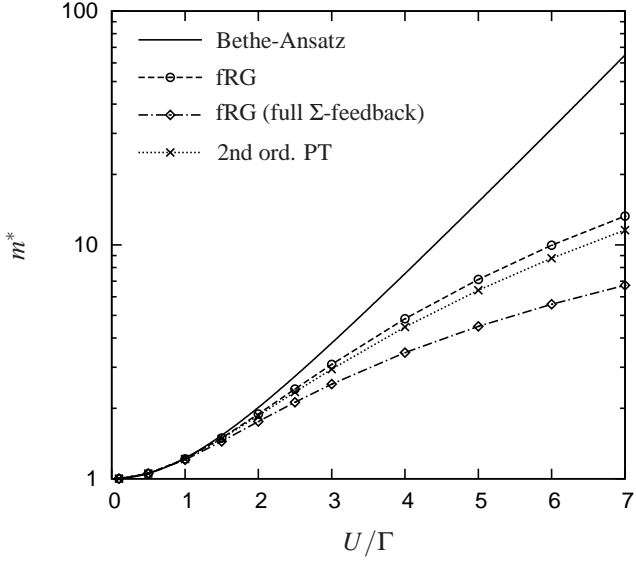}
  \end{center}
  \caption{The effective mass $m^\ast = 1 -
    {\Sigma^\ix{Ret}_\sigma}'\big|_{\omega = 0}$ as function of $U$
    for $V_\ix{g} = 0$, $T=0$, $B=0$, $V=0$. The exponential behaviour
    exhibited by the Bethe-Ansatz result obtained from
    Eqs.~\eqref{eq:eff_mass_chi} and~\eqref{eq:chi_exact} at
    large $U/\Gamma$ is neither reproduced by the fRG nor by second
    order perturbation theory. While the effective mass computed from
    the frequency dependent fRG with static self-energy feedback
    (labelled ``fRG'' and used as default throughout this section) is
    comparable to the one found in second order perturbation theory,
    the values obtained from the fRG with full self-energy feedback
    are clearly worse (compare section~\ref{subsec:feedback}).}
  \label{fig:eff_mass}
\end{figure}

Figure~\ref{fig:eff_mass} shows a comparison of the effective mass
computed by the fRG to that obtained from second order perturbation
theory and to the Bethe-Ansatz result. Since second order
perturbation theory will serve us several times as a reference
solution we briefly sketch how it is determined. When evaluating the
frequency-dependent second order diagram for the self-energy we use
propagators dressed with the restricted self-consistent Hartree-Fock
self-energy as internal propagators. The self-consistent Hartree-Fock
solution is obtained from the self-consistency equation
\begin{subequations}
  \begin{align}
    \Sigma^\ix{Ret}_\sigma
    &= U \left \langle n_{\overline \sigma} \right \rangle
    =
    - \frac{\im}{2 \pi} U \int \! \df \omega \, G^<_{\overline
      \sigma}(\omega)
    \\
    &=
    \frac{U}{2 \pi} \int \! \df \omega \, f_\ix{eff}(\omega)
    \frac{\Gamma}{(\omega - \epsilon_{\overline \sigma} -
      \Sigma^\ix{Ret}_{\overline \sigma})^2 + \Gamma^2/4}.
  \end{align}
\end{subequations}
This equation has a unique solution for small $U$, but there are three
solutions at larger values of $U$ (e.g. for $U > \pi \Gamma/2$ for
$T=0$, $B=0$). Two of them feature a local moment,
$\Sigma^\ix{Ret}_\uparrow \neq \Sigma^\ix{Ret}_\downarrow$, even for
$B=0$.\cite{anderson} The notion ``restricted Hartree-Fock'' refers to
the third solution which is energetically unfavorable as compared to
the other two ones but is nevertheless preferable because of
preserving spin symmetry, $\Sigma^\ix{Ret}_\uparrow =
\Sigma^\ix{Ret}_\downarrow$. In the particle-hole symmetric case, for
instance, the restricted Hartree-Fock self-energy
$\Sigma^\ix{Ret}_\uparrow = \Sigma^\ix{Ret}_\downarrow = U/2$ aligns
the single-particle levels with the chemical potential. We use this
perturbation theory only for $B=0$. For finite magnetic field it is
not clear which solution of the self-consistency equation ought to be
used to dress the single-particle propagators.

In Fig.~\ref{fig:eff_mass} we see that the effective mass computed
from the fRG does not increase exponentially for large $U$, but it is
comparable to the one computed with second order perturbation
theory. As a consequence the width of the Kondo resonance of the
spectral function resulting from the fRG approximation does not shrink
exponentially for increasing $U$. Figure~\ref{fig:eff_mass} also shows
the effective mass obtained from the fRG with full frequency dependent
back-coupling of the self-energy into the flow. As mentioned in
section~\ref{subsec:feedback} this feedback leads to an artificial
broadening of spectral features which becomes apparent here in the too
low values for the effective mass.
\begin{figure}
  \begin{center}
    \includegraphics{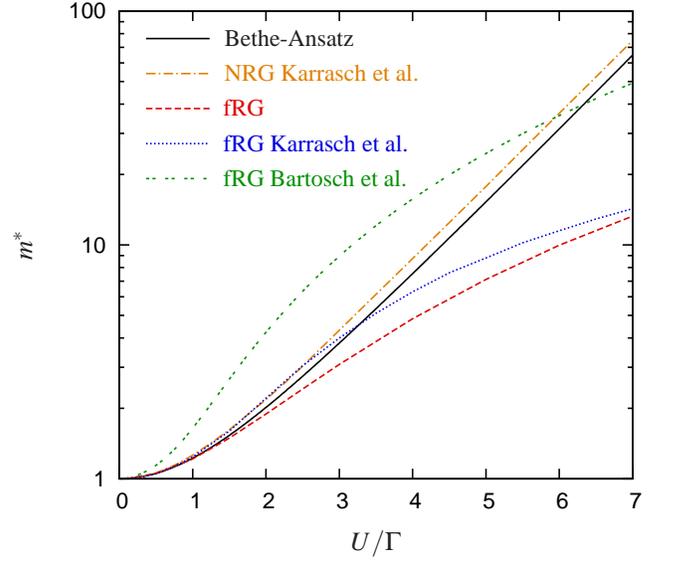}
  \end{center}
  \caption{The effective mass as function of $U$ for $V_\ix{g} = 0$,
    $T=0$, $B=0$, $V=0$. The figure compares the Bethe-Ansatz and fRG
    data of Fig.~\ref{fig:eff_mass} with results of the Matsubara
    fRG in ``approximation 1'' of Ref.~\onlinecite{karrasch2},
    the Matsubara fRG with partial bosonization in the spin-rotational
    invariant version of Ref.~\onlinecite{bartosch} (data
    extracted from the preprint arXiv:0811.2809v1), and with the NRG
    data used for comparison in Refs.~\onlinecite{karrasch2,
      bartosch}.}
  \label{fig:eff_mass_extra}
\end{figure}
In Fig.~\ref{fig:eff_mass_extra} a comparison with the results of
recent Matsubara fRG implementations is given.

\begin{figure}
  \begin{center}
    \includegraphics{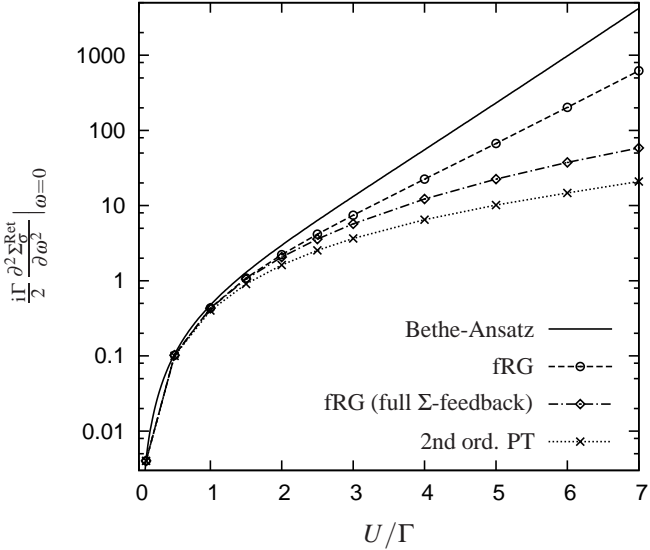}
  \end{center}
  \caption{The second derivative of the self-energy as function of $U$
    for $V_\ix{g} = 0$, $T=0$, $B=0$, $V=0$. The fRG produces an
    exponential increase with the exponent $4 U/\pi \Gamma$, whereas
    the exact exponent is $\pi U / 2 \Gamma$. Second order
    perturbation theory and the fRG with the full frequency dependent
    self-energy feedback (compare section~\ref{subsec:feedback}) do
    not yield an exponential increase.}
  \label{fig:2nd_deriv}
\end{figure}
From the expansion~\eqref{eq:SIAM_SE_expansion} and
from~\eqref{eq:sus_asmyptot} and~\eqref{eq:eff_mass_chi1} it also
follows that
\begin{equation}
  \frac{\im \Gamma}{2} \left. \frac{\partial^2
      \Sigma^\ix{Ret}_\sigma}{\partial \omega^2}\right|_{\omega = 0}
  =
  \frac{(\tilde \chi_\ix{s} - \tilde \chi_\ix{c})^2}{4}
  \; \simeq \; {m^\ast}^2, \qquad U \gtrsim \pi\Gamma,
    \label{eq:2nd_deriv_eff_mass}
  \end{equation}
increases exponentially with $U$ in the Kondo
regime. Figure~\ref{fig:2nd_deriv} presents the respective data
obtained by fRG and second order perturbation theory in comparison to
the Bethe-Ansatz result. In contrast to second order perturbation theory the
fRG produces an exponential scale as has been proven already in
Eq.~\eqref{eq:Sigma_second_scaling}.
Using Eqs.~\eqref{eq:T_K} and~\eqref{eq:2nd_deriv_eff_mass} we can relate
this exponential behaviour to a Kondo scale
\begin{equation}
  \left[T_\ix{K}\right]_\ix{fRG} = 4 \sqrt{\frac{U \Gamma}{3 \pi}}
  \exp\left(- \frac{2}{\pi} \frac{U}{\Gamma}\right).
\end{equation}
The identical exponential behaviour has been found to govern the
pinning of the spectral weight to the chemical potential in the
Matsubara fRG with frequency independent truncation
scheme~\cite{karrasch1, andergassen}. With $2/\pi \simeq
0.64$ the prefactor is slightly smaller than the correct one in
Eq.~\eqref{eq:T_K1}, $\pi/4 \simeq
0.79$. Figure~\ref{fig:2nd_deriv} also shows that the fRG with feedback
of the full frequency dependent self-energy into the flow does not
produce an exponential scale in the second derivative of the
self-energy. This illustrates again that the static feedback scheme
which we use as standard is advantageous, compare
section~\ref{subsec:feedback}.

\begin{figure}
  \begin{center}
    \includegraphics{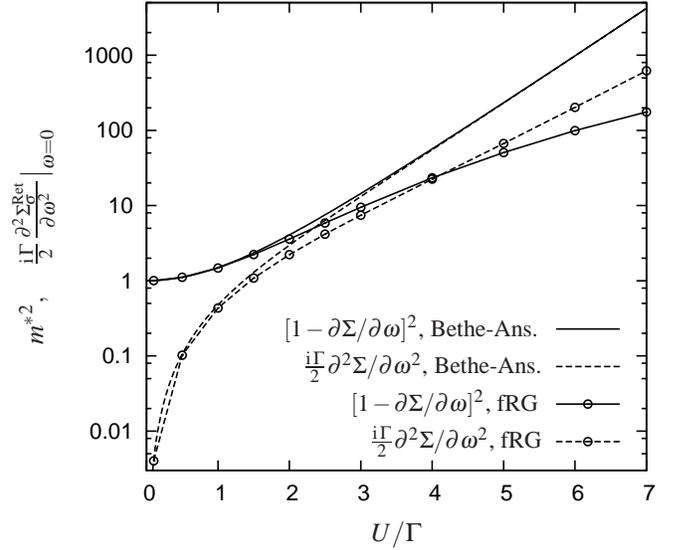}
  \end{center}
  \caption{Comparison of the square of the effective mass and of the
    second derivative of the self-energy as function of $U$ for
    $V_\ix{g} = 0$, $T=0$, $B=0$, $V=0$. In the exact solution both 
    have identical large $U$ asymptotics, ${m^\ast}^2$ being always
    the larger one of both. In the fRG data the curves cross at $U$
    slightly above $4 \Gamma$.}
  \label{fig:eff_mass_comp}
\end{figure}
The fact that the fRG produces an exponential behaviour in $\Imag
\Sigma''$ but not in the effective mass constitutes an inconsistency
which manifests that the method captures Kondo physics only
partly. The exact result satisfies
\begin{equation}
  {m^\ast}^2 = \frac{(\tilde \chi_\ix{s} + \tilde \chi_\ix{c})^2}{4}
  >
  \frac{(\tilde \chi_\ix{s} - \tilde \chi_\ix{c})^2}{4}
  = \frac{\im \Gamma}{2} \left. \frac{\partial^2
      \Sigma^\ix{Ret}_\sigma}{\partial \omega^2}\right|_{\omega = 0},
\end{equation}
where both sides asymptotically approach $\tilde \chi_\ix{s}^2/4$.
Figure~\ref{fig:eff_mass_comp} illustrates that the fRG result
violates this inequality for $U > 4 \Gamma$. Therefore we restrict the
following discussions to $U \le 4 \Gamma$.

\begin{figure}
  \begin{center}
    \includegraphics{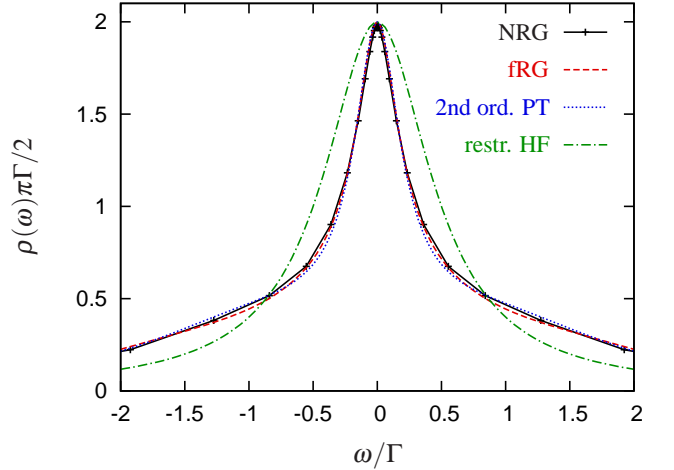}
  \end{center}
  \caption{Spectral function for $V_\ix{g} = 0$, $T=0$, $B=0$,
    $V=0$ and $U=2 \Gamma$. The results of fRG and second order
    perturbation theory agree very well with NRG data. The peak shape
    differs already significantly from the Lorentzian form produced by
    static approximations such as restricted Hartree-Fock.}
  \label{fig:spectral_dens1}
\end{figure}
\begin{figure}
  \begin{center}
    \includegraphics{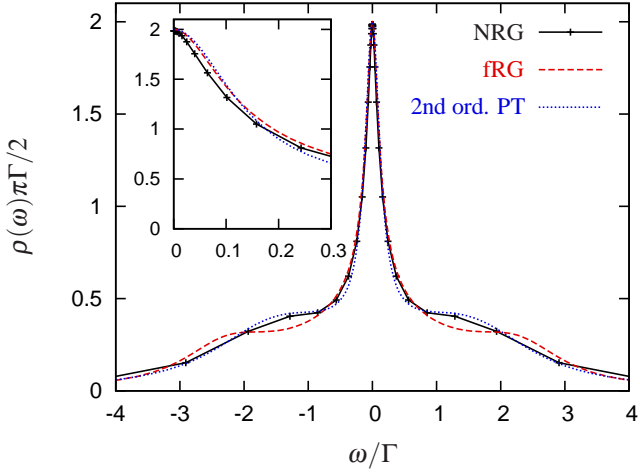}    
  \end{center}
  \caption{Spectral function for $V_\ix{g} = 0$, $T=0$, $B=0$,
    $V=0$ and $U=3 \Gamma$.}
  \label{fig:spectral_dens2}
\end{figure}
\begin{figure}
  \begin{center}
    \includegraphics{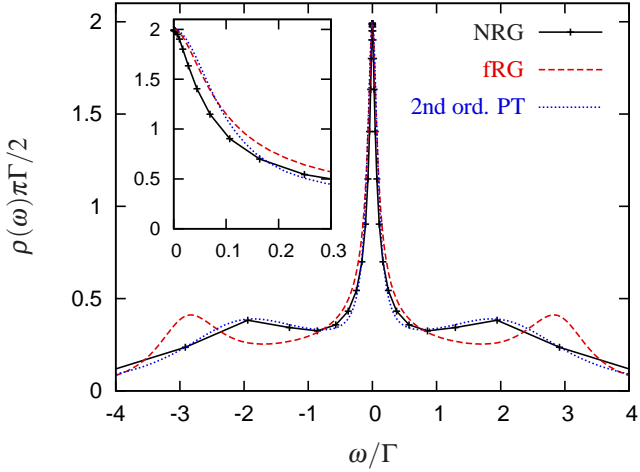}    
  \end{center}
  \caption{Spectral function for $V_\ix{g} = 0$, $T=0$, $B=0$,
    $V=0$ and $U=4 \Gamma$. The resonance at $\omega=0$ produced by
    fRG and by second order perturbation theory is too broad. The side
    peaks resulting from the fRG calculation are situated at too large
    $\abs{\omega}$.}
  \label{fig:spectral_dens3}
\end{figure}
Figures~\ref{fig:spectral_dens1}--\ref{fig:spectral_dens3} show
spectral functions in the particle-hole symmetric case at $B=0$,
$T=0$, in equilibrium. Results of the fRG and of second order
perturbation theory are compared to essentially exact NRG
data\cite{costi}. For $U \lesssim 2 \Gamma$ all three
methods yield nearly identical results. For $U = 2 \Gamma$ the shape
of the peak differs already significantly from the Lorentzian form
produced by static approximations such as restricted Hartree-Fock. At
this interaction strength the difference occurs almost exclusively due to
the second order self-energy diagrams which are also captured exactly
by the fRG.  For $U = 3\Gamma, 4\Gamma$ the resonance peaks produced
by fRG and second order perturbation theory are too broad, see insets
of Figs.~\ref{fig:spectral_dens2} and~\ref{fig:spectral_dens3},
which is a consequence of the effective mass being too small as
discussed above. The overall shape of the spectral function is
reproduced better by second order perturbation theory than by fRG. The
fRG has a tendency to shift spectral weight too far away from the
central peak; in Fig.~\ref{fig:spectral_dens3} the side peaks
computed by fRG are situated at $\abs{\omega} \simeq 3 \Gamma$ whereas
they are expected to be at $\abs{\omega} \simeq U/2 = 2 \Gamma$. For
the computation of equilibrium spectra it is hence preferable to
resort to the frequency dependent Matsubara fRG which has been found
to be clearly superior in this respect to second order perturbation
theory at $U = 2.5 \Gamma$.\cite{karrasch2}

\begin{figure}
  \begin{center}
    \includegraphics{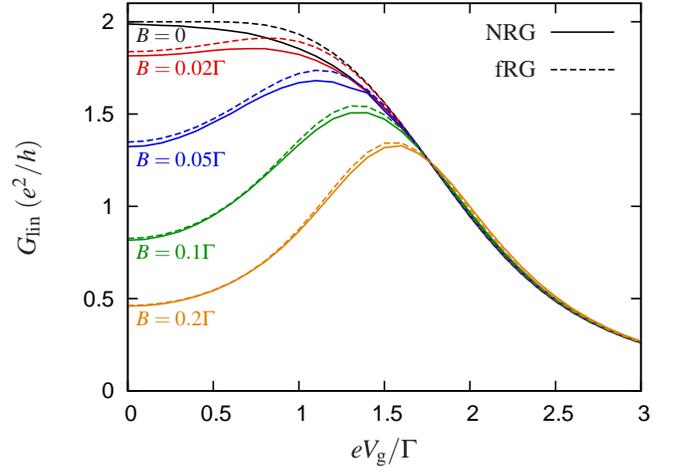}
  \end{center}
  \caption{Linear conductance as function of gate voltage for $T=0$,
    $U=4\Gamma$ and different magnetic fields.}
  \label{fig:lin_cond_VG_B}
\end{figure}
\begin{figure}
  \begin{center}
    \includegraphics{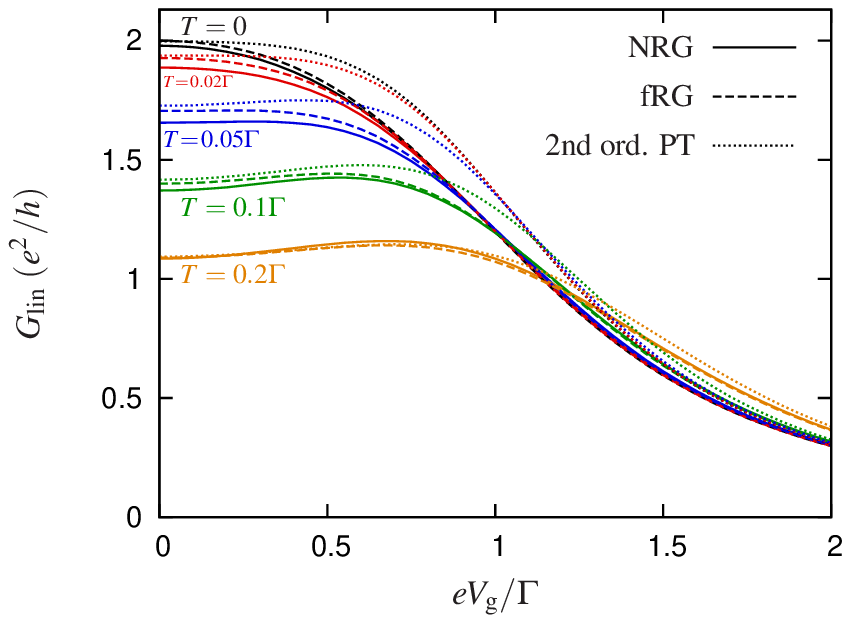}
  \end{center}
  \caption{Linear conductance as function of gate voltage for $B=0$,
  $U=2\Gamma$ and different temperatures.}
  \label{fig:lin_cond_VG_T_U2}
\end{figure}
\begin{figure}
  \begin{center}
    \includegraphics{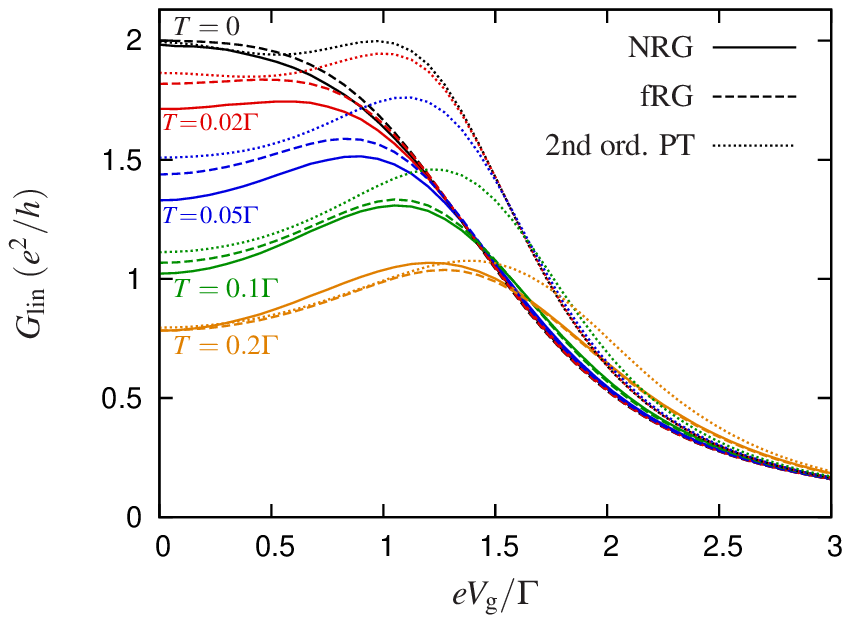}
  \end{center}
  \caption{Linear conductance as function of gate voltage for $B=0$,
    $U=3\Gamma$ and different temperatures.}
  \label{fig:lin_cond_VG_T_U3}
\end{figure}
The linear conductance as function of gate voltage for different
magnetic fields at $T=0$ and $U=4\Gamma$ is shown in
Fig.~\ref{fig:lin_cond_VG_B}. The fRG data agree very well with NRG
results; the frequency dependent Keldysh fRG obviously maintains the
high quality in describing $G_\ix{lin}(V_\ix{g}, B; T=0)$ which has
already been achieved by the fRG with frequency independent truncation
scheme~\cite{karrasch1, andergassen}.

Figures~\ref{fig:lin_cond_VG_T_U2} and~\ref{fig:lin_cond_VG_T_U3}
present the linear conductance as function of gate voltage for
different temperatures at $B=0$ and $U=2\Gamma,3\Gamma$. In contrast to second
order perturbation theory the fRG reproduces the NRG results for the
width of the plateau at $T=0$ and the position of the maxima for $T >
0$ quite accurately. The decrease of the conductance at $V_\ix{g} = 0$
with increasing temperature is captured not completely correct by the
fRG but distinctively better than by second order perturbation
theory. The fRG provides an altogether acceptable description of
$G_\ix{lin}(V_\ix{g}, T)$ for $U \lesssim 3$. This is an important
improvement compared to previous fRG approaches. The static
fRG~\cite{karrasch1, andergassen} is in principle unable to
reproduce the minimum of $G_\ix{lin}$ at $V_\ix{g}=0$.  The reason is
that the linear conductance is given by
\begin{equation}
  G_\ix{lin} = e^2 \frac{\Gamma^\ix{L} \Gamma^\ix{R}}{\Gamma} 
  \int \! \df \omega \, \left(- \frac{\partial f(\omega)}{\partial
      \omega}
  \right) \rho(\omega),
  \label{eq:SIAM_lin_cond}
\end{equation}
as follows from the current formula~\eqref{eq:SIAM_current}. In any
static approximation at $B=0$ the spectral function $\rho(\omega)$ is
represented by a Lorentzian peak with fixed width and height, centered
at the renormalized level position. Obviously
Eq.~\eqref{eq:SIAM_lin_cond} can produce a single maximum of
$G_\ix{lin}(V_\ix{g})$ when the renormalized level is aligned with the
chemical potential, but no local minimum.  For the frequency dependent
Matsubara fRG on the other hand, problems with the analytic
continuation from imaginary to real frequencies obstructed the
computation of the linear conductance.\cite{karrasch2} Although it has
been later found possible to circumvent this obstacle, the results for
$G_\ix{lin}(V_\ix{g})$ are less accurate than those obtained from the
Keldysh fRG.\cite{karrasch3} A static fRG scheme based on a real
frequency cut-off in Keldysh formalism has also been found to
reproduce qualitatively the shape of $G_\ix{lin}(V_\ix{g},
T)$.\cite{gezzi_thesis} In view of the argument presented after
Eq.~\eqref{eq:SIAM_lin_cond} this has to be the consequence of a
renormalization of the level broadening. The latter however can be
achieved only in a dynamic approximation as can be seen from the
fluctuation dissipation theorem~\eqref{eq:fluc_diss_Sigma} that
manifestly requires a frequency dependent Keldysh component of the
self-energy if $\Sigma^\ix{Ret}$ has nonvanishing imaginary
part. Therefore we suspect that the corresponding result of the static
flow scheme used in Ref.~\onlinecite{gezzi_thesis} is an artifact
connected to the violation of causality as consequence of the real
frequency cut-off.

The results of our Keldysh fRG for the current as function of bias
voltage at $V_\ix{g} = 0$, $B=0$, $T=0$ have been compared in
Ref.~\onlinecite{heidrich-meisner} to data obtained by a TD-DMRG
treatment. Excellent agreement between both methods has been found for
$U \le 4 \Gamma$. A more sensitive quantity for comparisons is however
the differential conductance.  Unfortunately, the TD-DMRG
  conductance data existing by now are not enough for a meaningful
  comparison.
\begin{figure}
  \begin{center}
    \includegraphics{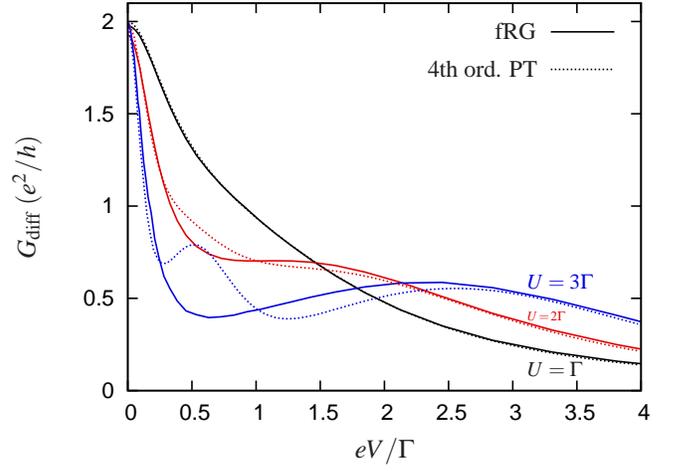}
  \end{center}
  \caption{Differential conductance as function of bias voltage for
    $V_\ix{g} = 0$, $B=0$ and different values of $U$. The temperature
    is zero except for the two fRG curves with $U=\Gamma,2\Gamma$
    where $T=0.02\Gamma$; the corresponding two data sets for the
    current at $T=0$ have not been sufficiently smooth to allow for
    numerical differentiation. The fourth order perturbation theory
    results have been extracted from the
    preprint~arXiv:cond-mat/0211616v1 to
    Ref.~\onlinecite{fujii}.}
  \label{fig:diff_cond_FU}
\end{figure}
In Fig.~\ref{fig:diff_cond_FU} we compare the differential
conductance as computed by fRG to the fourth order perturbation theory
results of Ref.~\onlinecite{fujii}. The agreement for small and
large voltages is rather good. A discrepancy is found at intermediate
$V$ where the fRG does not reproduce the anomalous peak found by
Ref.~\onlinecite{fujii} for sufficiently large $U$. Probably the
anomalous peak is an artifact of fourth order perturbation theory;
this conclusion is supported by recent nonequilibrium QMC results for
the SIAM.\cite{werner}

\begin{figure}
  \begin{center}
    \includegraphics{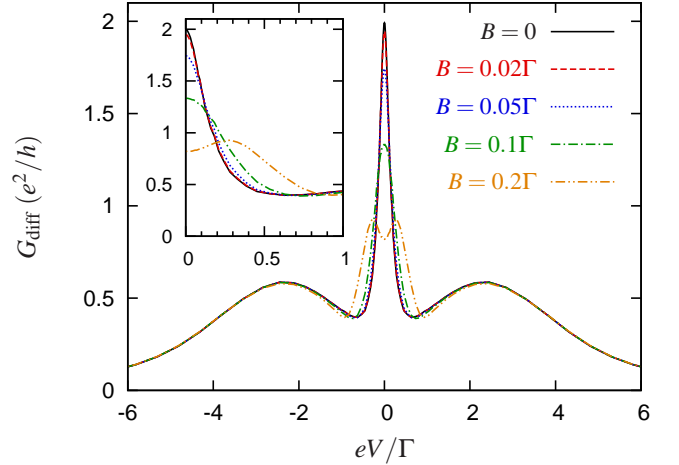}
  \end{center}
  \caption{FRG results for the differential conductance as function of
    bias voltage for $V_\ix{g} = 0$, $T=0$, $U=3\Gamma$ and different
    values of the magnetic field.}
  \label{fig:diff_cond_B_U3}
\end{figure}
\begin{figure}
  \begin{center}
    \includegraphics{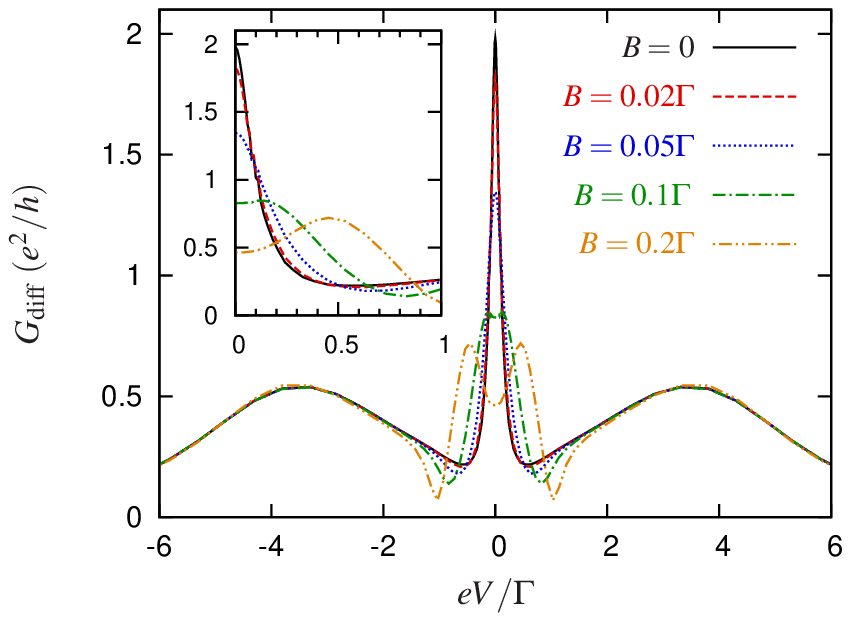}
  \end{center}
  \caption{FRG results for the differential conductance as function of
    bias voltage for $V_\ix{g} = 0$, $T=0$, $U=4\Gamma$ and different
    values of the magnetic field.}
  \label{fig:diff_cond_B_U4}
\end{figure}
\begin{figure}
  \begin{center}
    \includegraphics{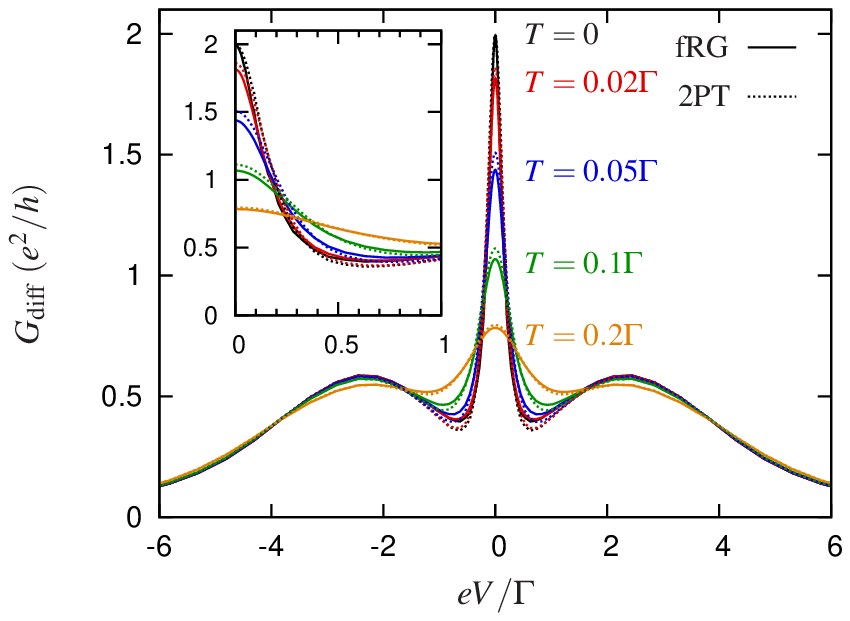}
  \end{center}
  \caption{FRG results for the differential conductance as function of
    bias voltage for $V_\ix{g} = 0$, $B=0$, $U=3\Gamma$ and different
    values of temperature, in comparison with second order
      perturbation theory.}
  \label{fig:diff_cond_T_U3}
\end{figure}
\begin{figure}
  \begin{center}
    \includegraphics{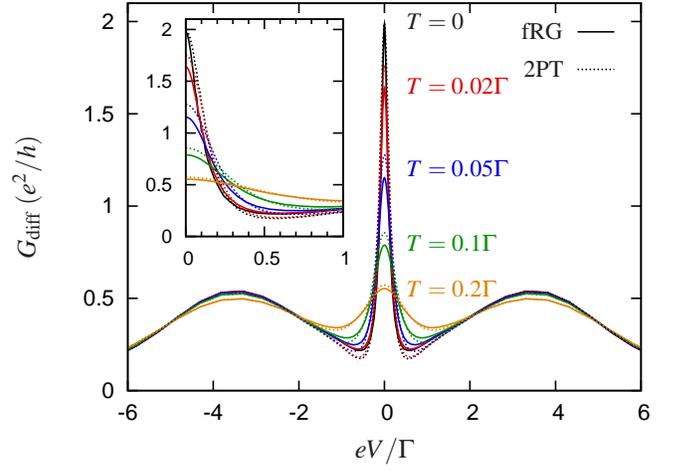}
  \end{center}
  \caption{FRG results for the differential conductance as function of
    bias voltage for $V_\ix{g} = 0$, $B=0$, $U=4\Gamma$ and different
    values of temperature, in comparison with second order
      perturbation theory.}
  \label{fig:diff_cond_T_U4}
\end{figure}
The data for $G_\ix{diff}(V)$ at different magnetic fields presented
in Figs.~\ref{fig:diff_cond_B_U3} and~\ref{fig:diff_cond_B_U4} could
in principle be compared to results of the recently introduced
scattering states NRG~\cite{anders}. The data of the latter approach
is however is still too noisy to allow for definite
conclusions. Finally, Figs.~\ref{fig:diff_cond_T_U3}
and~\ref{fig:diff_cond_T_U4} present the differential conductance for
different temperatures in comparison to second order perturbation
theory.

A detailed comparison of nonequilibrium results of the fRG with that
of recent numerical methods is given in Ref.~\onlinecite{eckel}.


\section{Conclusion}
\label{sec:SIAM_conclusion}

In this paper we studied a frequency dependent fRG approximation for
the Anderson impurity model. In equilibrium this model provides the
possibility to compare our results to reliable data from Bethe-Ansatz
or the NRG. Existing studies of the model within Matsubara fRG give
indication on how to approach the problem within Keldysh
formalism. The non-equilibrium properties of the model are subject of
numerous investigations in recent time; however, a consistent picture
did not yet emerge.

In order to preserve the Fermi liquid properties of the model we have
chosen as flow parameter the level broadening $\Gamma/2=\Delta$. In
section~\ref{sec:basic_approx} we studied the basic approximation
scheme with static vertex renormalization and reproduced the results
of the static Matsubara fRG. For the more advanced dynamic second
order truncation scheme it was necessary to parameterize the frequency
dependence of the vertex in order to obtain tractable equations.  The
self-energy which we feed back into the flow is static, and does not
contain interaction induced quasiparticle decay rates.  Nevertheless,
it turned out that the results can be trusted up to values of the
Coulomb interaction $U\lesssim 3\Gamma=6\Delta$, which are typical
parameter regimes also used in diverse recent numerical studies of the
model, see, e.g., Ref.~\onlinecite{eckel}.  Furthermore, for technical
reasons we restricted our study to $T_\ix{L} = T_\ix{R}$ and
$\Gamma_\ix{L} = \Gamma_\ix{R}$.

At the particle-hole symmetric point the description of the spectral
function obtained by our fRG is comparable to second order
perturbation theory at weak interactions, while getting worse for
increasing values of $U$. But in contrast to perturbation theory the
fRG generates exponential behaviour in certain quantities. The
exponent of the Kondo temperature extracted from the second order
derivative of the self-energy is identical to that found from the
pinning mechanism in the static fRG, which is not far from the exact
value. The fact that the effective mass derived from the fRG does not
exhibit an exponential scale is related to the missing renormalization
of the static part of the two-particle vertex in the present
second-order truncation scheme. An extension which takes into account
contributions from the flow of the three-particle vertex could
probably cure this deficiency. In that case we expect that also the
feedback of the full frequency dependent self-energy into the flow
will be favorable compared to the static feedback used here.

A big success of the method is the very good description of the linear
conductance as function of gate voltage, temperature and magnetic
field.  At zero temperature excellent conductance data are obtained
even for $U = 4 \Gamma$ while for finite $T$ acceptable quality is
achieved for $U \lesssim 3 \Gamma$ with much better results than in
perturbation theory. In this respect the method is a utile complement
to existing fRG approaches: the static fRG is applicable only at
$T=0$, and the frequency dependent Matsubara fRG does not produce
conductance data of comparable quality.

Concerning non-equilibrium properties the results of the fRG compare
rather good to those of other methods for $U \lesssim 3 \Gamma$ as far
as comparisons are possible.  In particular, we obtained very good
agreement for the nonlinear current $I(V)$ with TD-DMRG data at the
most critical point $V_g=T=B=0$. Together with the fact that our
equilibrium results agree well with NRG at finite magnetic field and
finite temperature, we expect our results for the nonlinear
conductance $G(V)$ to be reliable at finite $T$ and $B$, although no
firm benchmark has yet been obtained by other techniques.

An issue to be addressed for future research is the problem of current
conservation by the fRG. For the finite bias data which we presented
in section~\ref{sec:SIAM_results} current conservation is given as a
consequence of the complete symmetry established by $eV_\ix{g} =
(\mu_\ix{L} + \mu_\ix{R})/2$, $T_\ix{L} = T_\ix{R}$, $\Gamma_\ix{L} =
\Gamma_\ix{R}$. In more general cases the fRG is not expected to be a
current conserving method. Similar problems are known for other
methods, for example perturbation theory~\cite{hershfield}. 
Only when reliable benchmarks become available, it can be concluded,
whether the violation of current conservation generates serious
errors in the nonlinear conductance $G(V)$, and whether more elaborate
parametrizations within nonequilibrium fRG are necessary.

The comparison of the frequency dependent Keldysh and Matsubara fRG
reveals that in spite of being closely related, the two methods are
not equivalent in equilibrium. The choice of an imaginary frequency
cut-off as used in Matsubara fRG is not possible in the Keldysh fRG
without destroying the Fermi liquid property at the particle-hole
symmetric point. For the Keldysh fRG the static self-energy feedback
into the flow is preferable to the feedback of the full one, as
opposed to the Matsubara fRG. And while the Matsubara fRG yields
better results for the form of the spectral function at the
particle-hole symmetric point, the Keldysh fRG produces better data
for the linear conductance as function of gate voltage, making
accessible finite temperatures where the Matsubara fRG is handicapped
by the problem of analytic continuation.

In total our investigation gave insight on how a frequency dependent
approximation to the Keldysh fRG can be constructed, and produced a
flexible method to compute properties of the Anderson impurity model
for moderate interactions. A fundamental problem of applying the fRG
in the considered truncation scheme to this model is that the
justification for the performed approximations can only be given by
perturbative arguments. This difficulty is rooted in the regular
perturbative expansion of the model. This contrasts with the situation
in models with low energy divergencies which typically allow to
distinguish relevant and irrelevant contributions to the RG flow by
power counting. It would be interesting to figure out whether
perturbative RG techniques are still capable to capture the
exponentially small scale of the Kondo temperature in the effective
mass for large Coulomb interactions. Within our proposed approximation
scheme, this scale already appeared in the second order derivative of
the self-energy, showing that our perturbative RG scheme has the
potential to extract such scales. Therefore, it might appear possible
that certain contributions from the 3-particle vertex can help to
reveal the same scale for the effective mass as well. We expect that
the approximation can be systematically improved by including the
influence of higher order vertices on the flow, if care is taken to
formulate the approximation scheme in accordance with Ward
identities. It might be also interesting to access the Kondo limit of
the Anderson model using an approach based on Hubbard-Stratonovich
fields decomposition as proposed in Ref.~~\onlinecite{bartosch} in
combination with nonequilibrium fRG.

In summary, the method proposed in the present paper seems to be
sufficient to describe Coulomb interactions up to the order of the
band width $D$ of the local system. For a single-level quantum dot,
the band width is given by $D\sim\Gamma$, whereas in quantum wires it
is given by $D\sim t$, where $t$ is the hopping matrix element. In
both cases, the nonequilibrium fRG schemes of this paper and of
Ref.~\onlinecite{jakobs2} seem to be reliable for $U\lesssim D$,
providing the hope for a unified approach describing the whole
crossover from local to extended quantum systems.


\section*{Acknowledgments}

We are especially grateful to Theo Costi for providing us with a
program to produce NRG data for comparison with the results of the
fRG. We thank Christoph Karrasch and Volker Meden for numerous
valueable discussion on applying the fRG to the SIAM and for providing
data for comparison. We thank Sabine Andergassen, Frithjof Anders,
Johannes Bauer, Fabian Heidrich-Meisner, and Frank Reininghaus
for helpful discussions. This work was supported by the
DFG-Forschergruppe 723.


\appendix

\section{Identification of independent components of
  $\varphi^{\ix{p},\ix{x},\ix{d}}_\lambda$}
\label{sec:structure}

The numerous Keldysh and spin components of the functions
$\varphi^{\ix{p},\ix{x},\ix{d}}_\lambda$ are not independent of each
other. In this section we identify a set of independent components
from which all the others can be determined. For that purpose we make
extensive use of the relations for the vertex functions which
originate from permutation of particles, complex conjugation,
causality and the KMS conditions and are described in
Ref.~\onlinecite{jakobs1}.  As explained in that reference, the
hybridization flow parameter preserves these relations during the
truncated RG flow. This is as well the case under the additional
approximation induced by the replacement~\eqref{eq:replacement} since
the static renormalized vertices appearing in the flow
equation~\eqref{eq:approxB} have the same spin and Keldysh structure
as the original bare vertex. The relations can be even used for the
three individual functions $\varphi^{\ix{p},\ix{x},\ix{d}}_\lambda$
since the diagrammatic structure of the three corresponding flow
equations is invariant under the manipulations done in the proof of
those relations, cf.~Ref.~\onlinecite{jakobs1}. The only
exception is that permutation of either the incoming or the outgoing
indices maps $\varphi^{\ix{x}}$ onto $\varphi^{\ix{d}}$ and vice
versa.


\subsection{Analysis of the spin structure}
\label{subsec:spin_structure}

Due to the spin structure of $\overline v$,
$(\Phi^\ix{p,x,d}_\lambda)$, and $(I^\ix{\,pp,xph}_\lambda)$ certain
spin components of $\varphi^\ix{p,x,d}_\lambda$ vanish: Knowing that
\begin{subequations}
  \begin{align}
    \left.
      \begin{aligned}
        \overline v_{\sigma'_1 \sigma'_2|\sigma_1 \sigma_2}
        &=0
        \\
        (\Phi^\ix{p}_\lambda)_{\sigma'_1 \sigma'_2|\sigma_1 \sigma_2}
        &= 0
        \\
        (\Phi^\ix{x}_\lambda)_{\sigma'_1 \sigma'_2|\sigma_1 \sigma_2} +
        (\Phi^\ix{d}_\lambda)_{\sigma'_1 \sigma'_2|\sigma_1 \sigma_2}
        &= 0
      \end{aligned}
    \right\} & \text{if $\sigma_1 \neq \overline \sigma_2$ or $\sigma'_1
      \neq \overline \sigma'_2$},
    \\
    (\Phi^\ix{x,d}_\lambda)_{\sigma'_1 \sigma'_2|\sigma_1 \sigma_2} =0
    \quad &
    \text{if $\sigma'_1 + \sigma'_2 \neq \sigma_1 + \sigma_2$},
  \end{align}
\end{subequations}
we conclude from the flow equation~\eqref{eq:approxB} that
\begin{subequations}
  \begin{align}
    \label{eq:phi_vanish_pp}
    (\varphi^\ix{p}_\lambda)_{\sigma'_1 \sigma'_2|\sigma_1 \sigma_2} &=
    0, \quad\text{if $\sigma_1 \neq \overline \sigma_2$ or $\sigma'_1
      \neq \overline \sigma'_2$},
    \\
    (\varphi^\ix{x}_\lambda)_{\sigma'_1 \sigma'_2|\sigma_1 \sigma_2} &=
    0, \quad\text{if $\sigma_1 + \sigma_2 \neq \sigma'_1 + \sigma'_2$},
    \\
    (\varphi^\ix{d}_\lambda)_{\sigma'_1 \sigma'_2|\sigma_1 \sigma_2} &=
    0, \quad\text{if $\sigma_1 + \sigma_2 \neq \sigma'_1 + \sigma'_2$}.
  \end{align}
\end{subequations}
Concerning the remaining spin components, we can restrict our study to
\begin{equation}
  \label{eq:spin_comp}
  (\varphi^\ix{p}_\lambda)_{\sigma \overline \sigma|\sigma \overline \sigma},\;
  (\varphi^\ix{x}_\lambda)_{\sigma \overline \sigma|\sigma \overline \sigma},\;
  (\varphi^\ix{d}_\lambda)_{\sigma \overline \sigma|\sigma \overline \sigma},\;
  (\varphi^\ix{d}_\lambda)_{\sigma \sigma|\sigma \sigma},
\end{equation}
because the other ones can be derived from these by permutations of
particle indices,
\begin{subequations}
  \label{eq:phi_permut_prop}
  \begin{align}
    \label{eq:phi_permut_prop_p}
    (\varphi^\ix{p}_\lambda)^{\alpha'_1 \alpha'_2|\alpha_1 \alpha_2}_{\sigma \overline
      \sigma|\overline \sigma \sigma}(\Pi)
    &=
    - (\varphi^\ix{p}_\lambda)^{\alpha'_1 \alpha'_2|\alpha_2 \alpha_1}_{\sigma \overline
      \sigma|\sigma \overline \sigma}(\Pi),
    \\
    \label{eq:phi_permut_prop_x}
    (\varphi^\ix{x}_\lambda)^{\alpha'_1 \alpha'_2|\alpha_1 \alpha_2}_{\sigma \overline
      \sigma|\overline \sigma \sigma}(X)
    &=
    - (\varphi^\ix{d}_\lambda)^{\alpha'_1 \alpha'_2|\alpha_2 \alpha_1}_{\sigma \overline
      \sigma|\sigma \overline \sigma}(-X),
    \\
    \label{eq:phi_permut_prop_d}
    (\varphi^\ix{d}_\lambda)^{\alpha'_1 \alpha'_2|\alpha_1 \alpha_2}_{\sigma \overline
      \sigma|\overline \sigma \sigma}(\Delta)
    &=
    - (\varphi^\ix{x}_\lambda)^{\alpha'_1 \alpha'_2|\alpha_2 \alpha_1}_{\sigma \overline
      \sigma|\sigma \overline \sigma}(-\Delta),
    \\
    \label{eq:phi_permut_prop_dd}
    (\varphi^\ix{x}_\lambda)^{\alpha'_1 \alpha'_2|\alpha_1 \alpha_2}_{\sigma
      \sigma|\sigma \sigma}(X)
    &=
    - (\varphi^\ix{d}_\lambda)^{\alpha'_1 \alpha'_2|\alpha_2 \alpha_1}_{\sigma
      \sigma|\sigma \sigma}(-X).
  \end{align}
\end{subequations}


\subsection{Analysis of the Keldysh structure}
\label{subsec:Keldysh_structure}

In the following it is convenient to describe the Keldysh components
of two-particle functions in terms of block matrices, in which they
are arranged according to the table
\begin{equation}
  \left(
    \begin{array}{cc|cc}
      (11|11) & (11|21) & (11|12) & (11|22)
      \\
      (21|11) & (21|21) & (21|12) & (21|22)
      \\
      \hline
      (12|11) & (12|21) & (12|12) & (12|22)
      \\
      (22|11) & (22|21) & (22|12) & (22|22)
    \end{array}
  \right).
\end{equation}
The indices $\alpha'_2, \alpha_2$ of an index tuple $(\alpha'_1
\alpha'_2|\alpha_1 \alpha_2)$ indicate which of the blocks is to be
chosen, while $\alpha'_1, \alpha_1$ defines the position inside a
block. The bare interaction vertex~\eqref{eq:bare_vertex_model} for
example is described by the block matrix
\begin{equation}
  \label{eq:vert_Keldysh_struct}
  \overline v^{\alpha'_1 \alpha'_2|\alpha_1 \alpha_2}_{\sigma'_1
    \sigma'_2|\sigma_1 \sigma_2} = 
  \frac{1}{2} \overline v_{\sigma'_1 \sigma'_2|\sigma_1 \sigma_2}
  \left(
    \begin{array}{cc|cc}
      0 & 1 & 1 & 0
      \\
      1 & 0 & 0 & 1
      \\
      \hline
      1 & 0 & 0 & 1
      \\
      0 & 1 & 1 & 0
    \end{array}
  \right)_{\alpha'_1 \alpha'_2|\alpha_1 \alpha_2}.
\end{equation}

From the flow equation~\eqref{eq:approxB_pp} it follows that
$\varphi^\ix{p}_\lambda(\Pi)$ has the Keldysh structure
\begin{equation}
  \label{eq:pp_struct}
  \varphi^\ix{p}_\lambda
  =
  \left(
    \begin{array}{cc|cc}
      c & d & d & c
      \\
      a & b & b & a
      \\
      \hline
      a & b & b & a
      \\
      c & d & d & c
    \end{array}
  \right)_\lambda,
\end{equation}
which can be seen as follows: the initial value
$\varphi^\ix{p}_{\lambda =\infty} \equiv 0$ is consistent with the
structure~\eqref{eq:pp_struct}. Since $\varphi^\ix{p}_\lambda(\Pi)$
has this structure, so does $[\overline v +
\varphi^\ix{p}_\lambda(\Pi) + \Phi^\ix{x}_\lambda +
\Phi^\ix{d}_\lambda]$. From Eq.~\eqref{eq:approxB_pp} it can then
be derived that $\df \varphi^\ix{p}_\lambda / \df \lambda$ has the
structure~\eqref{eq:pp_struct} as well. Therefore this structure is
conserved during the flow.

Exploiting the causality relation $(\varphi^\ix{p}_\lambda)^{22|22}
\equiv 0$ and the transformation properties with respect to complex
conjugation we find
\begin{equation}
  \renewcommand{\arraystretch}{1.3}
  \label{eq:phi_pp_Keldysh_structure}
  (\varphi^\ix{p}_\lambda)_{\sigma \overline \sigma | \sigma \overline \sigma}(\Pi)
  =
  \left(
    \begin{array}{cc|cc}
      0 & {a^\ix{p}_\lambda}^\ast & {a^\ix{p}_\lambda}^\ast & 0
      \\
      a^\ix{p}_\lambda & b^\ix{p}_\lambda & b^\ix{p}_\lambda & a^\ix{p}_\lambda
      \\
      \hline
      a^\ix{p}_\lambda & b^\ix{p}_\lambda & b^\ix{p}_\lambda & a^\ix{p}_\lambda
      \\
      0 & {a^\ix{p}_\lambda}^\ast & {a^\ix{p}_\lambda}^\ast & 0
    \end{array}
  \right)_{\sigma \overline \sigma}(\Pi),
\end{equation}
where $(a^\ix{p}_\lambda)_{\sigma \overline \sigma}(\Pi)$ is a
complex valued function while $(b^\ix{p}_\lambda)_{\sigma \overline
  \sigma}(\Pi)$ is purely imaginary. As a consequence of causality,
$(a^\ix{p}_\lambda)_{\sigma \overline \sigma}(\Pi)$ is analytic in
the upper half plane of $\Pi$. The relations for exchange of
particle indices yield
\begin{subequations}
\label{eq:abp}
  \begin{align}
    (a^\ix{p}_\lambda)_{\sigma \overline \sigma}(\Pi) &=
    (a^\ix{p}_\lambda)_{\overline \sigma \sigma}(\Pi),
    \\
    (b^\ix{p}_\lambda)_{\sigma \overline \sigma}(\Pi) &=
    (b^\ix{p}_\lambda)_{\overline \sigma \sigma}(\Pi).
  \end{align}
\end{subequations}
In case of thermal equilibrium at temperature $T$ and chemical
potential $\mu=0$ the only nontrivial component of the generalized
fluctuation dissipation theorem of Ref.~\onlinecite{jakobs1}
reads
\begin{equation}
  \label{eq:p_KMS}
  (b^\ix{p}_\lambda)_{\sigma \overline \sigma}(\Pi)
  =
  \im 2 \coth \left( \frac{\Pi}{2T} \right)
  \Imag (a^\ix{p}_\lambda)_{\sigma \overline \sigma}(\Pi).
\end{equation}

The same reasoning as used for $(\varphi^\ix{p}_\lambda)_{\sigma \overline
  \sigma|\sigma \overline \sigma}(\Pi)$ shows that the Keldysh
structure of $(\varphi^\ix{x}_\lambda)_{\sigma \overline \sigma | \sigma
  \overline \sigma}(X)$ is
\begin{equation}
  \renewcommand{\arraystretch}{1.3}
  (\varphi^\ix{x}_\lambda)_{\sigma \overline \sigma | \sigma \overline \sigma}(X)
  =
  \left(
    \begin{array}{cc|cc}
      0 & {a^\ix{x}_\lambda}^\ast & a^\ix{x}_\lambda & b^\ix{x}_\lambda
      \\
      a^\ix{x}_\lambda & b^\ix{x}_\lambda & 0 & {a^\ix{x}_\lambda}^\ast
      \\
      \hline
      {a^\ix{x}_\lambda}^\ast & 0 & b^\ix{x}_\lambda & a^\ix{x}_\lambda
      \\
      b^\ix{x}_\lambda & a^\ix{x}_\lambda & {a^\ix{x}_\lambda}^\ast & 0
    \end{array}
  \right)_{\sigma \overline \sigma}(X),
\end{equation}
with a complex valued function $(a^\ix{x}_\lambda)_{\sigma \overline
  \sigma}(X)$ and a purely imaginary function
$(b^\ix{x}_\lambda)_{\sigma \overline \sigma}(X)$.
$(a^\ix{x}_\lambda)_{\sigma \overline \sigma}(X)$ is analytic in the
lower half plane of $X$. The behaviour of $a^\ix{x}_\lambda$ and
$b^\ix{x}_\lambda$ under exchange of particle indices is
\begin{subequations}
  \label{eq:abx}
  \begin{align}
    (a^\ix{x}_\lambda)_{\sigma \overline \sigma}(X)
    &=
    (a^\ix{x}_\lambda)_{\overline \sigma \sigma}(-X)^\ast,
    \\
    (b^\ix{x}_\lambda)_{\sigma \overline \sigma}(X)
    &=
    (b^\ix{x}_\lambda)_{\overline \sigma \sigma}(-X).
  \end{align}
\end{subequations}
In case of thermal equilibrium the KMS conditions are
\begin{equation}
  \label{eq:x_KMS}
  (b^\ix{x}_\lambda)_{\sigma \overline \sigma}(X)
  =
  - \im 2 \coth\left(\frac{X}{2T}\right)
  \Imag (a^\ix{x}_\lambda)_{\sigma \overline \sigma}(X).
\end{equation}

The Keldysh structure of $(\varphi^\ix{d}_\lambda)_{\sigma \overline \sigma |
  \sigma \overline \sigma}(\Delta)$ has the form
\begin{equation}
  \renewcommand{\arraystretch}{1.3}
  (\varphi^\ix{d}_\lambda)_{\sigma \overline \sigma | \sigma \overline
    \sigma}(\Delta) =
  \\
  \left(
    \begin{array}{cc|cc}
      0 & (a^\ix{d}_\lambda)_{\sigma \overline \sigma} &
      (a^\ix{d}_\lambda)^\ast_{\overline \sigma \sigma} &
      (b^\ix{d}_\lambda)_{\sigma \overline \sigma}
      \\
      (a^\ix{d}_\lambda)_{\sigma \overline \sigma} & 0 &
      (b^\ix{d}_\lambda)_{\sigma \overline \sigma} &
      (a^\ix{d}_\lambda)^\ast_{\overline \sigma \sigma} 
      \\
      \hline
      (a^\ix{d}_\lambda)^\ast_{\overline \sigma \sigma} &
      (b^\ix{d}_\lambda)_{\sigma \overline \sigma} & 0 &
      (a^\ix{d}_\lambda)_{\sigma \overline \sigma}
      \\
      (b^\ix{d}_\lambda)_{\sigma \overline \sigma} &
      (a^\ix{d}_\lambda)^\ast_{\overline \sigma \sigma} &
      (a^\ix{d}_\lambda)_{\sigma \overline \sigma} & 0
    \end{array}
  \right)(\Delta),
\end{equation}
where $(a^\ix{d}_\lambda)_{\sigma \overline \sigma}(\Delta)$ and
$(b^\ix{d}_\lambda)_{\sigma \overline \sigma}(\Delta)$ are complex
valued functions which satisfy
\begin{subequations}
\label{eq:adud}
  \begin{align}
    \label{eq:ad_symm}
    (a^\ix{d}_\lambda)_{\sigma \overline \sigma}(\Delta) &=
    (a^\ix{d}_\lambda)_{\sigma \overline \sigma}(-\Delta)^\ast,
    \\
    (b^\ix{d}_\lambda)_{\sigma \overline \sigma}(\Delta) &=
    -(b^\ix{d}_\lambda)_{\sigma \overline \sigma}(-\Delta)^\ast,
    \\
    (b^\ix{d}_\lambda)_{\sigma \overline \sigma}(\Delta)
    &=
    (b^\ix{d}_\lambda)_{\overline \sigma \sigma}(-\Delta).
  \end{align}
\end{subequations}
The function $(a^\ix{d}_\lambda)_{\sigma \overline \sigma}(\Delta)$ is
analytic in the upper half plane of $\Delta$. In case of thermal
equilibrium the KMS conditions demand
\begin{subequations}
  \label{eq:d_KMS}
  \begin{align}
    \Real (b^\ix{d}_\lambda)_{\sigma \overline \sigma}(\Delta)
    &=
    \tanh\left(\frac{\Delta}{2T}\right)
    \Real \left[(a^\ix{d}_\lambda)_{\sigma \overline \sigma}(\Delta)
      - (a^\ix{d}_\lambda)_{\overline \sigma \sigma}(\Delta)\right],
    \\
    \Imag (b^\ix{d}_\lambda)_{\sigma \overline \sigma}(\Delta)
    &=
    \coth \left(\frac{\Delta}{2T}\right)
    \Imag \left[(a^\ix{d}_\lambda)_{\sigma \overline \sigma}(\Delta)
      + (a^\ix{d}_\lambda)_{\overline \sigma \sigma}(\Delta)\right].
  \end{align}
\end{subequations}

Finally the Keldysh structure of $(\varphi^\ix{d}_\lambda)_{\sigma \sigma
  | \sigma \sigma}(\Delta)$ is
\begin{equation}
  \renewcommand{\arraystretch}{1.3}
  (\varphi^\ix{d}_\lambda)_{\sigma \sigma | \sigma \sigma}(\Delta)
  =
  \left(
    \begin{array}{cc|cc}
      0 & a^\ix{d}_\lambda & {a^\ix{d}_\lambda}^\ast & b^\ix{d}_\lambda
      \\
      a^\ix{d}_\lambda & 0 & b^\ix{d}_\lambda & {a^\ix{d}_\lambda}^\ast
      \\
      \hline
      {a^\ix{d}_\lambda}^\ast & b^\ix{d}_\lambda & 0 & a^\ix{d}_\lambda
      \\
      b^\ix{d}_\lambda & {a^\ix{d}_\lambda}^\ast & a^\ix{d}_\lambda & 0
    \end{array}
  \right)_{\sigma \sigma}(\Delta),
\end{equation}
where $(a^\ix{d}_\lambda)_{\sigma \sigma}(\Delta)$ is a complex valued and
$(b^\ix{d}_\lambda)_{\sigma \sigma}(\Delta)$ a purely imaginary
function; they satisfy
\begin{subequations}
\label{eq:aduu}
  \begin{align}
    \label{eq:add_symm}
    (a^\ix{d}_\lambda)_{\sigma \sigma}(\Delta)
    &=
    (a^\ix{d}_\lambda)_{\sigma \sigma}(-\Delta)^\ast,
    \\
    (b^\ix{d}_\lambda)_{\sigma \sigma}(\Delta)
    &=
    (b^\ix{d}_\lambda)_{\sigma \sigma}(-\Delta).
  \end{align}
\end{subequations}
$(a^\ix{d}_\lambda)_{\sigma \sigma}(\Delta)$ is analytic in the upper
half plane of $\Delta$.  In case of thermal equilibrium the KMS
conditions are
\begin{equation}
  \label{eq:dd_KMS}
  (b^\ix{d}_\lambda)_{\sigma \sigma}(\Delta)
  =
  \im 2 \coth \left(\frac{\Delta}{2T}\right)
  \Imag (a^\ix{d}_\lambda)_{\sigma \sigma}(\Delta).
\end{equation}

In summary we conclude that all spin and Keldysh components of the
functions $\varphi^\ix{p,x,d}$ can be determined from the selection
\begin{equation}
  \label{eq:indep_comp}
  (a^{\ix{p},\ix{x},\ix{d}}_\lambda)_{\uparrow \downarrow},
  \;
  (b^{\ix{p},\ix{x},\ix{d}}_\lambda)_{\uparrow \downarrow},
  \;
  (a^\ix{d}_\lambda)_{\downarrow \uparrow},
  \;
  (a^\ix{d}_\lambda)_{\sigma \sigma},
  \;
  (b^\ix{d}_\lambda)_{\sigma \sigma}, \quad \sigma = \uparrow, \downarrow.
\end{equation}
The precise form of the flow equations shows, that
$(a^\ix{d}_\lambda)_{\downarrow \uparrow}$ can be expressed through
$(a^\ix{d}_\lambda)_{\uparrow \downarrow}$, see
Eq.~\eqref{eq:ad_symm1} below.


\begin{widetext}
  \section{Flow equation for the self-energy}
  \label{sec:flow_eq_SE}

  In our approximation the two-particle vertex acquires a special
  structure described in section~\ref{subsec:approx_mix} and
  Appendix~\ref{sec:structure}. When we insert this structure into
  Eq.~\eqref{eq:flow_self-energy} we obtain as flow equation for
  the retarded component of the self-energy
  \begin{align}
    \label{eq:flow_SigmaRet}
    \frac{\df \Sigma^{\ix{Ret} \,\lambda}_\sigma(\omega)}{\df \lambda}
    =
    -\frac{\im}{2 \pi}
    \int \! \df \omega' \,
    \bigg\{&
    \left[
      (b^\ix{x}_\lambda)_{\sigma \overline \sigma}(\omega'-\omega) 
      + (b^\ix{d}_\lambda)_{\sigma \overline \sigma}(0)
    \right] S^{\ix{Ret} \, \lambda}_{\overline \sigma}(\omega')
    -
    \left[
      (b^\ix{d}_\lambda)_{\sigma \sigma}(\omega-\omega') 
      - (b^\ix{d}_\lambda)_{\sigma \sigma}(0)
    \right] S^{\ix{Ret} \, \lambda}_{\sigma}(\omega')
    \nonumber
    \\ & +
    \left[
      (b^\ix{p}_\lambda)_{\sigma \overline \sigma}(\omega'+\omega) 
      + (b^\ix{d}_\lambda)_{\sigma \overline \sigma}(0)
    \right] S^{\ix{Av} \, \lambda}_{\overline \sigma}(\omega')
    +
    (b^\ix{d}_\lambda)_{\sigma \sigma}(0)
    S^{\ix{Av} \, \lambda}_{\sigma}(\omega')
    \nonumber
    \\ & \hspace{-8em}
    + \left[ 
      \frac{U}{2}
      + (a^\ix{p}_\lambda)_{\sigma \overline \sigma}(\omega'+\omega) 
      + (a^\ix{x}_\lambda)_{\sigma \overline \sigma}(\omega'-\omega) 
      + (a^\ix{d}_\lambda)_{\sigma \overline \sigma}(0)
    \right] S^{\ix{K} \, \lambda}_{\overline \sigma}(\omega')
    -
    \left[ 
      (a^\ix{d}_\lambda)_{\sigma \sigma}(\omega-\omega') 
      - (a^\ix{d}_\lambda)_{\sigma \sigma}(0)
    \right] S^{\ix{K} \, \lambda}_{\sigma}(\omega') \bigg\}.
  \end{align}

  As a consequence of Eq.~\eqref{eq:single_scale_prop}
  $S^\ix{Ret}(\omega')$ [$S^\ix{Av}(\omega')$] is analytic in the upper
  [lower] half plane of $\omega'$ and vanishes as $\omega'^{-2}$ for
  $\abs{\omega'} \rightarrow \infty$. Hence,
  \begin{equation}
    \int \! \df \omega' \, S^\ix{Ret, Av}(\omega') = 0,
  \end{equation}
  and Eq.~\eqref{eq:flow_SigmaRet} is reduced to
  \begin{align}
    \frac{\df \Sigma^{\ix{Ret} \,\lambda}_\sigma(\omega)}{\df \lambda}
    =
    -\frac{\im}{2 \pi}
    \int \! \df \omega' \,
    \bigg\{&
    (b^\ix{x}_\lambda)_{\sigma \overline \sigma}(\omega'-\omega) 
    S^{\ix{Ret} \, \lambda}_{\overline \sigma}(\omega')
    -
    (b^\ix{d}_\lambda)_{\sigma \sigma}(\omega-\omega') 
    S^{\ix{Ret} \, \lambda}_{\sigma}(\omega')
    +
    (b^\ix{p}_\lambda)_{\sigma \overline \sigma}(\omega'+\omega) 
    S^{\ix{Av} \, \lambda}_{\overline \sigma}(\omega')
    \nonumber
    \\ & \hspace{-8em}
    +
    \left[ 
      \frac{U}{2}
      + (a^\ix{p}_\lambda)_{\sigma \overline \sigma}(\omega'+\omega) 
      + (a^\ix{x}_\lambda)_{\sigma \overline \sigma}(\omega'-\omega) 
      + (a^\ix{d}_\lambda)_{\sigma \overline \sigma}(0)
    \right] S^{\ix{K} \, \lambda}_{\overline \sigma}(\omega')
    -
    \left[ 
      (a^\ix{d}_\lambda)_{\sigma \sigma}(\omega-\omega') 
      - (a^\ix{d}_\lambda)_{\sigma \sigma}(0)
    \right] S^{\ix{K} \, \lambda}_{\sigma}(\omega')
    \bigg\}.
  \end{align}

  Since the relation $\Sigma^{\ix{Av} \, \lambda}_\sigma(\omega) =
  \Sigma^{\ix{Ret} \, \lambda}_\sigma(\omega)^\ast$ is maintained
  during the flow, it is not necessary to compute the flow of
  $\Sigma^{\ix{Av} \, \lambda}_\sigma(\omega)$ separately. Following
  similar steps as for $\Sigma^\ix{Ret}$ we acquire the flow equation
  for $\Sigma^\ix{K}$,
  \begin{align}
    \frac{\df \Sigma^{\ix{K} \, \lambda}_\sigma(\omega)}{\df \lambda}
    =
    -\frac{\im}{2 \pi}
    \int \! \df \omega' \,
    \bigg\{&
    2 \Real 
    \left[
      \left(
        \frac{U}{2}
        + (a^\ix{p}_\lambda)_{\sigma \overline \sigma}(\omega'+\omega)^\ast
        + (a^\ix{x}_\lambda)_{\sigma \overline \sigma}(\omega'-\omega) 
      \right)
      S^{\ix{Ret} \,\lambda}_{\overline \sigma}(\omega')
    \right]
    -
    2 \Real 
    \left[
      (a^\ix{d}_\lambda)_{\sigma \sigma}(\omega - \omega')
      S^{\ix{Ret} \, \lambda}_{\sigma}(\omega')
    \right]
    \nonumber
    \\ &
    +
    \left[
      (b^\ix{p}_\lambda)_{\sigma \overline \sigma}(\omega'+\omega) 
      + (b^\ix{x}_\lambda)_{\sigma \overline \sigma}(\omega'-\omega)
    \right] S^{\ix{K} \, \lambda}_{\overline \sigma}(\omega')
    -
    (b^\ix{d}_\lambda)_{\sigma \sigma}(\omega-\omega')
    S^{\ix{K} \, \lambda}_{\sigma}(\omega')
    \bigg\}.
  \end{align}


  \section{Flow equation for the two-particle vertex}
  \label{sec:flow_eq_vert}

  Due to the structure of the two-particle vertex function described
  in section~\ref{subsec:approx_mix} and Appendix~\ref{sec:structure},
  its flow is determined completely by the flow of the
  components~\eqref{eq:indep_comp}.  The flow equations of these
  components can be formulated as a closed set, if the other
  components are eliminated by use of the relations found in
  Appendix~\ref{sec:structure}. Setting up these flow equations is
  cumbersome but straightforward. This section is devoted to a brief
  sketch of some simplifications which can be made during the
  calculation and indicating the result.

  The flow equations for $(a^\ix{p}_\lambda)_{\sigma \overline \sigma} =
  (\varphi^\ix{p}_\lambda)^{12|22}_{\sigma \overline \sigma | \sigma
    \overline \sigma}$ and $(b^\ix{p}_\lambda)_{\sigma \overline \sigma}
  = (\varphi^\ix{p}_\lambda)^{12|21}_{\sigma \overline \sigma | \sigma
    \overline \sigma}$ follow from Eq.~\eqref{eq:approxB_pp}. Due
  to Eqs.~\eqref{eq:eff_vertex}, \eqref{eq:Integral_Indices},
  and~\eqref{eq:phi_vanish_pp} the implicit summation over spin indices
  in Eq.~\eqref{eq:approxB_pp} is reduced to the two contributions
  $(\sigma_3 \sigma_4|\sigma'_3 \sigma'_4) = (\sigma \overline
  \sigma|\sigma \overline \sigma), (\overline \sigma \sigma| \overline
  \sigma \sigma)$. The summation over Keldysh indices is also largely
  reduced: first, because of the vanishing components in
  Eq.~\eqref{eq:phi_pp_Keldysh_structure}, and, second, because of
  \begin{subequations}
    \label{eq:I_pp_vanish}
    \begin{gather}
      \label{eq:I_pp_vanish1}
      (I^\ix{pp}_\lambda)^{\alpha_1 \alpha_2|\alpha'_1 \alpha'_2} = 0,
      \quad \text{if $\alpha_1 = \alpha'_1 = 1$ or $\alpha_2 = \alpha'_2
        = 1$},
      \\
      \label{eq:I_pp_vanish2}
      (I^\ix{pp}_\lambda)^{12|21} 
      = (I^\ix{pp}_\lambda)^{21|12} 
      = 0.
    \end{gather}
  \end{subequations}
  Equation~\eqref{eq:I_pp_vanish1} is a consequence of $G^{1|1} = 0$,
  $S^{1|1} = 0$, while Eq.~\eqref{eq:I_pp_vanish2} follows from
  $G^\ix{Ret}(\omega)$, $S^\ix{Ret}(\omega)$ [$G^\ix{Av}(\omega)$,
  $S^\ix{Av}(\omega)$] being analytic in the upper [lower] half plane of
  $\omega$. Making further use of 
  \begin{equation}
    \label{eq:I_pp_conj}
    (I^\ix{pp}_\lambda)^{\alpha'_1 \alpha'_2|\alpha_1 \alpha_2}(\omega) 
    =
    -(-1)^{\alpha'_1 + \alpha'_2 + \alpha_1 + \alpha_2}
    (I^\ix{pp}_\lambda)^{\alpha_1 \alpha_2|\alpha'_1 \alpha'_2}(\omega)^\ast
  \end{equation}
  for the remaining components, which follows form
  $G^\ix{Ret}_\sigma(\omega)^\ast = G^\ix{Av}_\sigma(\omega)$,
  $S^\ix{Ret}_\sigma(\omega)^\ast = S^\ix{Av}_\sigma(\omega)$ and
  $G^\ix{K}_\sigma(\omega)^\ast = -G^\ix{K}_\sigma(\omega)$,
  $S^\ix{K}_\sigma(\omega)^\ast = -S^\ix{K}_\sigma(\omega)$, we find
  \begin{subequations}
    \label{eq:p_floweq_B}
    \begin{align}
      \label{eq:ap_floweq_B}
      \frac{\df (a^\ix{p}_\lambda)_{\sigma \overline
          \sigma}(\Pi)}{\df \lambda}
      =&
      \left[
        \frac{U + U^\ix{x}_\lambda + U^\ix{d}_\lambda}{2} +
        (a^\ix{p}_\lambda)_{\sigma \overline \sigma}(\Pi)
      \right]^2
      \left[
        (I^\ix{\,pp}_\lambda)_{\sigma \overline \sigma}^{22|12}(\Pi) 
        + (I^\ix{\,pp}_\lambda)_{\sigma \overline \sigma}^{22|21}(\Pi) 
      \right],
      \\[1ex] \label{eq:bp_floweq_B}
      \frac{\df (b^\ix{p}_\lambda)_{\sigma \overline
          \sigma}(\Pi)}{\df \lambda} 
      =& \;
      2 \im \Imag 
      \bigg\{
      \bigg|\frac{U + U^\ix{x}_\lambda + U^\ix{d}_\lambda}{2} +
      (a^\ix{p}_\lambda)_{\sigma \overline \sigma}(\Pi)\bigg|^2
      \bigg[
      (I^\ix{\,pp}_\lambda)_{\sigma \overline \sigma}^{22|11}(\Pi)
      + \frac{1}{2} (I^\ix{\,pp}_\lambda)_{\sigma \overline \sigma}^{22|22}(\Pi)
      \bigg]
      \nonumber
      \\
      & \; \phantom{2 \im \Imag \bigg\{} +
      \bigg[
      \frac{U + U^\ix{x}_\lambda + U^\ix{d}_\lambda}{2} +
      (a^\ix{p}_\lambda)_{\sigma \overline \sigma}(\Pi)
      \bigg]
      (b^\ix{p}_\lambda)_{\sigma \overline
        \sigma}(\Pi)
      \Big[ 
      (I^\ix{\,pp}_\lambda)_{\sigma \overline \sigma}^{22|12}(\Pi)
      + (I^\ix{\,pp}_\lambda)_{\sigma \overline \sigma}^{22|21}(\Pi)
      \Big]
      \bigg\}.
    \end{align}
  \end{subequations}
  
  The flow equations for $(a^\ix{x}_\lambda)_{\sigma \overline
    \sigma}(X)$ and $(b^\ix{x}_\lambda)_{\sigma \overline \sigma}(X)$
  can be derived in an analogous way. Instead of
  Eqs.~\eqref{eq:I_pp_vanish} and~\eqref{eq:I_pp_conj} we use
  \begin{subequations}
    \label{eq:I_ph_vanish}
    \begin{gather}
      \label{eq:I_ph_vanish1}
      (I^\ix{ph}_\lambda)^{\alpha_1 \alpha_2|\alpha'_1 \alpha'_2} = 0,
      \quad \text{if $\alpha_1 = \alpha'_1 = 1$ or $\alpha_2 = \alpha'_2
        = 1$},
      \\
      \label{eq:I_ph_vanish2}
      (I^\ix{ph}_\lambda)^{11|22} 
      = (I^\ix{ph}_\lambda)^{22|11} 
      = 0,
    \end{gather}
  \end{subequations}
  and
  \begin{equation}
    \label{eq:I_ph_conj}
    (I^\ix{ph}_\lambda)^{\alpha'_1 \alpha'_2|\alpha_1 \alpha_2}(\omega) 
    =
    -(-1)^{\alpha'_1 + \alpha'_2 + \alpha_1 + \alpha_2}
    (I^\ix{ph}_\lambda)^{\alpha_1 \alpha_2|\alpha'_1 \alpha'_2}(\omega)^\ast,
  \end{equation}
  and find
  \begin{subequations}
    \label{eq:x_floweq_B}
    \begin{align}
      \label{eq:ax_floweq_B}
      \frac{\df (a^\ix{x}_\lambda)_{\sigma \overline
          \sigma}(X)}{\df \lambda} 
      =&
      \left[
        \frac{U + U^\ix{p}_\lambda + U^\ix{d}_\lambda}{2} +
        (a^\ix{x}_\lambda)_{\sigma \overline \sigma}(X)
      \right]^2
      \left[
        (I^\ix{\,ph}_\lambda)_{\sigma \overline \sigma}^{21|22}(X)
        + (I^\ix{\,ph}_\lambda)_{\sigma \overline \sigma}^{22|12}(X)
      \right],
      \\[1ex] \label{eq:bx_floweq_B}
      \frac{\df (b^\ix{x}_\lambda)_{\sigma \overline
          \sigma}(X)}{\df \lambda} 
      =&
      \; 2 \im \Imag \bigg\{
      \bigg|
      \frac{U + U^\ix{p}_\lambda + U^\ix{d}_\lambda}{2} +
      (a^\ix{x}_\lambda)_{\sigma \overline \sigma}(X)
      \bigg|^2
      \bigg[
      (I^\ix{\,ph}_\lambda)_{\sigma \overline \sigma}^{12|21}(X)
      + \frac{1}{2} (I^\ix{\,ph}_\lambda)_{\sigma \overline \sigma}^{22|22}(X)
      \bigg]
      \nonumber
      \\
      & \; \phantom{2 \im \Imag \bigg\{} +
      \bigg[
      \frac{U + U^\ix{p}_\lambda + U^\ix{d}_\lambda}{2} +
      (a^\ix{x}_\lambda)_{\sigma \overline \sigma}(X)
      \bigg]
      (b^\ix{p}_\lambda)_{\sigma \overline \sigma}(X)
      \bigg[ 
      (I^\ix{\,ph}_\lambda)_{\sigma \overline \sigma}^{21|22}(X) 
      + (I^\ix{\,ph}_\lambda)_{\sigma \overline \sigma}^{22|12}(X)
      \bigg]
      \bigg\}.
    \end{align}
  \end{subequations}

  The flow equations for $(a^\ix{d}_\lambda)_{\sigma \overline
    \sigma}(\Delta)$ and $(b^\ix{d}_\lambda)_{\sigma \overline
    \sigma}(\Delta)$ are
  \begin{subequations}
    \label{eq:d_floweq_B}
    \begin{align}
      \label{eq:a_d_floweq_B}
      \frac{\df (a^\ix{d}_\lambda)_{\sigma \overline
          \sigma}(\Delta)}{\df \lambda} 
      =&
      -
      \bigg[
      \frac{U + U^\ix{p}_\lambda + U^\ix{x}_\lambda}{2} +
      (a^\ix{d}_\lambda)_{\sigma \overline \sigma}(\Delta)
      \bigg]
      \sum_{s = \uparrow, \downarrow}
      \bigg[
      (a^\ix{d}_\lambda)_{ss}(\Delta)
      - \frac{W^\ix{d}_{\lambda \, s}}{2}
      \bigg]
      \left[
        (I^\ix{\,ph}_\lambda)_{ss}^{22|21}(\Delta)
        + (I^\ix{\,ph}_\lambda)_{ss}^{12|22}(\Delta)
      \right],
      \\[1ex]
      \frac{\df (b^\ix{d}_\lambda)_{\sigma \overline
          \sigma}(\Delta)}{\df \lambda} =& -\left[ \frac{U +
          U^\ix{p}_\lambda + U^\ix{x}_\lambda}{2} +
        (a^\ix{d}_\lambda)_{\sigma \overline \sigma}(\Delta) \right]
      \bigg[ (a^\ix{d}_\lambda)_{\overline \sigma \overline
        \sigma}(\Delta)^\ast - \frac{W^\ix{d}_{\lambda \, \overline
          \sigma}}{2} \bigg] \left[
        (I^\ix{\,ph}_\lambda)^{12|21}_{\overline \sigma \overline
          \sigma}(\Delta) + (I^\ix{\,ph}_\lambda)^{21|12}_{\overline
          \sigma \overline \sigma}(\Delta) +
        (I^\ix{\,ph}_\lambda)^{22|22}_{\overline \sigma \overline
          \sigma}(\Delta) \right] \nonumber
      \\
      & -\left[ \frac{U + U^\ix{p}_\lambda + U^\ix{x}_\lambda}{2} +
        (a^\ix{d}_\lambda)_{\overline \sigma \sigma}(\Delta)^\ast
      \right] \bigg[ (a^\ix{d}_\lambda)_{\sigma \sigma}(\Delta) -
      \frac{W^\ix{d}_{\lambda \, \sigma}}{2} \bigg] \left[
        (I^\ix{\,ph}_\lambda)^{12|21}_{\sigma \sigma}(\Delta) +
        (I^\ix{\,ph}_\lambda)^{21|12}_{\sigma \sigma}(\Delta) +
        (I^\ix{\,ph}_\lambda)^{22|22}_{\sigma \sigma}(\Delta) \right]
      \nonumber
      \\
      & -\left[ \frac{U + U^\ix{p}_\lambda + U^\ix{x}_\lambda}{2} +
        (a^\ix{d}_\lambda)_{\sigma \overline \sigma}(\Delta) \right]
      (b^\ix{d}_\lambda)_{\overline \sigma \overline \sigma}(\Delta)
      \left[ (I^\ix{\,ph}_\lambda)^{22|21}_{\overline \sigma \overline
          \sigma}(\Delta) + (I^\ix{\,ph}_\lambda)^{12|22}_{\overline
          \sigma \overline \sigma}(\Delta) \right] \nonumber
      \\
      & -\left[ \frac{U + U^\ix{p}_\lambda + U^\ix{x}_\lambda}{2} +
        (a^\ix{d}_\lambda)_{\overline \sigma \sigma}(\Delta)^\ast
      \right] (b^\ix{d}_\lambda)_{\sigma \sigma}(\Delta) \left[
        (I^\ix{\,ph}_\lambda)^{21|22}_{\sigma \sigma}(\Delta) +
        (I^\ix{\,ph}_\lambda)^{22|12}_{\sigma \sigma}(\Delta) \right]
      \nonumber
      \\
      & - \bigg[ (a^\ix{d}_\lambda)_{\overline \sigma \overline
        \sigma}(\Delta)^\ast - \frac{W^\ix{d}_{\lambda \, \overline
          \sigma}}{2} \bigg] (b^\ix{d}_\lambda)_{\sigma \overline
        \sigma}(\Delta) \left[
        (I^\ix{\,ph}_\lambda)^{21|22}_{\overline \sigma \overline
          \sigma}(\Delta) + (I^\ix{\,ph}_\lambda)^{22|12}_{\overline
          \sigma \overline \sigma}(\Delta) \right] \nonumber
      \\
      & - \bigg[ (a^\ix{d}_\lambda)_{\sigma \sigma}(\Delta) -
      \frac{W^\ix{d}_{\lambda \, \sigma}}{2} \bigg]
      (b^\ix{d}_\lambda)_{\sigma \overline \sigma}(\Delta) \left[
        (I^\ix{\,ph}_\lambda)^{22|21}_{\sigma \sigma}(\Delta) +
        (I^\ix{\,ph}_\lambda)^{12|22}_{\sigma \sigma}(\Delta) \right].
    \end{align}
  \end{subequations}  
  Since $(a^\ix{d}_\lambda)_{\sigma \overline \sigma}$ and
  $(a^\ix{d}_\lambda)_{\overline \sigma \sigma}$ have the same (zero)
  initial value at $\lambda=\infty$ we conclude from
  Eq.~\eqref{eq:a_d_floweq_B} that
  \begin{equation}
    \label{eq:ad_symm1}
    (a^\ix{d}_\lambda)_{\sigma \overline \sigma}(\Delta)
    =
    (a^\ix{d}_\lambda)_{\overline \sigma \sigma}(\Delta).
  \end{equation}
  
  Finally, the flow equations for $(a^\ix{d}_\lambda)_{\sigma
    \sigma}(\Delta)$ and $(b^\ix{d}_\lambda)_{\sigma \sigma}(\Delta)$
  turn out to be
  \begin{subequations}
    \label{eq:dd_floweq_B}
    \begin{align}
      \frac{\df (a^\ix{d}_\lambda)_{\sigma
          \sigma}(\Delta)}{\df \lambda} 
      =&
      -\left[
        \frac{U + U^\ix{p}_\lambda + U^\ix{x}_\lambda}{2} +
        (a^\ix{d}_\lambda)_{\sigma \overline \sigma}(\Delta)
      \right]^2
      \left[
        (I^\ix{\,ph}_\lambda)^{22|21}_{\overline \sigma \overline \sigma}(\Delta) 
        + (I^\ix{\,ph}_\lambda)^{12|22}_{\overline \sigma \overline \sigma}(\Delta)
      \right]
      \nonumber
      \\
      &
      - \left[
        (a^\ix{d}_\lambda)_{\sigma \sigma}(\Delta)
        - \frac{W^\ix{d}_{\lambda \, \sigma}}{2}
      \right]^2
      \left[
        (I^\ix{\,ph}_\lambda)^{22|21}_{\sigma \sigma}(\Delta) 
        + (I^\ix{\,ph}_\lambda)^{12|22}_{\sigma \sigma}(\Delta)
      \right],
      \\[1ex]
      \frac{\df (b^\ix{d}_\lambda)_{\sigma
          \sigma}(\Delta)}{\df \lambda}  
      =&
      \; - 2 \im \Imag \bigg\{
      \bigg|\frac{U + U^\ix{p}_\lambda + U^\ix{x}_\lambda}{2} +
        (a^\ix{d}_\lambda)_{\sigma \overline \sigma}(\Delta)\bigg|^2
      \bigg[
        (I^\ix{\,ph}_\lambda)_{\overline \sigma \overline \sigma}^{12|21}(\Delta)
        + \frac{1}{2}(I^\ix{\,ph}_\lambda)_{\overline \sigma \overline \sigma}^{22|22}(\Delta)
      \bigg]
      \nonumber
      \\ &
      \; \phantom{-2 \im \Imag \bigg\{}
      +
      \bigg|(a^\ix{d}_\lambda)_{\sigma
        \sigma}(\Delta)-\frac{W^\ix{d}_{\lambda \, \sigma}}{2}\bigg|^2
      \bigg[
      (I^\ix{\,ph}_\lambda)_{\sigma \sigma}^{12|21}(\Delta)
      + \frac{1}{2}(I^\ix{\,ph}_\lambda)_{\sigma \sigma}^{22|22}(\Delta)
      \bigg]
      \nonumber
      \\ &
      \; \phantom{-2 \im \Imag \bigg\{}
      +
      \bigg[\frac{U + U^\ix{p}_\lambda + U^\ix{x}_\lambda}{2} +
        (a^\ix{d}_\lambda)_{\sigma \overline
          \sigma}(\Delta)^\ast \bigg]
      (b^\ix{d}_\lambda)_{\sigma \overline \sigma}(\Delta)
      \Big[
      (I^\ix{\,ph}_\lambda)_{\overline \sigma \overline \sigma}^{21|22}(\Delta)
      + (I^\ix{\,ph}_\lambda)_{\overline \sigma \overline \sigma}^{22|12}(\Delta)
      \Big]
      \nonumber
      \\ &
      \; \phantom{-2 \im \Imag \bigg\{}
      +
      \bigg[(a^\ix{d}_\lambda)_{\sigma \sigma}(\Delta)^\ast -
      \frac{W^\ix{d}_{\lambda \, \sigma}}{2}\bigg]
      (b^\ix{d}_\lambda)_{\sigma \sigma}(\Delta)
      \Big[
        (I^\ix{\,ph}_\lambda)_{\sigma \sigma}^{21|22}(\Delta)
        + (I^\ix{\,ph}_\lambda)_{\sigma \sigma}^{22|12}(\Delta)
      \Big] \bigg\}.
    \end{align}
  \end{subequations}
\end{widetext}


\section{Nonequilibrium  Fermi-liquid coefficient}
\label{app:fermi_liquid}

In this evaluation we make the same assumptions as in the section
\ref{sec:apprb}, i.e. $\Gamma_L = \Gamma_R$ and $\mu_L = - \mu_R =
\frac{eV}{2}$, $B = e V_g =0$. We remind that these conditions are
important to ensure the real value of the static component of the
vertex as well as the current conservation.

Let us define a nonequilibrium distribution function
\begin{eqnarray}
  h_F^{neq} (\omega) &=& \alpha_L h_{F,L} (\omega ) 
  + \alpha_R h_{F,R} (\omega ) \nonumber \\
  &=&  \frac12 h_F \left(\omega -\frac{eV}{2} \right) 
  + \frac12 h_F \left(\omega +\frac{eV}{2} \right),
  \label{eq:neqdf}
\end{eqnarray}
where $\alpha_L + \alpha_R =1$.

The nonequilibrium equation for the self-energy
\eqref{eq:flow_SigmaRet} in the particle-hole symmetric case reads
\begin{eqnarray}
  & & \frac{\df}{\df \lambda} \Sigma_{\sigma}^{\ix{Ret}} (\omega) = 
  \frac{1}{\pi} \int d \tilde{\omega} \left[ - \gamma_0 h_F^{neq}
    (\tilde{\omega}) 
    \Imag s_{\overline{\sigma}}^{\ix{Ret}} (\tilde{\omega})\right.
  \label{eq:se6} \\
  & & + h_F^{neq} (\tilde{\omega}) a_1 (\tilde{\omega} +\omega) 
  \Imag s_{\overline{\sigma}}^{\ix{Ret}} (\tilde{\omega})+ b_1 (\tilde{\omega}
  +\omega)  
  s_{\overline{\sigma}}^{\ix{Av}} (\tilde{\omega}) \nonumber \\
  & & + h_F^{neq} (\tilde{\omega}) a_3 (\tilde{\omega} -\omega) 
  \Imag s_{\overline{\sigma}}^{\ix{Ret}} (\tilde{\omega}) + b_3 (\tilde{\omega}
  -\omega) 
  s_{\overline{\sigma}}^{\ix{Ret}}  (\tilde{\omega}) \nonumber \\
  & & \left.  + h_F^{neq} (\tilde{\omega}) 
    \bar{a}_{2 \sigma} (\tilde{\omega}-\omega) 
    \Imag s_{\sigma}^{\ix{Ret}} (\tilde{\omega})
    + \bar{b}_{2 \sigma} (\tilde{\omega} -\omega) s_{\sigma}^{\ix{Ret}}
    (\tilde{\omega})\right], \nonumber 
\end{eqnarray}
where the vertex functions $a$ and $b$ components are no longer
related to each other by the KMS equalities, and where $\bar{b}_{2
  \sigma}$ is defined on the analogy of Eq.~\eqref{eq:bar_2sigma}.

In the following we will need an expression for a product of the 
two distribution
functions \eqref{eq:neqdf} evaluated at different frequency arguments
$\omega_1$ and $\omega_2$. At small $eV$ such a product can be represented as 
\begin{eqnarray}
  & & \left[\alpha_L h_{F,L} (\omega_1) + \alpha_R h_{F,R} (\omega_1)\right]
  \left[\alpha_L h_{F,L} (\omega_2) + \alpha_R h_{F,R} (\omega_2) \right] 
  \nonumber \\ 
  & &\qquad = \alpha_L  h_{F,L} (\omega_1) h_{F,L} (\omega_2) 
            + \alpha_R  h_{F,R} (\omega_1) h_{F,R} (\omega_2) \nonumber \\ 
  & & \qquad - \alpha_L \alpha_R 
              \left[ h_{F,L} (\omega_1) - h_{F,R} (\omega_1) \right]  
              \left[ h_{F,L} (\omega_2) - h_{F,R} (\omega_2) \right] \nonumber \\
  & & \qquad \approx \alpha_L  h_{F,L} (\omega_1) h_{F,L} (\omega_2) 
                   + \alpha_R  h_{F,R} (\omega_1) h_{F,R} (\omega_2)  
  \nonumber \\
  & & \qquad - 4 \alpha_L \alpha_R (e V)^2 \delta (\omega_1) \delta
  (\omega_2).  
  \label{eq:prodds1}
\end{eqnarray}

Let us consequently analyze voltage-induced corrections to the KMS relations
occurring in each channel.


\subsection{Particle-particle  channel}

In order to evaluate \eqref{eq:h1gg} we set $\omega_1 = \frac{\Pi}{2}
+\omega'$, $\omega_2 = \frac{\Pi}{2} - \omega'$ and use for the PP
channel the identity
\begin{equation}
  h_{F,\eta} (\omega_1) h_{F,\eta} (\omega_2) 
   = - 1 + h_{B,\eta} (\omega_1 +\omega_2)
  \left[h_{F,\eta} (\omega_1) + h_{F,\eta} (\omega_2) \right],  
\end{equation}
where $h_{B,\eta} (\Pi)= h_B (\Pi - \eta e V)$, $\eta = \pm$ for $\eta
=L/R$, and $h_B(\omega) = \coth \frac{\omega}{2T} \stackrel{T \to
  0}{=} {\rm sign} (\omega)$. It leads to
\begin{eqnarray}
  & & \sum_{\eta = \pm} \alpha_{\eta} h_{F, \eta} (\omega_1) h_{F, \eta} (\omega_2)   = \nonumber \\
& & = -1 + \frac12 \sum_{\eta=\pm} h_{B,\eta} (\Pi) \left[
    h_F^{neq} (\omega_1) + h_F^{neq} (\omega_2 )\right]  
  \nonumber \\
  & & \quad+  \frac12 \sum_{\eta=\pm} \eta h_{B,\eta}
      (\Pi) \left\{ \alpha_L [h_{F,L} (\omega_1 ) + h_{F,L} (\omega_2)]
    \right. \nonumber \\
& & \qquad \left.-\alpha_R [h_{F,R} (\omega_1 ) + h_{F,R} (\omega_2 )] \right\}, 
  \label{eq:prodds2}
\end{eqnarray}
where
\begin{equation}
  \sum_{\eta=\pm} \frac{\eta h_{B,\eta} (\Pi)}{2} \approx  
  -2 eV \delta (\Pi),
\end{equation}
as well as 
\begin{eqnarray}
 & &  \alpha_L [h_{F,L} (\omega_1 ) + h_{F,L} (\omega_2) ] 
  -\alpha_R [h_{F,R} (\omega_1 ) + h_{F,R} (\omega_2)] \nonumber \\
 & &  \approx - e V [ \delta (\omega_1) + \delta (\omega_2)]
\end{eqnarray}
for $\alpha_L = \alpha_R =\frac12$. Therefore, the second sum in
\eqref{eq:prodds2} approximately equals $\approx 4 (eV)^2 \delta
(\omega') \delta (\Pi)$.

We can represent
\begin{equation}
  H_1 (\Pi) = H_1^{I} (\Pi) + H_1^{II} (\Pi),
\end{equation}
where
\begin{eqnarray}
  H_1^{I} (\Pi) &=& \frac12 \left[ \sum_{\eta=\pm} h_{B,\eta} (\Pi)\right] 
                   \left[ F_1 (\Pi) -    F_1^* (\Pi)\right], \\ 
  H_1^{II} (\Pi) & \approx & i (s_1^a + s_1^b)\delta (\Pi), 
\end{eqnarray}
and
\begin{eqnarray}
  s_1^a &=&  \frac{8}{\pi} \alpha_L \alpha_R (eV)^2 
  \frac{\df}{\df \lambda} 
  \left[ \Imag g^{\ix{Ret}}_{\sigma} (0) \right]^2 , \\ 
  s_1^b &=& - 4 s_1^a.
\end{eqnarray}

Therefore the solution of Eq.~\eqref{eq:eqb1} has the form
\begin{equation}
  b_1 (\Pi) = b_1^{I} (\Pi) + i (b_1^{IIa} + b_1^{IIb})  \delta
  (\Pi), 
\end{equation}
where
\begin{equation}
  b_1 (\Pi ) = \frac12  \left[\sum_{\eta=\pm} h_{B,\eta} (\Pi) \right]
  [ a_1 (\Pi) - a^*_1 (\Pi)], 
\end{equation}
and $b_1^{IIa}$ and $b_1^{IIb}$ correspond to the inhomogeneity terms $\propto (eV)^2$
in Eqs.~\eqref{eq:prodds1} and \eqref{eq:prodds2}, respectively. 
They can be found from the equation
\begin{equation}
  \frac{\df b_1^{II,a/b}}{\df \lambda} = |a_1 (0) |^2
  s_1^{a/b} + 2 a_1 (0) F_1 (0) b_1^{II,a/b}. \label{eq:bII1} 
\end{equation}


\subsection{Particle-hole channel}

In order to evaluate \eqref{eq:h3gg} we set $\omega_1 = \omega' -\frac{X}{2}$, $\omega_2 = \omega' +\frac{X}{2}$ and use for the PH channel
the identity
\begin{equation}
  h_{F,\eta} (\omega_1) h_{F,\eta} (\omega_2) = 1 - h_{B} (\omega_1 -\omega_2)
  \left[ h_{F,\eta} (\omega_1) - h_{F,\eta} (\omega_2) \right].
\end{equation}
Then
\begin{equation}
  H_3 (X) = H_3^{I} (X) + H_3^{II} (X),
\end{equation}
where
\begin{eqnarray}
  H_3^I (X) &=& - h_B (X) \{ F_3 (X) - F_3^* (X) \}, \\
  H_3^{II} (X) &=& i s_3 \delta (X),
\end{eqnarray}
and
\begin{equation}
  s_3 = \frac{8}{\pi} \alpha_L \alpha_R (eV)^2 
  \frac{\df}{\df \lambda} 
  \left[ \Imag g^{\ix{Ret}}_{\sigma} (0)\right]^2.
\end{equation}

The solution of Eq.~\eqref{eq:eqb3} has the form
\begin{equation}
  b_3 (X) = b_3^{I} (X) + i b_3^{II} \delta (X),
\end{equation}
where
\begin{equation}
  b_3^{I} (X) = - h_B (X) \{ a_3 (X) - a_3^* (X)\},
\end{equation}
and $b_3^{II} (X)$ obeys the equation
\begin{equation}
  \frac{\df b_3^{II}}{\df \lambda} = |a_3 (0) |^2 s_3 + 2
  a_3 (0) F_3 (0) b_3^{II}. \label{eq:bII3} 
\end{equation}

Analogously, the function $ H_{2\sigma} (\Delta )$ occurring in the
Eqs.~\eqref{eq:eqb20} and \eqref{eq:eqb2s} can be represented as
\begin{equation}
H_{2\sigma} (\Delta ) = H_{2\sigma}^{I} (\Delta ) + H_{2\sigma}^{II} (\Delta ),
\end{equation}
where
\begin{eqnarray}
  H_{2\sigma}^I (\Delta ) &=&  h_B (\Delta ) \{ F_{2 \sigma} (\Delta )
  - F_{2 \sigma}^* (\Delta ) \}, \\ 
  H_{2\sigma}^{II} (\Delta ) &=& i s_{2\sigma} \delta (\Delta), 
\end{eqnarray}
and
\begin{equation}
  s_{2 \sigma} =  \frac{8}{\pi} \alpha_L \alpha_R (eV)^2 
  \frac{\df}{\df \lambda} 
  \left[ \Imag g^{\ix{Ret}}_{\sigma} (0)\right]^2.
\end{equation}
Note that from \eqref{eq:fprime0} it follows
\begin{equation}
  s_1^a = s_3 = s_{2 \sigma} = - (eV)^2 \Imag F'_3 (0).
  \label{eq:srel}
\end{equation}

The solution of Eqs.~\eqref{eq:eqb20} and
\eqref{eq:eqb2s} then takes the form
\begin{equation}
  b_{20/2 \sigma} (\Delta) = b_{20/2 \sigma}^I (\Delta) + i b_{20/2
    \sigma}^{II} \delta (\Delta), 
\end{equation}
where
\begin{equation}
  b_{20/2 \sigma}^I (\Delta) = h_B (\Delta) \{ F_{20/2\sigma} (\Delta)
  - F_{20/2\sigma}^* (\Delta) \},  
\end{equation}
and
\begin{eqnarray}
  \frac{\df b_{20}^{II}}{\df \lambda} &=& 
- 2 a_{20} (0) F_{2 \sigma} (0)
  b_{2 \sigma}^{II}, \label{eq:bII20} \\
  \frac{\df b_{2 \sigma}^{II}}{\df \lambda} &=& 
- |a_{20} (0)|^2 s_{2 \sigma}
  - 2 a_{20} (0) F_{2 \sigma} (0) b_{20}^{II}. \label{eq:bII2s}
\end{eqnarray}
Comparing  the latter relations with Eqs.~\eqref{eq:bII1} and \eqref{eq:bII3} and using \eqref{eq:srel} we establish that
\begin{equation}
  b_{2\sigma}^{II} =- \frac{b_1^{IIa} +b_3^{II}}{2}, \quad b_{20}^{II}
  =\frac{b_3^{II} -b_1^{IIa}}{2} .
\end{equation}


\begin{widetext}

  \subsection{Self-energy}

  An equation for the imaginary part of the self-energy in
  nonequilibrium reads
  \begin{eqnarray}
    \frac{\df}{\df \lambda} \Imag \Sigma_{\sigma}^{\ix{Ret}} (0) 
    &=& 
    \frac{1}{\pi} \int d \tilde{\omega}
    \Big[ h_F^{neq} (\tilde{\omega}) \Imag a_1
      (\tilde{\omega} ) 
      \Imag s_{\overline{\sigma}}^{\ix{Ret}} (\tilde{\omega})
      -\frac{i}{2} b_1^I (\tilde{\omega} ) 
      \Imag s_{\overline{\sigma}}^{\ix{Av}} (\tilde{\omega})
      + h_F^{neq} (\tilde{\omega}) \Imag a_3 (\tilde{\omega} ) 
    \Imag s_{\overline{\sigma}}^{\ix{Ret}} (\tilde{\omega}) 
    \nonumber \\
    & &
    - \frac{i}{2} b_3^I (\tilde{\omega})
    \Imag s_{\overline{\sigma}}^{\ix{Ret}}  (\tilde{\omega})
    + h_F^{neq} (\tilde{\omega}) \Imag \bar{a}_{2 \sigma}
      (\tilde{\omega}) 
      \Imag s_{\sigma}^{\ix{Ret}} (\tilde{\omega})
      -\frac{i}{2} \bar{b}_{2 \sigma}^I (\tilde{\omega}) \Imag
      s_{\sigma}^{\ix{Ret}}  (\tilde{\omega}) \Big]\nonumber \\ 
    & & + \frac{1}{2 \pi} \big[ (b_1^{IIa}+b_1^{IIb}) \Imag s^{\ix{Av}} (0) +
    (b_3^{II} -b_{2 \sigma}^{II})\Imag s^{\ix{Ret}} (0) \big], \nonumber \\  
    \label{eq:se7}
  \end{eqnarray}
  which simplifies due to the particle-hole symmetry to the form 
  \begin{eqnarray}
    \frac{\df}{\df \lambda} \Imag \Sigma_{\sigma}^{\ix{Ret}} (0) 
    &=& 
    \frac{1}{\pi} \int d \tilde{\omega}
    \Bigg\{ \left[ h_F^{neq} (\tilde{\omega}) 
        - \frac{1}{2} \sum_{\eta = \pm} h_{B, \eta}(\tilde{\omega})
      \right]
      \Imag \bar{a}_1 (\tilde{\omega} ) 
      \Imag s_{\overline{\sigma}}^{\ix{Ret}} (\tilde{\omega})\nonumber
      \\
      && + \left[h_F^{neq} (\tilde{\omega}) - h_B (\tilde{\omega})\right]
      \Imag [ a_3 (\tilde{\omega} ) + \bar{a}_{2 \sigma} (\tilde{\omega}) ] 
      \Imag s_{\overline{\sigma}}^{\ix{Ret}} (\tilde{\omega}) \Bigg\}
      \nonumber
      \\
      &&+ \frac{1}{2 \pi} \left[ - 2 b_1^{IIa} - b_1^{IIb}  + \frac32
      (b_1^{IIa} + b_{3}^{II})  \right] \Imag s^{\ix{Ret}} (0) . 
    \label{eq:se8}
  \end{eqnarray}
  Expanding it in $eV$, we obtain
  \begin{eqnarray}
    \frac{\df}{\df \lambda} \Imag \Sigma_{\sigma}^{\ix{Ret}} (0) 
    &=& 
    \frac{3}{2 \pi} \left( \frac{eV}{2} \right)^2 
    \int d \tilde{\omega} \delta' (\tilde{\omega}) \Imag [
    \bar{a}_1 (\tilde{\omega}) + a_3 (\tilde{\omega})] \Imag s^{\ix{Ret}}
    (\tilde{\omega})
    -\frac{1}{\pi} (eV)^2 
    \int d \tilde{\omega}  \delta' (\tilde{\omega}) 
    \Imag \bar{a}_1 (\tilde{\omega} ) 
    \Imag s^{\ix{Ret}} (\tilde{\omega}) \nonumber \\
    & & + \frac{1}{2 \pi} \left[ - 2 b_1^{IIa} - b_1^{IIb} + \frac32
      (b_1^{IIa} + b_{3}^{II}) \right] \Imag s^{\ix{Ret}} (0)\nonumber \\ 
    &=&  - \frac{3}{2 \pi} \left( \frac{eV}{2} \right)^2  \Imag \left[
      \frac{\partial \bar{a}_1}{\partial \omega} + \frac{\partial
        a_3}{\partial \omega}\right]_{\omega=0}\Imag s^{\ix{Ret}} (0)
    + \frac{1}{\pi} (eV)^2  \Imag \left[ \frac{\partial
        \bar{a}_1}{\partial \omega}\right]_{\omega=0}\Imag s^{\ix{Ret}} (0)
    \nonumber \\ 
    & & + \frac{1}{2 \pi} \left[ - 2 b_1^{IIa} - b_1^{IIb} + \frac32
      (b_1^{IIa} + b_{3}^{II}) \right] \Imag s^{\ix{Ret}} (0). \nonumber \\
    \label{eq:se9} 
  \end{eqnarray}

\end{widetext}

Comparing Eqs.~\eqref{eq:bII1} and \eqref{eq:bII3} with \eqref{eq:ap1}
and \eqref{eq:ap3} we establish that
\begin{eqnarray}
  b_{1}^{IIa} &=& -(eV)^2 \Imag \left[ \frac{\partial
      \bar{a}_{1}}{\partial \omega} \right]_{\omega =0}, \\  
  b_{1}^{IIb} &=& - 4 b_{1}^{IIa}, \\
  b_{3}^{II} &=& -(eV)^2 \Imag \left[ \frac{\partial a_{3}}{\partial
      \omega} \right]_{\omega =0}. 
\end{eqnarray}
These relations  allow us to express \eqref{eq:se9} as 
\begin{eqnarray}
  & & \frac{\df}{\df \lambda} \Imag \Sigma_{\sigma}^{\ix{Ret}} (0)
  \nonumber \\ 
  &=& - \frac{3}{2 \pi} \frac{3 (eV)^2}{4}  \Imag \left[
    \frac{\partial \bar{a}_1}{\partial \omega} + \frac{\partial
      a_3}{\partial \omega}\right]_{\omega=0}\Imag s^{\ix{Ret}} (0). 
\end{eqnarray}
After comparison with \eqref{eq:sedd}  it becomes clear that
\begin{equation}
  \Imag \frac{\partial^2 \Sigma_{\sigma}^{\ix{Ret}}}{\partial V^2}
  \bigg|_{\omega,V=0} = \frac34 \Imag \Sigma'', 
\end{equation}
which agrees with \eqref{eq:SIAM_SE_expansion} and
Ref.~\onlinecite{oguri}.




\begin{thebibliography}{99}
\bibitem{costi} The NRG data presented here and in the following
  figures have been created with a ready-to-use program by courtesy of
  Dr. Theo Costi, Institute of Solid State Research, Forschungszentrum
  J\"ulich.
\end{thebibliography}
\end{document}